\pdfoutput=1  

\documentclass[aps,prd,superscriptaddress,floatfix,nofootinbib,preprintnumbers,eqsecnum,twocolumn]{revtex4-1}
\usepackage[utf8]{inputenc}
\usepackage{graphicx}
\usepackage{amssymb,amsmath,multirow}
\usepackage{comment,color,placeins,centernot,amsmath,amsfonts,enumerate,bm}
\usepackage{lipsum}
\usepackage{slashed}
\usepackage{mathtools}

\usepackage{hyperref}
\usepackage[usenames,dvipsnames]{xcolor}
\hypersetup
{
  colorlinks,
  linkcolor={blue!80!black},
  citecolor={blue!70!black},
  urlcolor={blue!70!black}
}

\usepackage{xparse}
\newsavebox{\fminipagebox}
\NewDocumentEnvironment{fminipage}{m O{\fboxsep}}
 {\par\kern#2\noindent\begin{lrbox}{\fminipagebox}
  \begin{minipage}{#1}\ignorespaces}
 {\end{minipage}\end{lrbox}%
  \makebox[#1]{%
    \kern\dimexpr-\fboxsep-\fboxrule\relax
    \fbox{\usebox{\fminipagebox}}%
    \kern\dimexpr-\fboxsep-\fboxrule\relax
  }\par\kern#2
}

\makeatletter
\renewcommand*\env@matrix[1][\arraystretch]{%
  \edef\arraystretch{#1}%
  \hskip -\arraycolsep
  \let\@ifnextchar\new@ifnextchar
  \array{*\c@MaxMatrixCols c}}
\makeatother

\newcommand{\alignedfrac}[2]{%
    \setbox0\hbox{$#1$}        
    \dimen0=\wd0               
    \setbox1\hbox{$#2$}        
    \dimen1=\wd1               
    \ifdim\wd0<\wd1            
        \dfrac{#1\hfill}{#2}   
    \else                      
        \dfrac{#1}{#2\hfill}   
    \fi
}

\def\beq{\begin{equation}}
\def\eeq{\end{equation}}

\newcommand{\il}[1]{\mbox{$#1$}} 

\newcommand{\nn}{\nonumber}
\newcommand{\abs}[1]{\left| #1 \right|}

\newcommand{\p}{\partial}
\newcommand{\expt}[1]{\left\langle #1 \right\rangle} 
\newcommand{\phidot}{\dot{\phi}}
\newcommand{\phiddot}{\ddot{\phi}}

\newcommand{\e}{$e$}

\def\ie{{\it i.e.}\/}
\def\eg{{\it e.g.}\/}
\def\etc{{\it etc}.\/}
\def\phican{\widetilde{\phi}}
\def\hcan{\widetilde{h}}
\def\phiamp{\overline{\phi}}
\def\varphiamp{\overline{\varphi}}

\newcommand{\GeV}{\text{GeV}}

\newcommand{\etal}{{\em et al.}}

\newcommand{\hvar}{\langle h^2 \rangle}
\newcommand{\phivar}{\langle \phi^2 \rangle}
\newcommand{\chvar}{\langle \mathcal{H}^2 \rangle}
\newcommand{\cphivar}{\langle \varphi^2 \rangle}
\newcommand{\xiphi}{\xi_{\phi}}
\newcommand{\xih}{\xi_{h}}
\newcommand{\xieff}{\xi}
\newcommand{\Lphi}{\lambda_{\phi}}
\newcommand{\Lh}{\lambda_{h}}
\newcommand{\RJ}{R_{\rm J}}
\newcommand{\RE}{R_{\rm E}}
\newcommand{\VJ}{V_{\rm J}}
\newcommand{\VE}{V_{\rm E}}
\newcommand{\ggOmm}{g^2\hspace{-0.65mm}/\hspace{-0.2mm}m_{\phi}^2}
\newcommand{\ggOL}{g^2\hspace{-0.65mm}/\hspace{-0.2mm}\lambda_{\phi}}
\newcommand{\ggOLexp}{g^2\hspace{-0.65mm}/\hspace{-0.4mm}\lambda_{\phi}}
\newcommand{\phiend}{\phi_{\rm end}}

\newcommand{\xend}{x_{\rm end}}
\newcommand{\xNL}{x_{\rm NL}}
\newcommand{\xdec}{x_{\rm dec}}
\newcommand{\mumax}{\mu_{\rm max}}
\newcommand{\Hk}{\mathcal{H}_k}
\newcommand{\omegaHk}{\omega_{\mathcal{H}_k}}
\newcommand{\hcrit}{h_{\rm crit}}
\newcommand{\xcross}{x_{\times}}
\newcommand{\muh}{\mu_h}
\newcommand{\xxi}{x_{\xi}}
\newcommand{\xg}{x_g}
\newcommand{\rh}{r_{h}}
\newcommand{\rphi}{r_{\phi}}
\newcommand{\rhotot}{\rho_{\rm tot}}
\newcommand{\mphi}{m_{\phi}}

\DeclareMathOperator{\sn}{sn} 
\DeclareMathOperator{\cn}{cn}

\begin{document}

\title{Massless Preheating and Electroweak Vacuum Metastability}

\def\andname{\hspace*{-0.5em}} 
\author{Jeff Kost}
\email[Email address: ]{j.d.kost@sussex.ac.uk}
\affiliation{\mbox{Department of Physics \& Astronomy, University of Sussex, Brighton BN1 9QH, United Kingdom}}
\affiliation{\mbox{Center for Theoretical Physics of the Universe, Institute for Basic Science, Daejeon 34126 Korea}}
\author{Chang Sub Shin}
\email[Email address: ]{csshin@cnu.ac.kr}
\affiliation{\mbox{Center for Theoretical Physics of the Universe, Institute for Basic Science, Daejeon 34126 Korea}}
\affiliation{\mbox{Department of Physics and Institute of Quantum Systems (IQS), Chungnam National University, Daejeon 34134, Korea}}
\author{Takahiro Terada}
\email[Email address: ]{takahiro@ibs.re.kr}
\affiliation{\mbox{Center for Theoretical Physics of the Universe, Institute for Basic Science, Daejeon 34126 Korea}}

\preprint{CTPU-PTC-21-20}

\begin{abstract}
    Current measurements of Standard Model parameters suggest that 
    the electroweak vacuum is metastable.  This metastability has important 
    cosmological implications because large fluctuations in the Higgs field
    could trigger vacuum decay in the early universe.
    For the false vacuum to survive, interactions which stabilize the 
    Higgs during inflation---\eg, inflaton-Higgs interactions
    or non-minimal couplings to gravity---are typically necessary.  
    However, the post-inflationary preheating dynamics of these same interactions 
    could also trigger vacuum decay, thereby recreating the problem we sought to avoid.
    This dynamics is often assumed catastrophic for models exhibiting
    scale invariance, since these generically allow for unimpeded growth of fluctuations.
    In this paper, we examine the dynamics of such ``massless preheating'' 
    scenarios and show that the competing threats to metastability 
    can nonetheless be balanced to ensure viability.
    We find that fully accounting for both the backreaction from particle 
    production and the effects of perturbative decays reveals a large number of disjoint 
    ``islands of (meta)stability''  over the parameter space of couplings.
    Ultimately, the interplay among Higgs-stabilizing interactions plays a significant
    role, leading to a sequence of dynamical phases that effectively 
    extend the metastable regions to large Higgs-curvature couplings.
\end{abstract}

\maketitle





\section{Introduction\label{sec:Intro}}


A remarkable implication of the currently measured Standard Model (SM)
parameters is that the electroweak vacuum is metastable.
Specifically, given the measured Higgs boson and top quark masses~\cite{Zyla:2020zbs},
one finds that at energy scales exceeding \il{\mu\approx 10^{10}\,\GeV}
the Higgs four-point coupling runs to negative values \il{\Lh(\mu) < 0}, 
signifying the existence of a lower-energy 
vacuum~\mbox{\cite{Degrassi:2012ry,Bezrukov:2012sa,Alekhin:2012py,Buttazzo:2013uya,Bednyakov:2015sca}}.
Although today the timescale for vacuum decay is much 
longer than the age of the universe~\mbox{\cite{Andreassen:2017rzq,Chigusa:2018uuj}}, 
dynamics earlier in the cosmological history could have 
significantly threatened destabilization.
That the false vacuum has persisted until the present day
may thus provide a window into early-universe dynamics
involving the Higgs~\cite{Markkanen:2018pdo}.

In this respect, the evolution of the Higgs field during inflation
is especially relevant.  During inflation, light scalar 
fields develop fluctuations proportional to the Hubble scale $H$.
Without some additional stabilizing interactions, the fluctuations 
of the Higgs would likewise grow and inevitably trigger decay of the 
false vacuum, unless the energy scale of inflation is 
sufficiently small~\cite{East:2016anr,Markkanen:2018pdo,Fumagalli:2019ohr}.

Metastability thus provides a strong motivation for
investigating non-SM interactions that could stabilize the Higgs
during inflation---\eg, non-minimal gravitational couplings~\cite{Espinosa:2007qp}, 
direct Higgs-inflaton interactions~\cite{Lebedev:2012sy}, \etc{}
However, the situation is actually more delicate, 
as one must also ensure metastability throughout the remaining cosmological history.
While interactions such as those listed above may stabilize the vacuum
during inflation, they often proceed to \emph{destabilize} 
it during the post-inflationary preheating epoch,
thereby recreating the problem we sought to avoid.
Indeed, a balancing is typically necessary between the destabilizing
effects of inflationary and post-inflationary dynamics.

To this end, a detailed understanding of the field dynamics
after inflation is essential in determining which interactions 
and couplings are well motivated overall.  
A decisive component of this analysis is the form of the 
inflaton potential $V(\phi)$ during preheating.
After inflation, the inflaton field $\phi$ oscillates coherently
about the minimum of its potential.  These oscillations
determine the properties of the background cosmology, but they 
also furnish the quantum fluctuations of the Higgs field 
with time-dependent, oscillatory effective masses.  
These modulations are the underlying mechanism 
for Higgs particle production, as they ultimately give 
rise to non-perturbative processes such as tachyonic 
instabilities~\mbox{\cite{Felder:2000hj,Felder:2001kt,Dufaux:2006ee}}
and parametric resonances~\mbox{\cite{Kofman:1994rk,Kofman:1997yn,Greene:1997fu}}.
These processes can enormously amplify the field fluctuations over
the course of merely a few inflaton oscillations.

For most inflaton potentials---such as those which are quadratic 
after inflation---particles are produced within 
bands of comoving momentum, and these bands evolve non-trivially
as the universe expands.  The rates of particle production 
for these bands also evolve and generally 
weaken as preheating unfolds.  As a rule, the particle production
terminates after some relatively short time,  and one can
classify a model that remains metastable over the full 
duration as phenomenologically 
viable~\cite{Herranen:2015ima,Ema:2016kpf,Kohri:2016wof,Enqvist:2016mqj,Ema:2017loe}.

However, there is a notable exception to this rule: models which
exhibit scale invariance.  Under a minimal set of assumptions, 
in which the inflaton is the only non-SM field, the
scalar potential during preheating is restricted to the following interactions:
\beq
    V(\phi,h) ~=~ \frac{1}{4}\Lphi\phi^4 + \frac{1}{2}g^2\phi^2h^2 + \frac{1}{4}\Lh h^4 \ ,
    \label{eq:introV}
\eeq
and the epoch is termed ``massless preheating.''
Note that even if the inflaton-Higgs interaction does not appear as
a direct coupling, it should be generated radiatively since inflaton-SM 
interactions are generally necessary to reheat the universe~\cite{Gross:2015bea}.
Additionally, we emphasize that we have made no assumptions regarding the 
potential in the inflationary regime, except that it 
smoothly interpolates to Eq.~\eqref{eq:introV} at the end of inflation.

Above all, the scale invariance implies that the 
system's dynamical properties are independent of the 
cosmological expansion, in stark contrast to other preheating scenarios.
As a result, the salient features for vacuum metastability, 
in general, do not evolve: particles are produced 
within momentum bands that are fixed with time, 
and the production rates for the modes do not change.  
The field fluctuations grow 
steadily and \emph{unimpeded}~\cite{Greene:1997fu}.

At first glance, the unimpeded growth in massless preheating 
appears catastrophic for electroweak vacuum metastability.  
That said, several considerations should be evaluated
more carefully before reaching this conclusion.  First, the Higgs 
particles produced during preheating have an effective mass 
proportional to the inflaton background value \il{m_h \simeq g|\phi|}.
This dependence can significantly enhance the perturbative decay rate 
of the Higgs to SM particles.\footnote{The interplay between non-perturbative
production and perturbative decays has been studied in a 
variety of contexts~\mbox{\cite{
    Kasuya:1996np,
    Felder:1998vq,
    Bezrukov:2008ut,
    GarciaBellido:2008ab,
    Mukaida:2012bz,
    Repond:2016sol,
    Fan:2021otj}.
}}
For large enough coupling, the decay rate and the 
production rate could be comparable,
thereby checking the growth of Higgs fluctuations and effectively 
stabilizing the vacuum.
Second, as particles are produced, their backreaction
modifies the effective masses of the field fluctuations.
If the vacuum does not decay first, the energy density of the
produced particles inevitably grows enough to disrupt
the inflaton oscillations, terminating the parametric resonance 
and ushering in the non-linear phase of dynamics that follows.

In this paper, we assess the viability of massless 
preheating in the context of electroweak vacuum metastability
and address each of the above questions.   But our analysis
includes an important generalization: we allow non-minimal 
gravitational couplings for both the inflaton and the Higgs.
In some sense, these couplings are unavoidable
since if we ignore them in the tree-level action,
they are generated at loop level~\cite{Buchbinder:1992rb};
even so, we are inclined to include them for several other reasons.
First, a non-minimal Higgs-curvature coupling provides an additional
stabilizing interaction for the Higgs during inflation
without introducing additional non-SM field content.
Second, a non-minimal inflaton-curvature coupling allows us 
to present a complete and self-contained model.
The effect of the curvature coupling is to flatten the inflaton
potential in the large-field region, restoring viability 
to the quartic inflaton potential in the inflationary regime,
which is otherwise excluded by observations of the 
cosmic microwave background (CMB)~\cite{Planck:2018jri}.
Furthermore, during inflation, the non-minimal gravitational 
interaction is effectively scale invariant,
so such couplings fit naturally within the purview of scale-invariant 
theories of inflation~\cite{
    Spokoiny:1984bd,
    Futamase:1987ua,
    Salopek:1988qh,
    Fakir:1990eg,
    Bezrukov:2007ep,
    Salvio:2014soa,
    Csaki:2014bua,
    Kannike:2015apa}.  
In this way, our study provides an analysis of the preheating
dynamics that emerges from this well-motivated class of inflationary models.  

Along with the generalization of non-minimal gravitational couplings,
we shall consider a generalization of the gravity formulation.
That is, in addition to the conventional metric formulation,
we consider the so-called Palatini formulation, in which the connection
and metric are independent degrees of freedom~\cite{Palatini:1919,Einstein:1925}.
In a minimally coupled theory, such a distinction is not pertinent, 
but with non-minimal curvature couplings, there 
is a physical distinction between these formulations~\cite{Bauer:2008zj};
we shall consider certain facets of both in this paper.

Undoubtedly, the presence of non-minimal curvature couplings can 
have a substantial impact on the preheating dynamics.  
Because the curvature interactions break scale invariance 
during preheating, particle production from these terms 
dissipates over time and terminates, unassisted 
by backreaction effects.  Consequently, if the initial curvature 
contribution is at least comparable to the quartic contribution, 
the system experiences a \emph{sequence of dynamical phases}: 
particle production is dominated by the former immediately after 
inflation and transitions to the latter after some relatively short 
duration.  Moreover, even if the curvature coupling is small, it 
can play a significant role.  The curvature interaction
contributes either constructively or destructively to the effective 
mass of the Higgs modes, depending on the sign of the coupling. 
As a result, the generated Higgs fluctuations can have
an orders-of-magnitude sensitivity to this sign, which is ultimately
reflected in the range of curvature couplings that are most vacuum-stable.

By broadly considering the dynamics of such massless preheating models,
we place constraints on the space of Higgs couplings for
which the metastability of the electroweak vacuum survives.
And contrary to constraints that arise in other scenarios,
we do not find a simply connected region.
Instead, we find a large number of disjoint ``islands of (meta)stability''
over the parameter space, which merge into a contiguous
metastable region at large quartic coupling.  
Accordingly, unlike other preheating scenarios---which typically lead to
an upper bound on the Higgs-inflaton coupling---we find 
that massless preheating requires a more complex constraint,
and a \emph{lower bound} ultimately describes the most favorable 
region of metastability.  In this way, the constraints necessary 
to stabilize the Higgs during inflation and stabilize it during 
preheating work in concert rather than in opposition.

This paper is organized as follows.  
We begin Sec.~\ref{sec:Model} with an overview of
the class of models we consider for our study
of massless preheating and our assumptions therein.
We show how the inclusion of 
non-minimal gravitational couplings allows for a 
self-contained model with a viable inflationary regime in
the large-field region.  The post-inflationary evolution 
of the inflaton and other background quantities is also discussed.
In Sec.~\ref{sec:NoBackreaction}, we examine the production
of Higgs and inflaton particles in this background
without yet considering the backreaction from these processes.
At the close of the section, we analyze the effect of perturbative 
decays on the growth of Higgs fluctuations.
In the penultimate Sec.~\ref{sec:Backreaction}, we investigate 
backreaction and the destabilization of the electroweak vacuum.
We delineate the resulting metastability constraints on models 
in which massless preheating emerges.
Finally, in Sec.~\ref{sec:Conclusions}, we summarize our main
results and possible directions for future work.

This paper also includes three appendixes.
In Appendix~\ref{sec:quadratic}, we provide an
overview of vacuum metastability in the context of ``massive preheating,''
where the inflaton potential is approximately quadratic after inflation
and the scale invariance of the inflaton potential broken.
These scenarios have been studied in the 
literature~\mbox{\cite{Herranen:2015ima,Ema:2016kpf,Enqvist:2016mqj,Ema:2017loe}},
and we reproduce their findings for the purpose 
of comparing and contrasting to our results.
Meanwhile, in Appendix~\ref{sec:tachyonic}, we include the analytical calculations 
for the tachyonic resonance which are not given in the main body of 
the paper, and in Appendix~\ref{sec:numerics}, we provide the relevant
details of our numerical methods.


\section{The Model\label{sec:Model}}


Let us consider a model in which the Higgs doublet $H$ is coupled to 
a real scalar inflaton $\phi$, and both of these fields 
have a non-minimal coupling to the Ricci scalar curvature $R$.  
The model is described most succinctly 
in the Jordan frame by the Lagrangian\footnote{
    We employ a system of units 
    in which \il{c \equiv \hbar \equiv 8\pi G \equiv 1}
    and adopt the $(+,+,+)$ sign convention of Misner \etal~\cite{Misner:1974qy}.  
    Additionally, we follow a sign convention for the 
    non-minimal couplings $\xi_X$ (\il{X = \phi, h}) 
    such that the conformal value is \il{\xi_X = 1/6}.
}
\begin{align}
    \label{eq:model}
    \mathcal{L}_{\rm J} ~=~  -&\frac{1}{2}(\p^{\mu}\phi)^2 - \frac{1}{2}(\p^{\mu}h)^2 - \VJ(\phi,h) \nn \\
    &\qquad + \frac{1}{2}\big(1 - \xiphi\phi^2 - \xih h^2\big)\RJ  \ ,
\end{align}
where we have used that \il{H=\begin{pmatrix}0 & h/\sqrt{2}\end{pmatrix}^{\hspace{-0.7mm}\rm T}} 
in the unitary gauge.\footnote{\label{fn:unitary_gauge}The initial state is symmetric under 
the $SU(2)_L\times U(1)_Y$ gauge transformation. However, once the fluctuation 
of the Higgs-doublet field is amplified by the instability, 
it easily becomes classical with a finite Higgs expectation value. 
Thus, the radial mode and the phase directions are well defined at each local position, 
and this justifies our use of the unitary gauge. 
Additionally, note that the initial quantum fluctuations cause the orientations 
of the Higgs field to be randomly distributed, resulting in nontrivial 
topological configurations (Chern-Simons number). 
This leads to gauge-field production~\cite{Dufaux:2010cf}, 
the effects of which are an interesting topic for future work.}  
The potential is restricted to the 
scale-invariant interactions motivated in Sec.~\ref{sec:Intro}:
\beq\label{eq:VJ}
\VJ(\phi,h) ~=~ \frac{1}{4}\Lphi\phi^4 + \frac{1}{2}g^2\phi^2h^2 + \frac{1}{4}\Lh h^4 
\eeq
and now interpreted in the Jordan frame.
The first term in Eq.~\eqref{eq:VJ} determines the evolution of the 
cosmological background, while the second term is
meant to stabilize the electroweak vacuum during inflation.  
The last term is the source of electroweak instability, 
with $\Lh$ running to negative values at energy scales exceeding
\il{\sim 10^{10}\,\GeV}.

Let us make some ancillary remarks on the 
gravitational formulation used in what follows.
In principle, one can formulate general relativity in 
two ways: (i) using the metric $g_{\mu\nu}$ and 
taking the connection $\Gamma^{\alpha}_{\beta\mu}$ 
to be given by the Christoffel symbols
(the ``metric formulation'') or (ii) using the metric
and the connection as independent 
degrees of freedom (the ``Palatini formulation'').
Of course, these formulations are physically equivalent 
for a minimally coupled theory, but this distinction carries weight
given the non-minimal couplings in our theory.
Not favoring one formulation over the other, we
address both of these possibilities by introducing a parameter 
\beq
    \theta ~\equiv~
    \begin{cases}
        0\ , & \text{Palatini formulation} \\
        1\ , & \text{metric formulation}
    \end{cases} \ ,
    \label{eq:theta}
\eeq
and when relevant, we shall state our results as functions of $\theta$.
Although the distinctions between these formulations can be realized in
both the preheating and inflationary dynamics, they figure most prominently
in the large-field region of the potential, including the inflationary regime;
let us now focus our discussion on this regime.

\subsection{The Inflationary Regime\label{subsec:InflationaryRegime}}

In principle, the model we have stated in Eq.~\eqref{eq:model} is agnostic 
to the specific form of the inflaton potential in the inflationary regime.
As long as basic criteria are satisfied---\eg, the vacuum remains 
stable during inflation, isocurvature perturbations are 
negligible, \etc---our study of the preheating epoch is insensitive to 
details of the inflation model.  Even so, we are motivated to 
examine whether Eq.~\eqref{eq:model} could consist of a 
phenomenologically viable inflation potential in the large-field region, 
as this would furnish a complete and self-contained model.

On the one hand, a minimal coupling \il{\xiphi = 0}
would be the simplest possible scenario.
Unfortunately, this yields the standard quartic inflation,
which is known to predict too large a tensor-to-scalar ratio 
to satisfy observations of the CMB~\cite{Planck:2018jri}.
On the other hand, a finite \il{\xiphi < 0} allows for a range 
of other possibilities~\cite{
    Spokoiny:1984bd,
    Futamase:1987ua,
    Salopek:1988qh,
    Fakir:1990eg,
    Bezrukov:2007ep}.  In considering these, 
it is beneficial to transform to the Einstein frame, 
defined by the Weyl transformation 
\il{g_{\mu\nu}\!\longrightarrow \Omega g_{\mu\nu}} in which\footnote{
A covariant framework in which the equivalence of the Jordan
and Einstein frames are manifest is discussed 
in Ref.~\cite{Karamitsos:2017elm}, where extension beyond the tree-level
equivalence is delineated.
}
\beq
\Omega ~\equiv~ 1 - \xiphi \phi^2 - \xih h^2 \ .
\eeq
While this restores the canonical graviton normalization,
it also transforms the potential to \il{\VE(\phi,h) = \VJ(\phi,h)/\Omega^2}
and generates non-canonical kinetic terms for our scalar fields \il{X = \{\phi, h\}} 
such that the Lagrangian becomes
\beq
\mathcal{L}_{\rm E} ~=~ \frac{1}{2}\RE - \frac{1}{2}\sum_{ij}(\p^{\mu}X_i)\mathbf{K}_{ij}(\p_{\mu}X_j) - \VE(\phi,h) 
\eeq
with a kinetic mixing matrix 
\beq
    \mathbf{K} ~\equiv~ 
    \frac{1}{\Omega^2}
    \begin{bmatrix}
        \Omega  + \frac{3}{2}\theta(\p_{\phi}\Omega)^2 & \frac{3}{2}\theta\p_{\phi}\Omega\p_h\Omega \\  
        \frac{3}{2}\theta\p_{\phi}\Omega\p_h\Omega  & \Omega + \frac{3}{2}\theta(\p_h\Omega)^2 
    \end{bmatrix} \ .
\eeq

\begin{figure}[tb]
    \centering
    \includegraphics[keepaspectratio,width=0.49\textwidth]{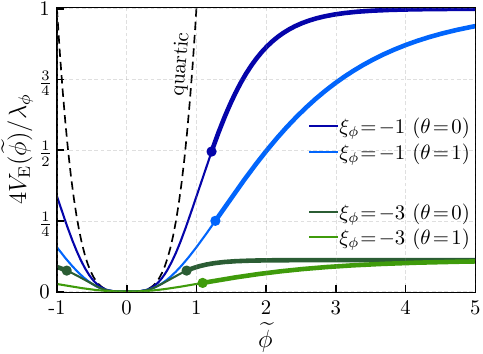}
    \caption{The inflaton potential for different curvature couplings $\xiphi$ 
    and gravity formulations $\theta$, all of which are approximately 
    quartic during preheating but vary substantially in the large-field region.
    The thick segment of each curve shows the inflationary regime 
    of the potential, with each point marker indicating 
    the end of inflation [coincident with Eq.~\eqref{eq:phiend}].
    The quartic potential in the absence of a non-minimal gravitational coupling 
    is shown (black-dashed curve) for reference.
    }
\label{fig:Vphi}
\end{figure}

In the large-field region \il{\phi\gg 1/\sqrt{-\xiphi}}, the inflaton potential
then takes the form 
\beq
    \VE(\phican) ~=~ \frac{\Lphi}{4 \xiphi^2} 
    \left\{
    \begin{array}{lll}
        \displaystyle{\tanh^4 \!\big( \sqrt{-\xiphi} \phican \big)} & \text{for} & \theta = 0 \\[2.5pt]
        \displaystyle{\big( 1 - e^{- \sqrt{\frac{2}{3}}\phican}\big)^{\!2}} & \text{for} & \theta = 1 
    \end{array}
    \right. \ ,
\eeq
where $\phican$ is the canonically normalized inflaton field.
In this region, the theory acquires an approximate scale invariance
and effectively approaches so-called ``attractor models'' which are 
phenomenologically viable~\cite{Kallosh:2013yoa,Galante:2014ifa,Carrasco:2015pla}.
In particular, for an inflationary epoch of $N$ \e-folds, one finds 
a spectral index \il{n_\text{s} = 1 - 2/N} and tensor-to-scalar ratio \il{r = 12\alpha/N^2}, 
where we have defined \il{\alpha \equiv \theta - 1/(6\xiphi)}.

A plot which illustrates the inflaton potential $\VE(\phican)$
for a number of possible curvature couplings $\xiphi$ and gravity 
formulations $\theta$, all of which lead to massless preheating
after inflation, is shown in Fig.~\ref{fig:Vphi}.
The inflationary trajectory is shown by the thick part of each curve,
and inflation ends at the location of the point marker.

There are several constraints on our model parameters 
necessary to achieve viability.
First, matching the observed amplitude of primordial curvature 
perturbations~\cite{Planck:2018jri} fixes the relationship between $\Lphi$ and $\xiphi$:
\beq
    \Lphi ~=~ 4.9\times 10^{-10} \left(\frac{55}{N}\right)^2 \alpha \xiphi^2 \ .
    \label{eq:Lphiconstraint}
\eeq
Second, we must ensure that the electroweak vacuum remains stable throughout inflation.
This requirement bounds the effective mass squared of the Higgs as
\begin{align}
    \left.\frac{\p^2 \VE}{\p \hcan^2}\right|_{\hcan=0} 
    \!\!\simeq~ \frac{g^2 \phi^2}{1 - \xiphi \phi^2} + \frac{\xih \Lphi \phi^4}{(1 -\xiphi \phi^2)^2} ~>~ 0 
    \label{eq:higgscurvature}
\end{align}
under the slow-roll approximation, 
where $\hcan$ is the canonically normalized Higgs field.
The slow-roll-suppressed contributions to the Higgs mass
neglected above can become more important toward the end of inflation,
so the precise condition may be modified.  In some inflation models,
this causes inflation to end prematurely~\cite{Renaux-Petel:2015mga};
this may happen for \il{\xi_\phi \ll -1} in the Palatini formalism, which
is outside the scope of this paper.
The constraint in Eq.~\eqref{eq:higgscurvature} is simplified if we consider that during 
inflation \il{1 \ll -\xiphi \phi^2} and thus Eq.~\eqref{eq:higgscurvature}
reduces to \il{\Lphi\xih - g^2 \xiphi > 0}.  
We shall assume that this constraint is satisfied to the 
extent that the Higgs is stabilized strongly at the origin during
inflation.  In other words, we assume that the Higgs mass is larger 
than the Hubble scale $H$ during inflation, and imposing 
this assumption, we obtain 
\begin{align}    
    \xih - \frac{g^2}{\Lphi}\xiphi ~\gg~ \frac{1}{12} \ .
    \label{eq:stabilityinflation}
\end{align}

Third, we must also ensure that quantum corrections to the 
scalar potential from Higgs loops are controlled and do not 
ruin the above predictions for inflationary observables.
To this end, if we require that \il{g^2 \ll \sqrt{\lambda_\phi}},
then the Higgs-loop contribution is subdominant. 
Taking a value \il{\Lphi=10^{-10}} [based on Eq.~\eqref{eq:Lphiconstraint}]
gives the constraint explicitly as
\il{\ggOL \ll 10^5}, which is well within the 
parameter space that we shall consider.

Finally, the accelerated expansion of the universe
ends once the energy density of the inflaton background 
falls below \il{\rho_{\phi} = 3V(\phi)/2}, \ie, when the 
field falls below
\beq
    \phiend ~\equiv~ \frac{4}{\sqrt{1 + \sqrt{1 - 32\xiphi(1 - 6\theta\xiphi)}}} \ .
    \label{eq:phiend}
\eeq
Afterward, $\phi$ begins to oscillate about the minimum of the potential
and we identify this point in time as the beginning of the preheating 
era; we focus our discussion throughout the rest of the paper on this epoch.

\subsection{Cosmological Background After Inflation\label{subsec:InflatonEvolution}}

Let us now discuss the evolution of the cosmological background after 
the end of inflation.  Assuming the vacuum has been sufficiently stabilized,
the Higgs field is negligible,\footnote{
The two-field evolution that one finds in breaking from this assumption is non-trivial
and has been investigated in Ref.~\cite{Bond:2009xx}.
}
and the cosmological background is determined solely by the 
dynamics of the inflaton and its potential \il{\VE(\phi)\simeq V(\phi)} 
in the region \il{\phi < \phiend}, where the notation $V(\phi)$
is introduced for the inflaton potential approximated in the small-field region.

The form of the inflaton potential in this region may still be sensitive 
to $\xiphi$.  Notably, for \il{|\xiphi|\gg 1} an interesting distinction 
appears between metric and Palatini gravity.  For metric gravity,
the potential is quartic at sufficiently small \il{\phi \ll 1/\smash{|\xiphi|}} 
but becomes \emph{quadratic} in the intermediate region 
\il{1/|\xiphi| \ll \phi \ll 1/\smash{\sqrt{|\xiphi|}}}.  
It follows that Higgs fluctuations are amplified by two different types of
parametric resonance depending on the field region.\footnote{In this context,
we refer the reader to Ref.~\cite{Rusak:2018kel},
in which the electroweak instability was studied (with \il{g\neq 0} and \il{\xih=0})
in the large non-minimal inflaton coupling regime \il{-\xiphi \gg 1}.
}
By contrast, in Palatini gravity the potential is purely
quartic and has no quadratic region~\cite{Fu:2017iqg,Karam:2020rpa}.
Our model therefore offers an explicit example of \emph{purely massless 
preheating that is consistent with inflation}, even for
a large non-minimal coupling $\xiphi$.
Furthermore, we can achieve the same effect with \il{\xiphi = \mathcal{O}(1)},
as the intermediate quadratic region vanishes.  
Given that the study of massless preheating is our main interest
and that this occurs in the small-field region, we shall 
henceforth assume that \il{|\xiphi| \lesssim 1} (with \il{\xiphi < 0}).

After the end of inflation, the evolution of the background inflaton 
field is then well approximated by
\beq\label{eq:inflatoneqnofmotion1}
    \phiddot + 3H\phidot + \Lphi\phi^3 ~=~ 0 \ ,
\eeq
in which the dots correspond to time derivatives and \il{H \equiv \dot{a}/a} 
is the Hubble parameter given in terms of the scale factor \il{a=a(t)}.
As discussed in Sec.~\ref{sec:Intro}, the approximate scale invariance 
of the system makes its field dynamics and resonance structure rather unique.  
These features have been studied extensively~\cite{Greene:1997fu} and 
here we briefly review them.  The scale invariance is made transparent 
by writing Eq.~\eqref{eq:inflatoneqnofmotion1} 
in terms of the conformal time \il{\eta \equiv \int^t dt'/a(t')} and conformal inflaton 
field \il{\varphi \equiv a\phi}:
\beq\label{eq:inflatoneqnofmotion2}
\varphi''+ \Lphi \varphi^3 ~=~ 0 \ ,
\eeq
where the prime notation corresponds to $\eta$ derivatives.
Note that we have ignored a term proportional to \il{\phi^2 R},
as this term negligibly impacts the inflaton evolution.

The approximate scale invariance of the theory is manifested
by the fact that the inflaton equation of motion in Eq.~\eqref{eq:inflatoneqnofmotion2}
is independent of the cosmological expansion.
The solutions carry a constant amplitude $\varphiamp$ of
oscillations and are given in terms of a Jacobi elliptic function\footnote{
We define the Jacobi elliptic sine \il{\sn(X,Y)=\sin Z}
and cosine \il{\cn(X,Y)=\cos Z} functions through the relation
\begin{align}
    X ~=~ \int_0^Z \frac{d\theta}{\sqrt{1 - Y^2 \sin^2 \theta}} \ .
\end{align} 
}
\begin{align}
    \label{eq:varphi}
    \varphi(x) ~=~ \varphiamp \cn\hspace{-0.8mm}\left( x - x_0 , \frac{1}{\sqrt{2}} \right),
\end{align}
in which \il{x\equiv\smash{\sqrt{\Lphi}}\varphiamp \eta} is the
conformal time measured in units of the effective inflaton mass
$\sqrt{\Lphi} \varphiamp$.  The constant $x_0$ is used to match to 
the inflaton configuration at the end of inflation, and the 
period of oscillations in $x$ is
\beq
T ~\equiv~ 4 K\left(\frac{1}{\sqrt{2}}\right) ~\approx~ 7.42 \ ,
\eeq
with \il{\smash{K(X)\equiv \int_0^{\pi/2}\!d\theta (1-X^2\sin^2\!\theta)^{-1/2}}} 
defined as the complete elliptic integral of the first kind. 

A cosmological energy density which is dominated by the coherent oscillations 
of a scalar field may behave in a variety of ways depending on 
the scalar potential.  For example, the class of potentials 
\il{V(\phi)\propto \phi^{2n}} (for integer \il{n>0}) yield a cosmological 
background that behaves as a fluid with the equation-of-state parameter~\cite{Turner:1983he}
\beq\label{eq:eqnofstate}
w ~\equiv~ \frac{P}{\rho} ~=~ \frac{n-1}{n+1} \ ,
\eeq
in which $P$ is the pressure and $\rho$ is the energy density,
averaged over several oscillations.
For the quadratic (\il{n=1}) and quartic (\il{n=2}) potentials
this demonstrates the well-known result that 
scalar-field oscillations in these potentials correspond to
perfect fluids with matter-like (\il{w=0}) and 
radiation-like (\il{w=1/3}) equations of state, respectively.
This behavior reflects our observation in Eq.~\eqref{eq:inflatoneqnofmotion2} 
that $\varphi$ evolves independently of the cosmological expansion.
Namely, since the inflaton energy density 
scales like radiation \il{\rho_{\phi} \propto 1/a^4}, the amplitude 
of inflaton oscillations scale as \il{\phiamp \propto 1/a}. 
Therefore, the corresponding conformal amplitude $\varphiamp$ is fixed.

It also follows that in a radiation-like background the scale factor is 
proportional to $x$ and grows according to
\beq
    a(x) ~=~ \frac{\varphiamp\, x}{\sqrt{12}} \ .
\eeq

Inflation ends once the kinetic energy grows 
sufficiently to have \il{3V(\phi)/2 \leq \rho_{\phi}},
which corresponds to the time \il{\xend \equiv \sqrt{12}/\phiend}.
Note that chronologically one always has \il{x_0 < \xend}
and these are related explicitly by
\beq
\xend - x_0 ~=~ \text{arccn}\!\left(\!\left(\frac{2}{3}\right)^{\!\!\frac{1}{4}}\!\!,\frac{1}{2}\right) ~\approx~ 0.45 \ ;
\eeq
for simplicity we have used \il{\xiphi = 0} in this expression.

Finally, in addition to the inflaton background, it is 
important that we examine the scalar curvature after inflation. 
In general, the curvature is a frame-dependent quantity with
\il{\Omega R_\text{J} = 4\VJ(\phi) - \dot{\phi}^2} 
in the Jordan frame.  Nevertheless, at sufficiently small $\phi$
we have \il{\Omega \approx 1} and 
\begin{align}
    a^4R ~&=~ \Lphi\varphi^4 - \varphi'^2 \nn \\
    ~&=~ \frac{\Lphi}{2}\varphiamp{}^4\left[3\left(\frac{\varphi}{\varphiamp}\right)^{\!\!4} - 1\right] \ ,
    \label{eq:curvature}
\end{align}
where we have employed the solution in Eq.~\eqref{eq:varphi}.
The scalar curvature thus oscillates about zero and 
over several oscillations averages to \il{\langle R\rangle = 3H(1 - 3w)  = 0}.
However, as we shall find upon examining particle production,
neither the curvature terms nor their time dependence 
can be neglected, as they can impart a significant contribution to the 
preheating dynamics.


\section{Production of Higgs Particles\label{sec:NoBackreaction}}


Having established the evolution of the classical inflaton field in 
Sec.~\ref{subsec:InflatonEvolution}, we can now
discuss Higgs particle production in this background.
We write the quantized Higgs field $\widehat{h}$ in the Heisenberg picture 
as a function of the fluctuations $h_k(t)$ of comoving momenta $k$:
\beq
    \widehat{h}(\bm{x},t) =
    \!\int\! \frac{d^3k}{(2\pi)^{\frac{3}{2}}}\!\Big[\widehat{a}_kh_k(t)e^{+i\mathbf{k}\cdot\mathbf{x}} + \widehat{a}_k^{\dagger}h_k^{\!*}(t)e^{-i\mathbf{k}\cdot\mathbf{x}}\Big] \ ,
\eeq
where $\widehat{a}^{\dagger}_k$ and $\widehat{a}_k$ are creation and 
annihilation operators, respectively.
For a given comoving momentum, these fluctuations follow equations of motion
\beq
\ddot{h}_k + (3H + \Gamma_{h_k})\dot{h}_k + \omega^2_{h_k}h_k ~=~ 0
\eeq
where \il{\Gamma_{h_k}} is a phenomenological term accounting 
for perturbative decays of the Higgs~\mbox{\cite{Kofman:1997yn,GarciaBellido:2008ab,Repond:2016sol}} 
and $\omega_{h_k}$ is the energy of the mode.
These modes are coupled to both the oscillating inflaton 
background and the scalar curvature $R$, and therefore their
effective masses carry an implicit time dependence.  
In the Jordan frame,
\beq
\label{eq:omegahk}
\omega_{h_k}^2 ~=~ \frac{k^2}{a^2} + g^2\phi^2 + \xih R \ .
\eeq
However, in the Einstein frame, \il{\xiphi\neq 0} 
generates a kinetic mixing between the inflaton field and the Higgs modes, 
thereby producing an inflaton-dependent friction term.  We can absorb 
this friction into the effective mass term by the field redefinition 
\il{\mathcal{H}_k \equiv a \Omega^{-1/2}|_{h=0} h_k}~\cite{Rusak:2018kel},
yielding the equations of motion
\begin{align}
    \label{eq:Hkmodeeqn}
 \mathcal{H}''_k + a \Gamma_{h_k} \mathcal{H}'_k + \omega^2_{\mathcal{H}_k} \mathcal{H}_k = 0, 
\end{align}
in which the transformed modes are given by
\beq\label{eq:omegaHE}
    \omega^2_{\mathcal{H}_k} =~  k^2 + g^2 \varphi^2\!\left(\!1 + \xiphi \frac{\varphi^2}{a^2}\right) + \xieff a^2 R - a^2\Gamma_{h_k}H 
\eeq
and we have defined the effective non-minimal coupling
\beq\label{eq:xieff}
    \xieff ~\equiv~ \xih + \xiphi - 6 \theta \xih \xiphi - \frac{1}{6} \ .
\eeq
Note that we have neglected terms which are 
higher order in $a^{-1}$, such that the scalar curvature is given by
\il{a^4R = \Lphi\varphi^4 - (\varphi')^2}.
In this way, we have absorbed most $\xiphi$ effects and  
dependence on the gravity formulation into the single effective 
parameter $\xieff$.  Finally, we confirm that if we take \il{\xiphi = 0},
the scale invariance is restored for the conformal 
value \il{\xih = 1/6}, as one would expect.

Solving the equations of motion in Eq.~\eqref{eq:Hkmodeeqn}, we can
track the production of Higgs particles.  In particular,
the comoving phase-space density of particles associated with 
a mode of comoving momentum $k$ is given by
\beq
    n_{h_k} ~=~ \frac{\omegaHk}{2}\left( \frac{\abs{\mathcal{H}'}^2}{\omegaHk^2} + \abs{\mathcal{H}_k}^2\right) - \frac{1}{2} \ .
\eeq
The physical mechanism that drives this production differs considerably
between the $\xieff a^2 R$ and $g^2\varphi^2$ terms.  
For the former, when the inflaton field passes through the minimum
of its potential, one may find that a range of Higgs 
modes become tachyonic \il{\omega_{\mathcal{H}_k}^2 < 0}.
The tachyonic instability is strongest for the smaller-momentum modes, 
and these modes produce particles for longer durations.
By contrast, oscillations in $\varphi$ drive particle production 
from the latter.  The resulting time-dependent modulations 
of $\omegaHk$ give rise to parametric resonances for Higgs modes 
within certain momentum bands.

Another crucial distinction between these mechanisms 
is found by evaluating their overall scaling under the cosmological expansion.
The main term in Eq.~\eqref{eq:omegaHE} responsible for tachyonic production 
redshifts as \il{\xieff a^2 R \propto 1/a^2}, while the term responsible for 
the parametric resonance $g^2 \varphi^2$ does not dissipate at all.
Tachyonic production therefore always terminates after some finite time, 
even for the zero-momentum mode.
On the other hand, production from the parametric instability
continues unimpeded and ceases only once the evolution 
is disrupted by backreaction effects, as we cover in Sec.~\ref{sec:Backreaction}.
Indeed, this distinction is ultimately traced to the Higgs-curvature
interaction breaking the scale invariance that is preserved
by all of the other relevant terms.

We first break our analysis into two limiting cases: 
one in which the quartic inflaton-Higgs interaction is dominant 
and the other in which the curvature interaction is dominant.
Then, we explore the interplay between these interactions
and finally begin to analyze the impact of perturbative Higgs 
decays on the field dynamics.

\subsection{Production from Parametric Instability\label{subsec:Parametric}}

Let us first consider the case that the curvature coupling $\xieff$ is 
negligible and thus ignore the tachyonic production.  
Then, Higgs particles are produced purely from the 
parametric resonance associated with the $g^2\varphi^2$ term 
in Eq.~\eqref{eq:omegaHE}.  The dominant production in this case
arises from the fact that as the inflaton passes through the origin,
the effective masses may evolve non-adiabatically:
\beq\label{eq:adiabaticity}
    \frac{|\dot{\omega}_{\Hk}|}{\omegaHk^2} ~\gtrsim~ 1 \ ,
\eeq
which triggers a burst of particle production.\footnote{A number of
the results we present in this subsection are the subject of
Ref.~\cite{Greene:1997fu}; we summarize only the most relevant aspects.}

The growth of the number density for a given mode is exponential and 
follows \il{\log n_k \simeq 2\mu_k x} over several oscillations, where
\il{\mu_k} is the characteristic exponent.
Note the distinction between this regular exponential growth
and the stochastic growth one finds for theories without 
an approximate scale invariance, \eg, those with
a quadratic inflaton potential~\mbox{\cite{Kofman:1997yn, Greene:1997fu}}.
The stochastic nature of the resonance appears in these
scenarios because the accumulated phase of each mode evolves with 
the cosmic expansion, destroying the phase coherence.
In the scale-invariant theory, no such time dependence may arise
and phase coherence is maintained among the modes.
A more in-depth comparison of the quadratic and 
quartic theories, with an emphasis on the results 
of this paper, is provided in Appendix~\ref{sec:quadratic}.

\begin{figure}[tb]
    \centering
    \includegraphics[keepaspectratio,width=0.50\textwidth]{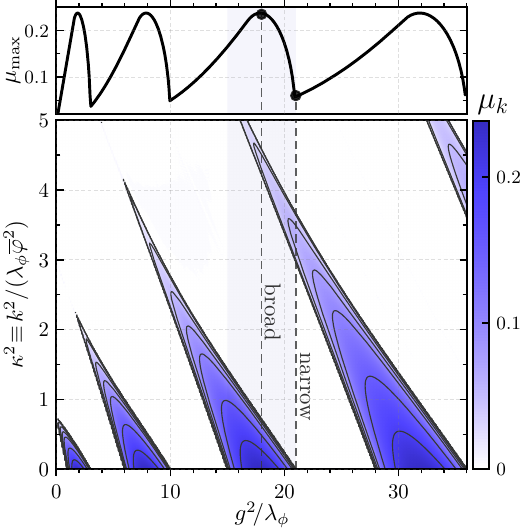}
    \caption{The instability bands of the parametric resonance 
    arising from the inflaton-Higgs interaction (bottom panel)
    and the corresponding maximum characteristic exponents \il{\mumax\equiv \max_k\mu_k}
    for each coupling (top panel). 
    The contours in the bottom panel show the value of the exponent $\mu_k$
    such that a given occupation number grows as \il{n_{h_k} \propto e^{2\mu_k x}}.
    The couplings which lead to the strongest growth are found at the center of bands 
    which have an unstable zero mode, \ie, for \il{\ggOL = 2, 8, \ldots, 2n^2}
    for \il{n \in \mathbb{N}}, while the weakest are found at the edge of
    these bands \il{\ggOL = 3, 10, \ldots, 2n^2+n}---we refer to these
    as the ``broad'' and ``narrow'' regimes, respectively.
    There is a universal \il{\mumax\approx 0.24} for the former,
    while for the latter $\mumax$ is a non-trivial function of $\ggOL$, 
    given in Eq.~\eqref{eq:narrowfit} and plotted over an extended range
    in Fig.~\ref{fig:narrowmumax}.
    }
\label{fig:bands}
\end{figure}

\begin{figure}[b]
    \centering
    \includegraphics[keepaspectratio,width=0.49\textwidth]{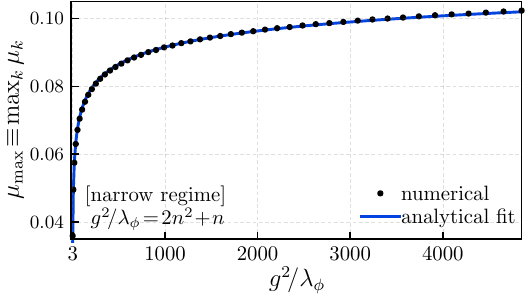}
    \caption{The maximum exponents \il{\mumax\equiv \max_k\mu_k}
    for the narrow-resonance couplings \il{\ggOL = 2n^2 + n} (for \il{n\in\mathbb{N}}).
    These are computed numerically (point markers) and compared to the 
    analytical fitting formula in Eq.~\eqref{eq:narrowfit} (solid curve).
    Note that the analytical function is evaluated only at the same
    discrete values \il{\ggOL = 2n^2 + n} in this figure, 
    \ie, the minima of the top panel of Fig.~\ref{fig:bands}.
    In the \il{\ggOL\rightarrow \infty} limit the
    broad/narrow regimes become degenerate with \il{\mumax \approx 0.24}, 
    but this asymptote is approached slowly.
    }
\label{fig:narrowmumax}
\end{figure}

The size of a particular growth exponent $\mu_k$ is determined by 
a combination of the comoving momentum $k$ and the quotient $\ggOL$.
In Fig.~\ref{fig:bands}, we have numerically solved the mode equations
and plotted contours of $\mu_k$ in the space of these two quantities, 
rescaling the momentum as \il{\kappa\equiv k/(\sqrt{\Lphi}\varphiamp)}.
It is natural to separate the resonances into two different classes.
The couplings with instability bands that include the zero-momentum 
mode, \ie, those between \il{2n^2-n < \ggOL < 2n^2 + n}, for 
\il{n\in \mathbb{N}}, contain the broadest resonances and generally 
give the most copious particle production---we refer to these 
collectively as the ``broad regime.''  Meanwhile, those couplings 
with only finite-momentum bands contain the most narrow resonances
and give generally weaker production, and we refer to these
as the ``narrow regime.''  The most weak and narrow 
bands occur at the boundaries \il{\ggOL = 2n^2 +n}.
The distinctions between these two coupling regimes 
have important dynamical implications and play a major role in this paper.

\begin{figure}[tb]
    \centering
    \includegraphics[keepaspectratio,width=0.49\textwidth]{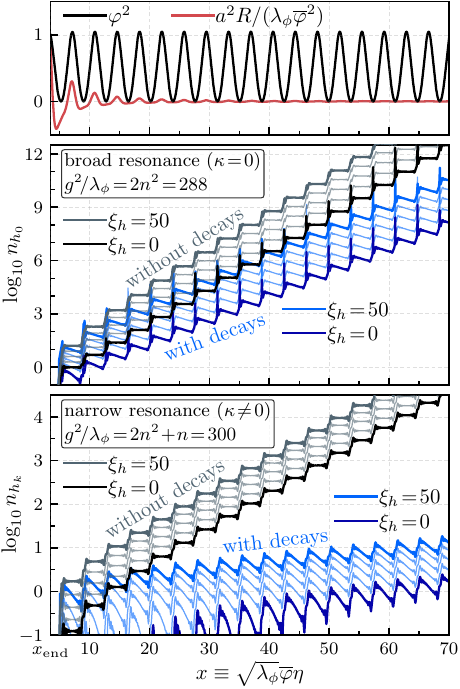}
    \caption{The evolution of background quantities $\varphi$, $a^2\!R$ (top panel)
    and Higgs phase-space density $n_{h_k}$ (center/bottom panels)
    during the early stage of preheating.
    The latter two panels show the broad/narrow regimes:
    taking \il{n=12}, the center panel uses a coupling 
    \il{\ggOL=2n^2} and has a broad range of resonant modes, 
    while the bottom panel uses the adjacent \il{\ggOL=2n^2+n} 
    and has only a narrow range of resonant modes;  
    the mode with the largest growth exponent is shown in each case.
    Additionally, these two panels show the effect of allowing for perturbative
    Higgs decays (first discussed in Sec.~\ref{subsec:PerturbativeDecays}, 
    shown in blue) and non-minimal gravitational couplings $\xih$, evenly 
    spaced over the range \il{0\leq \xih \leq 50}.
    }
\label{fig:ngrowth}
\end{figure}

Although, in principle, each mode with \il{\mu_k > 0} contributes to 
particle production, in practice the maximum exponent of the 
band \il{\mumax \equiv \max_k\mu_k} dominates.
In the broad regime, the maximum exponent is \il{\mumax \approx 0.24},
found at the central values \il{\ggOL = 2n^2}, and this
value is universal over the entire range of couplings.
Conversely, the $\mumax$ values in the narrow regime
are \emph{not} universal.  
These features are most easily observed in the 
top panel of Fig.~\ref{fig:bands}, where $\mumax$ is seen 
to oscillate back and forth between the broad and narrow regimes.
Indeed, the growth rate of particle production is \emph{highly 
non-monotonic} as one dials the inflaton-Higgs coupling.
In fact, the narrow resonances grow to join 
the broad resonance at \il{\mumax \approx 0.24} in the 
formal limit \il{\ggOL \rightarrow \infty}. 
The non-monotonicity of $\mumax$ thus diminishes as a function
of $\ggOL$, albeit at an extremely slow pace.  
In what follows it proves useful to quantify this observation, 
so we have numerically computed $\mumax$ for the discrete minima
of $\mumax$ in the narrow regime \il{\ggOL = 2n^2 +n} 
and found that they are well approximated by the function
\beq
    \displaystyle{\frac{\mumax}{0.24} ~\approx~ 
    \frac{\log(A\ggOL)}{1 + \log(B\ggOL)}} \ .
    \label{eq:narrowfit}
\eeq
The constants \il{A=2.82} and \il{B=3.34\times 10^5} 
reproduce the numerical results to better than \il{1\%} accuracy 
over the range \il{3\lesssim \ggOL \lesssim 10^4}, which spans the
full range of narrow resonances relevant to this paper.
We display the function in Eq.~\eqref{eq:narrowfit} and the numerical 
results together in Fig.~\ref{fig:narrowmumax} for comparison.
Note that the weakest resonances globally are found in the 
limit \il{\ggOL \rightarrow 0}, where the 
bands become increasingly narrow and 
follow \il{\mumax \approx 0.15\ggOL}~\cite{Greene:1997fu};
we shall discuss this small-coupling limit further
in Sec.~\ref{subsubsec:SmallCoupling}.

We have plotted numerical solutions of the phase-space
density $n_{h_k}$ in the broad and narrow regimes, 
for adjacent coupling bands, shown by the black curve 
in the top and bottom panels of Fig.~\ref{fig:ngrowth}, respectively.
The mode corresponding to the most rapid growth is shown in both cases:
for the broad regime this is the \il{\kappa = 0} mode, but for
the narrow regime $\mumax$ corresponds to a finite momentum.
We neglect the blue curves for the moment, as
these first enter our discussion in Sec.~\ref{subsec:PerturbativeDecays}.

For our purposes, the non-monotonic nature of $\mumax$ 
as a function of $\ggOL$ has extensive implications.  In contrast to 
many other preheating scenarios, the magnitude of our coupling has
no bearing on the growth rate of a given fluctuation---only the associated 
band within the repeating resonance structure is important.
That said, since the width of the momentum bands increases 
with $\ggOL$, the total number density $n_h$ does, in fact, depend on this coupling.
We obtain the total comoving number density of the produced Higgs particles
by using the saddlepoint approximation to integrate over each band:
\beq
n_h ~\simeq~ \frac{1}{2}\left(\!\frac{\sqrt{\Lphi}\varphiamp}{2\pi}\!\right)^{\!\!3} \left(\frac{g^2}{2\Lphi}\right)^{\!\!3/4} \!\frac{e^{2 \mumax x}}{\sqrt{\mumax x}} \ .
\label{eq:nhbroad}
\eeq
As this density partly determines the variance of the
Higgs fluctuations, it plays a major role in assessing 
the metastability of the electroweak vacuum.  Accordingly,
we shall continue to calculate $n_h$ in all of the regimes.

\subsection{Production from Tachyonic Instability\label{subsec:Tachyonic}}

Let us now consider the opposite coupling limit \il{g\rightarrow 0}, \ie, the limit
in which particle production is driven not by parametric resonance 
but by a tachyonic instability.
The effective masses $\omegaHk$ are given by 
\beq
    \frac{\omega_{\Hk}^2}{\Lphi \varphiamp{}^2} ~=~ \kappa^2 + \rh\left[3\left(\frac{\varphi}{\varphiamp}\right)^{\!\!4} - 1\right] \ ,
    \label{eq:omegaxi}
\eeq
where we have defined a quantity \il{\rh \equiv \xieff\varphiamp{}^2/(2a^2)}
that indicates the strength of the curvature term at a given time
and utilized Eq.~\eqref{eq:curvature}.
We recall from Sec.~\ref{subsec:InflationaryRegime} that
vacuum stability during inflation requires \il{\Lphi\xih > \xiphi g^2},
and, therefore, we consider only \il{\xih > 0} for the moment.
Indeed, for sufficiently small momenta, one finds modes that cross
the tachyonic threshold \il{\omegaHk^2 < 0} when the inflaton is near the minimum 
of its potential, triggering an exponential growth in the 
corresponding Higgs fluctuations.  Although this growth is typically short-lived
due to the redshifting of the curvature term \il{a^2\xieff R\propto 1/a^2}, 
these may still be a source of copious particle production soon after
inflation and thus serve as a legitimate threat to destabilize 
the electroweak vacuum.

There have been a number of studies devoted to calculating the rate of
tachyonic particle production in different 
settings~\mbox{\cite{Felder:2001kt, Dufaux:2006ee}},
and we employ several of those techniques in this section.
Similar to Sec.~\ref{subsec:Parametric}, near the turning
points \il{\omegaHk^2 = 0} the masses may change non-adiabatically 
and one must then compute the Bogoliubov coefficients to proceed.
However, unlike in the previous section, the $\omegaHk^2$ may experience
two distinct adiabatic segments of evolution.
As long as \il{\rh \gtrsim 1}, the effective masses 
change adiabatically away from the 
turning points in both the tachyonic and non-tachyonic segments.
Under this assumption, we can apply the WKB approximation 
by calculating the phase accumulated by modes both during
the tachyonic \il{X_k \equiv \int dt\,\Omega_{h_k}} 
(for \il{\smash{\Omega^2_k \equiv -\omega_{h_k}^2 > 0}})
and non-tachyonic \il{\Theta_k\equiv \int dt\,\omega_{h_k}} segments 
of evolution.

Applying these methods, one finds that after passing 
through a tachyonic region $j$ times, the phase-space density 
for a given mode is written generally as~\cite{Dufaux:2006ee}
\beq
    n_{h_k} ~=~ e^{2 j X_k} \left(2\cos\Theta_k\right)^{2(j-1)} \ .
    \label{eq:tachyonicformula}
\eeq
Hence, employing Eq.~\eqref{eq:omegaxi} we can compute the 
accumulated quantities $X_k$ and $\Theta_k$ for \il{g=0}.  
The details of this calculation
appear in Appendix~\ref{sec:tachyonic} and give
\begin{align}
    X_k ~&\simeq~ \frac{\sqrt{2\pi}\,\Gamma(\frac{5}{4})(\rh - \kappa^2)^{3/4}}{\Gamma(\frac{7}{4}) (3\rh)^{1/4}} \nn \\
    \Theta_k ~&\simeq~ \frac{4 \pi^{3/2} \sqrt{2 \rh + \kappa^2} }{\Gamma(\frac{1}{4})^2} \ .
    \label{eq:Xk}
\end{align}
After successive bursts of tachyonic production,
the growth exponent in Eq.~\eqref{eq:tachyonicformula} accumulates a value
\il{\sum_j 2j X_k \simeq 4\int dx X_k/T}, with the zero mode
receiving the greatest share of the number density.
The accumulated phases  $\Theta_k$ supply only oscillatory behavior 
or modify the distribution over momenta.  Given that our primary 
concern is the overall growth of the Higgs number density, we shall neglect
these quantities.  Then, we can estimate that
\begin{align}
    n_{h_k} ~\simeq~ \left(\frac{x}{x_0}\right)^{\!\!4\sqrt{\!\frac{2\xieff}{3\sqrt{3}}}}\!\exp\left[-\frac{2(x^2 - x_0^2)\kappa^2}{3^{9/4}\sqrt{\xieff}}\right] \ ,
    \label{eq:nhktachyonic}
\end{align}
which holds for the duration of the tachyonic instability.

Let us focus on the \il{\kappa = 0} mode, which experiences
the strongest growth in the tachyonic regime. 
At first glance, the growth may actually appear weak in 
comparison to the parametric resonance [in Eq.~\eqref{eq:nhbroad}]
since it is not exponential in time---it merely obeys a power law.
The difference is that the power scales with \il{\sqrt{\xieff}}
and has no upper bound, in line with studies 
of the tachyonic instability in other
settings~\mbox{\cite{Dufaux:2006ee, Ema:2016kpf}}.  This feature sharply
contrasts with the parametric resonance, in which the $\mumax$ growth rate 
is bounded \emph{universally} from above (as we observed in 
Fig.~\ref{fig:bands}), regardless of the coupling.  

Of course, as the universe expands the tachyonic production 
soon terminates.  In particular, as $\rh$ redshifts to values
below unity, the tachyonic masses $\Omega_{h_k}$ are suppressed
and the adiabatic assumption breaks down.
Using \il{|\dot{\Omega}_{h_k}|\gtrsim \Omega_{h_k}^2} as the 
threshold for where this breakdown occurs, we find
\il{\rh^2 \gtrsim (\rh - \kappa^2)^3}.
This condition implies that the span of modes exposed to the
instability is bounded above by \il{\kappa \lesssim \sqrt{\rh}}
and that a given mode is active for
\il{x \lesssim \sqrt{6\xieff}/\!\max(1,\kappa)}.
The modes shut down successively, starting with the largest-momentum
modes, such that tachyonic production ends at the time
\beq
    \xxi ~\simeq~ \sqrt{6\xieff} \ .
    \label{eq:xxi}
\eeq

The total number density of Higgs particles is given by 
integrating Eq.~\eqref{eq:nhktachyonic} over the phase space, but 
a cutoff \il{\kappa \lesssim \rh^2} should be imposed on each momentum
band per the discussion above.  We again use the saddlepoint approximation 
to perform the integration and obtain a comoving number density
\beq
    n_h ~\simeq~ \frac{1}{8}\Big(\sqrt{\Lphi}\varphiamp\Big)^3
    \left(\frac{3^{9\hspace{-0.2mm}/\hspace{-0.2mm}4}\!\sqrt{\xieff}}{2\pi x^2}\right)^{\!\!3/2}
    \!\!\left(\frac{x}{x_0}\right)^{\!\!4\sqrt{\frac{2\xieff}{3\sqrt{3}}}} \ ,
    \label{eq:nhtachyonic}
\eeq
which is applicable for \il{x \lesssim \xxi}.
Although there is a brief transient phase of non-adiabatic 
production for \il{x \gtrsim \xxi}, particle production from 
the curvature term proceeds only through the substantially weaker narrow 
resonance, which shuts down entirely soon thereafter.
We cover the details of this regime in Sec.~\ref{subsubsec:SmallCoupling} below.

\subsection{Production in the Mixed Case\label{subsec:MixedCase}}

Let us now promote our discussion to the mixed case, in which both couplings
$\ggOL$ and $\xieff$ are non-zero. As such, the Higgs modes evolve 
with the effective masses
\beq
    \frac{\omega_{\Hk}^2}{\Lphi \varphiamp{}^2} 
    = \kappa^2 + \frac{g^2}{\Lphi}\hspace*{-0.7mm}\left(\frac{\varphi}{\varphiamp}\right)^{\!\!2}\!\left[1 \!-\! 2\rphi\!\left(\frac{\varphi}{\varphiamp}\right)^{\!\!2}\right]
    \!+ \rh\!\left[3\left(\frac{\varphi}{\varphiamp}\right)^{\!\!4} \!-\! 1\right]
    \label{eq:omegamixed}
\eeq
that follow directly from Eq.~\eqref{eq:omegaHE}, and by analogy to $\rh$
we have defined \il{\rphi \equiv -\xiphi\varphiamp{}^2/(2a^2)}.

There are several immediate implications; let us
discuss these with a focus on the effect of the Higgs-curvature coupling.
First, since only the curvature term dissipates with the cosmological 
expansion, the dynamics may progress through several distinct phases,
most noticeably if \il{\rh \gg \ggOL} at early times.  
Second, the reintroduction of the \il{g^2\varphi^2} term lifts
the effective masses and thereby opens the \il{\xieff < 0} region 
to viability.  Indeed, the \il{\xieff < 0} region was excluded 
in Sec.~\ref{subsec:Tachyonic} only because the inflationary constraints 
[in Eq.~\eqref{eq:higgscurvature} and Eq.~\eqref{eq:stabilityinflation}] 
would have been violated, but for mixed couplings this half of 
parameter space can be reincorporated.
Third, the small-coupling regime (\il{\ggOL\lesssim 1} and 
\il{\rh \lesssim 1}) stands apart from most of our discussion thus far.
The particle production in this regime is driven entirely by narrow
parametric resonances arising from two different
types of modulations $(\varphi/\varphiamp)^2$ and $(\varphi/\varphiamp)^4$.
If these modulations can be approximated as sinusoidal, then
a perturbative treatment is likely effective.
We investigate all of these possible scenarios below.

\subsubsection{Dominant $g^2\phi^2$ \emph{(}with \il{\ggOL \gg 1}\emph{)}}

The simplest possibility is an initially small curvature term \il{|\rh| \lesssim \ggOL},
since this does not allow a tachyonic instability to develop during preheating.  
(We confine our discussion to \il{\ggOL \gg 1} for the moment
and cover the small-coupling regime separately.)

Nonetheless, the curvature term in this case has an impact
on the Higgs dynamics.  Depending on the sign of $\xieff$, these terms may 
add constructively or destructively, enhancing or suppressing the effective 
masses $\omegaHk$, respectively.  
For \il{\xieff > 0}, the effect is constructive and the resonance bands
are widened; \eg, the broad instability bands are extended 
to \il{\kappa ^2 \leq \sqrt{2g^2/(\pi^2\Lphi) + \rh}}.
We have seen that Fig.~\ref{fig:ngrowth} falls within 
this category, and this extension explains the variation among different $\xieff$.
Conversely, for \il{\xieff < 0} the effect is \emph{destructive}: 
not only do the bands narrow, but the non-adiabaticity that drives 
the parametric resonance is weakened.
Explicitly, the maximum characteristic exponent is reduced to
\beq
    \max_{\ggOL}\mumax ~=~ \frac{2}{T}\log\!\left(\!e^{-\frac{\pi\rh^2}{\ggOLexp}} \!+ \sqrt{1 + e^{-\frac{2\pi\rh^2}{\ggOLexp}}}\hspace{0.7mm}\right) \ .
    \label{eq:mumaxaltered}
\eeq

Surely, after a sufficient duration one has \il{\rh \lesssim \sqrt{\ggOL}}
and the exponent \il{\mumax \approx 0.24} is restored.
In other words, the full capacity of the parametric resonance 
is \emph{delayed} until a time $\xg$, which is given approximately by
\beq
    \xg ~\simeq~ \sqrt{-6\xieff}\left(2\pi\frac{\Lphi}{g^2}\right)^{\!\!1/4} \ .
    \label{eq:xg}
\eeq
After this time the curvature term continues to dissipate and has 
an increasingly negligible influence.  The growth of 
fluctuations then proceeds according to Sec.~\ref{subsec:Parametric}.

\subsubsection{Dominant \il{\xieff R h^2} \emph{(}with \il{\ggOL\gg 1}\emph{)}}

\begin{figure}[tb]
    \centering
    \includegraphics[keepaspectratio,width=0.49\textwidth]{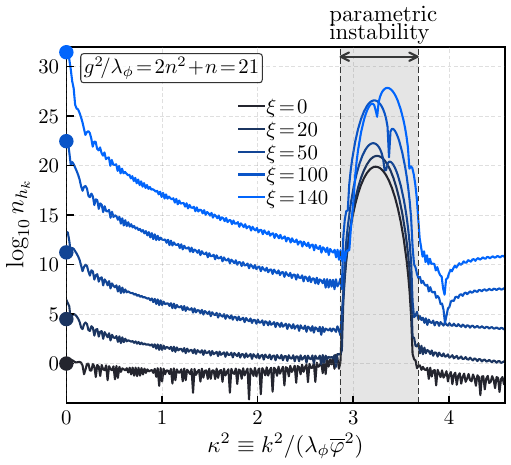}
    \caption{The phase-space density of Higgs particles $n_{h_k}$ produced 
    during preheating, computed numerically and evaluated at the time \il{x = 400}.  
    The point markers along the vertical axis show our analytical
    result for $n_{h_0}$, found by evaluating Eq.~\eqref{eq:mixednk} at zero momentum. 
    The Higgs-inflaton coupling chosen \il{\ggOL = 21} corresponds to a narrow
    resonance, and the different curves show various choices for 
    the curvature coupling $\xieff$.  The only production 
    from the quartic interaction occurs at finite momenta (gray region) 
    and the curvature interaction dominates the production at small momenta.  
    }
\label{fig:spectrum}
\end{figure}

An initially large \il{\rh \gg \ggOL \gg 1} ensures the Higgs dynamics are
dominated by tachyonic production until a time $\xxi$.
Unlike the converse situation above, where we could ignore the 
curvature interaction after some duration, there is no regime 
in which we can ignore the $g^2\phi^2$ interaction
entirely.  Aside from the fact that the parametric resonance from this 
term will always play a role, its presence also alters the size 
of the tachyonic masses.  We can incorporate this effect by performing an analysis 
along the lines of Sec.~\ref{subsec:Tachyonic} with \il{\ggOL \neq 0}.
Then, for a given mode, we find that when 
\il{\sqrt{\ggOL} \lesssim \rh \lesssim \ggOL + \kappa^2}
is satisfied the tachyonic instability is active and 
proceeds adiabatically.  In order to compute the integral 
analytically, we assume that \il{(\ggOL)^2 \gg (\rh - \kappa^2)(3\rh - 2\rphi\ggOL)}.
Then, the accumulated quantities [from Eq.~\eqref{eq:tachyonicformula}] 
are found to be
\beq
    X_k \simeq \frac{\pi (\rh - \kappa^2)}{\sqrt{ 2 \ggOL}} \ , \quad \ 
    \Theta_k \simeq  \frac{\pi \sqrt{\ggOL - \rh + \kappa^2}}{\sqrt{2}}  \ .
    \label{eq:Xkmixed}
\eeq
As expected, Eq.~\eqref{eq:Xkmixed} shows that tachyonic production is 
still concentrated in the small-momentum modes.  However, the evolution of $n_{h_k}$
in the presence of \il{\ggOL\neq 0} is markedly different.  Rather than a power law,
the phase-space density rapidly asymptotes to a constant as
\begin{align}
    \log n_{h_k} ~\simeq~ \frac{12\sqrt{2}\pi\xieff}{T\sqrt{\ggOL}}\left[\frac{1}{x_0} - \frac{1}{x} - \frac{(x-x_0)\kappa^2}{6\xieff}\right]  \ ,
    \label{eq:mixednk}
\end{align}
again neglecting the oscillatory component $\cos\Theta_k$.
The termination of the tachyonic resonance is pushed to an earlier time
depending on the quartic couplings:
\beq
    \xxi ~\simeq~ \sqrt{\!\frac{6\xieff}{\sqrt{\ggOL}}} \ .
    \label{eq:xximixed}
\eeq
Afterward, the modes grow as \il{n_{h_k}\propto e^{2\mu_k x}},
driven by the parametric resonance.  The phase-space density can be 
found at these times by matching to Eq.~\eqref{eq:mixednk} at $\xxi$.

Finally, the number density of Higgs particles produced 
during this tachyonic phase is found by integrating Eq.~\eqref{eq:mixednk}.
Using the saddlepoint approximation we find
\beq
    n_h ~\simeq~ \frac{(\!\sqrt{\Lphi}\varphiamp)^3}{8\pi^3}
    \!\left(\!\frac{T}{2x}\sqrt{\frac{g^2}{2\Lphi}}\right)^{\!\!\!3/2}
    \!\!\!e^{\frac{12\sqrt{2}\pi\xieff}{T\sqrt{\ggOL}}\left(\frac{1}{x_0} - \frac{1}{x}\right)} \ .
    \label{eq:mixedn}
\eeq

In Fig.~\ref{fig:spectrum} we have plotted the phase-space 
density $n_{h_k}$, calculated numerically over the momentum 
space of the Higgs modes.  The quartic coupling chosen \il{\ggOL = 21}
sits on a narrow resonance and therefore produces particles only at finite
momenta (highlighted by the gray region).  Meanwhile, the different curves
show various non-minimal couplings $\xieff$.  
Because the parametric resonance yields unimpeded particle
production, the finite-momentum peak in $n_{h_k}$ continues to grow with time,
while production for the other modes ceases once \il{x\gtrsim \xxi}.
The large point markers along the vertical axis show 
agreement with our analytical result for $n_{h_0}$ 
where the tachyonic production is dominant, found by 
evaluating Eq.~\eqref{eq:mixednk} at zero momentum. 

The same logic as above may be applied to the \il{\xieff < 0} region
of parameter space. As this region is dominated by tachyonic particle 
production, we may again employ Eq.~\eqref{eq:tachyonicformula} to
calculate the generated Higgs spectrum.  While this calculation 
is straightforward in principle, it requires some additional 
technical details that we include in Appendix~\ref{sec:tachyonic}, 
but we summarize the results here.  
The phase-space density for the Higgs is given by
\beq
    \log n_{h_k} \simeq \frac{4H_1(3,0)}{T} \!\!\int \!dx\, \frac{\kappa^2\! - 2\rh + \frac{g^2}{\Lphi}\!\left(2\rphi\!-\!1\right)}{\sqrt{\kappa^2 - \rh}} \ ,
\eeq
where \il{H_1(3,0)\approx 0.76} corresponds to the function
defined in Eq.~\eqref{eq:Hfuncs} of Appendix~\ref{sec:tachyonic}.
The small-momentum modes provide the largest contribution, 
and we can approximate
\begin{align}
    \log n_{h_0} ~\simeq~ &\frac{8H_1(3,0)}{T}\!\left(\!1 + \frac{g^2}{\Lphi}\frac{\xiphi}{\xieff}\!\right)\!\sqrt{6|\xieff|}\log\!\left(\frac{x}{x_0}\right)\nn \\
    \ &- \frac{H_1(3,0)}{T}\sqrt{\frac{2}{-3\xieff}}\frac{g^2}{\Lphi}(x^2 - x_0^2)
    \label{eq:nh0mixedN}
\end{align}
in this limit.  Compared to \il{\xieff > 0}, 
the power indices are a similar magnitude if we neglect 
the effect of $\xiphi$.

\begin{figure}[t]
    \centering
    \includegraphics[keepaspectratio,width=0.49\textwidth]{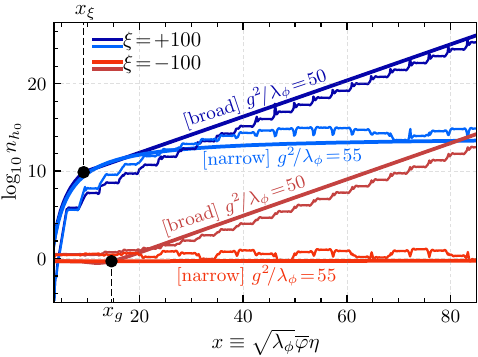}
    \caption{The growth of the zero-mode phase-space density 
    in the mixed-coupling case, exploring both signs \il{\xieff = \pm 100}
    and the broad and narrow regimes of the parametric resonance.
    The thin (thick) curves show our numerical (analytical)
    results, respectively.  For \il{\xieff > 0}, the tachyonic instability 
    drives particle production early on [following Eq.~\eqref{eq:mixednk}], 
    but once \il{x \gtrsim \xxi} it is driven by the parametric 
    resonance \il{n_k \propto e^{2\mu_k x}}.  The quartic and curvature
    contributions add constructively and enhance the instability.
    Meanwhile, for \il{\xieff < 0}, these contributions add destructively, and the 
    growth exponent $\mu_k$ is effectively shut off
    [following Eq.~\eqref{eq:mumaxaltered}] until the time $x_g$.
    }
\label{fig:mixedcase}
\end{figure}

In Fig.~\ref{fig:mixedcase}, we demonstrate the growth of 
the zero-mode $n_{h_0}$ in the mixed-coupling case, showing 
both our numerical (thin curves) and analytical (thick curves) results.  
For comparison, we have included both the positive (blue curves) 
and negative (red curves) $\xieff$ regimes, as well as the 
narrow and broad resonance regimes.  The destructive effect of 
\il{\xieff < 0} is evident, as \emph{changing the sign of $\xieff$ 
induces a gap of many orders of magnitude} in the zero-mode density.
As a result, we should expect that electroweak vacuum metastability 
shows a preference for the \il{\xieff < 0} half of parameter space.
We shall return to this topic when we examine Higgs
destabilization in Sec.~\ref{sec:Backreaction}.

\subsubsection{Small-Coupling Regime\label{subsubsec:SmallCoupling}}

Let us now focus on the small-coupling regime, \ie, where both \il{\ggOL \lesssim 1} 
and \il{|\rh| \lesssim 1}.  The latter inequality is always imminent,
so we are free to also interpret this regime as the late-time behavior of a 
model with \il{\ggOL \lesssim 1} and an 
arbitrary curvature coupling.  Indeed, as soon as \il{|\rh| \lesssim 1} 
the tachyonic instability transitions into a narrow resonance.
In fact, both of the instabilities in this regime experience a narrow 
parametric resonance, and our analysis should follow a different approach.
Namely, the oscillatory terms in Eq.~\eqref{eq:omegamixed} 
act as small modulations to $\omegaHk^2$, and we can treat these 
terms perturbatively~\cite{Greene:1997fu}.  We expand the 
elliptic functions as
\begin{align}
    \left(\frac{\varphi}{\varphiamp}\right)^{\!2} \!\!~&=~ \sum_{\ell=0}^{\infty} F_\ell \cos\frac{4\pi\ell (x-x_0)}{T} \nn \\
    \left(\frac{\varphi}{\varphiamp}\right)^{\!4} \!\!~&=~ \sum_{\ell=0}^{\infty} G_\ell \cos\frac{4\pi\ell (x-x_0)}{T} \ ,
    \label{eq:smallcouplingexpansion}
\end{align}
with the leading five coefficients given in Table~\ref{tab:coefficients}.
Noting that the series converges quickly after the first three terms, we truncate
at order \il{\ell=2} and the expansion serves as a good approximation.  
Moreover, given that $\rh$ are slowly varying compared to the 
timescale of inflaton oscillations,
the Higgs mode equations take the form of the
Whittaker-Hill equation:
\begin{align}
    \frac{d^2\mathcal{H}_k}{dz^2} + \left(A_k + 2 p \cos 2z + 2 q \cos 4z\right) \mathcal{H}_k ~=~ 0 \ ,
    \label{eq:smallcouplingeqn}
\end{align}
where we have defined \il{z\equiv 2\pi (x-x_0)/T}.  The coefficients of the frequency terms are given by
\beq
    A_k = \left(\frac{T}{2\pi}\right)^{\!2} \!\Big[\kappa^2 + \frac{g^2}{\Lphi} (F_0  - 2 \rphi G_0 ) + \rh (3 G_0 - 1)\Big] 
\eeq
in which we have defined the quantities
\begin{align}
    p ~&=~ - \frac{1}{2}\left(\frac{T}{2\pi}\right)^{\!2} \left[\frac{g^2}{\Lphi} (F_1 - 2 \rphi G_1)  + 3\rh G_1\right] \nn \\
    q ~&=~ +\frac{1}{2}\left(\frac{T}{2\pi}\right)^{\!2} \left[\frac{g^2}{\Lphi} (F_2 - 2 \rphi G_2)  + 3\rh G_2\right]  \ .
\end{align}

\begin{figure}[tb]
    \centering
    \includegraphics[keepaspectratio,width=0.49\textwidth]{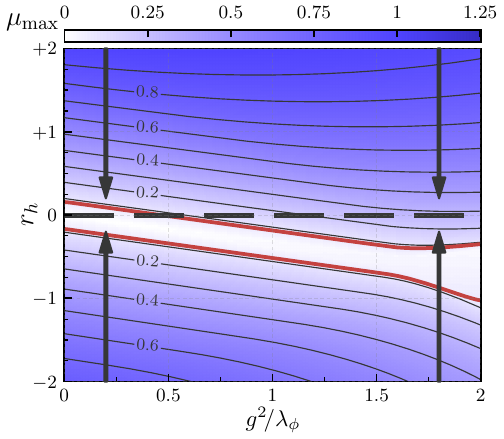}
    \caption{Contours of the maximum characteristic exponent 
    \il{\mumax\equiv \max_k\mu_k} shown over the space of small couplings
    \il{\{\ggOL,\rh\}} with \il{\xiphi = 0} for simplicity.
    Because the curvature parameter redshifts as 
    \il{|\rh| \propto 1/a^2}, this figure may be interpreted in a
    time-dependent way.  From some initial point, 
    the $\mumax$ value flows along lines of fixed $\ggOL$ toward 
    \il{|\rh|\rightarrow 0} (illustrated 
    by gray arrows) before finally landing along the narrow quartic resonance 
    \il{\mumax \approx 0.15\ggOL} (gray-dashed line).  
    }
\label{fig:WH}
\end{figure}

Using Floquet theory it is relatively straightforward to 
calculate the characteristic exponents $\mu_k$
for the unstable modes~\mbox{\cite{whittaker_watson_1996,Lachapelle:2008sy}}.
We can therefore find the maximum exponents \il{\mumax\equiv \max_k\mu_k}
over the small-coupling parameter space.  
These results are displayed in Fig.~\ref{fig:WH}.

\begin{table}[t]
\begin{center}
    \caption{Coefficients for the expansions in Eq.~\eqref{eq:smallcouplingexpansion}.}
    \begin{ruledtabular}
    \begin{tabular}{ccccccc}
        $\ell$ & $0$ & $1$ & $2$ & $3$ & $4$\\
        $F_\ell$ & $0.457$ & $0.497$ & $4.29\times 10^{-2}$ & $2.78\times 10^{-3}$ & $1.60\times 10^{-4}$\\
        $G_\ell$ & $0.333$ & $0.476$ & $0.164$ & $2.39\times 10^{-2}$ & $2.45\times 10^{-3}$
    \end{tabular}
    \end{ruledtabular}
    \label{tab:coefficients}
\end{center}
\end{table}

An interesting feature of Fig.~\ref{fig:WH} is that,
given the scale-factor dependence of $\rh$, we can
interpret the figure as showing the flow of $\mumax$
as a function of time.  For some initial \il{\rh \neq 0} value,
$\mumax$ flows along contours of constant $\ggOL$ (\eg, the
gray arrows) until it reaches the \il{\rh = 0} axis.  Notably,
for \il{\xieff < 0} the evolution of $\mumax$ over this time 
is not necessarily monotonic. For larger quartic couplings,
one may enter the \il{\mumax = 0} region (enclosed by red curves) 
and then exit it while approaching \il{\rh = 0}.

Note that in the regime of negligible curvature couplings,
Eq.~\eqref{eq:smallcouplingeqn} reduces to the form of the Mathieu equation.
The Mathieu equation is familiar from models of massive preheating,
but unlike such models the parametric instability from 
the $g^2\phi^2$ interaction does not terminate due to 
the approximate scale invariance of the theory.
The fact that the curvature contributions are neither long-lived 
nor tachyonic means that the quartic contribution becomes 
the most salient in the small-coupling regime.
Even for couplings as large as \il{\rh=\mathcal{O}(\ggOL)},
the resonance bands are widened, but their influence only 
induces a logarithmic correction to the growth rates.

For these reasons, the \il{\rh\rightarrow 0} limit is 
usually the most pertinent in the context of vacuum metastability.
Here, the modes grow as \il{n_{h_k}\approx e^{2\mu_k x}/2} 
for momenta around the narrow bands.   The primary resonance 
is centered at \il{\kappa^2 = 2\pi/T}, and one finds the maximum 
exponent \il{\mu_{\rm max} = F_1 T g^2\!/(8\pi\Lphi) \approx 0.15 \ggOL}~\cite{Greene:1997fu}.
Using the saddlepoint approximation, we integrate to find
\beq
    n_h ~\simeq~ \frac{\pi\Lphi\varphiamp{}^3}{T^2}\sqrt{\frac{2\Lphi\mumax}{T x}}e^{2\mumax x} \ .
    \label{eq:nhnarrow}
\eeq

\subsection{Perturbative Higgs Decays\label{subsec:PerturbativeDecays}}

Thus far, for simplicity we have neglected the perturbative 
decay of Higgs particles after their production.  
These decays may not appear important when considering the 
substantial rates of steady non-perturbative particle production.
After all, perturbative decays are known to have only a
minor effect in preheating with massive inflaton 
potentials (see Appendix~\ref{sec:quadratic}).
However, given the unique properties that appear
for massless preheating, it is worth examining 
the Higgs decays in closer detail.

Before continuing on this path, let us briefly 
address the other ways in which the Higgs may transfer its energy density,
most notably the non-perturbative production of gauge bosons.
Even though the Higgs mass term oscillates coherently
due to the inflaton-field motion, the gauge-field mass terms do not oscillate uniformly.
Therefore, in the early stages of preheating, we expect that resonant production 
of $W$ and $Z$ gauge bosons from the Higgs evolution is less significant than 
resonant production of the Higgs field itself.  Still, the averaged value of the gauge-boson 
mass terms are modulated by the time dependence of $\langle h^2 \rangle$, 
so a substantial amount of gauge bosons could be produced.  As long as the 
Higgs field is much smaller than the critical value in Eq.~\eqref{eq:hcrit}, 
a simple analysis shows that the gauge-field production from fast oscillations of 
$\langle h^2\rangle$ is in the narrow regime. More precise estimation
of gauge-field production requires a dedicated study (see also footnote~\ref{fn:unitary_gauge}). 
In the following, we shall therefore assume that non-perturbative gauge-field production
induced by the Higgs field can be ignored and that perturbative decays are
the dominant mechanism through which energy is transferred out of the Higgs field.

In the early stages of preheating, we expect that perturbative decay 
channels for the Higgs are kinematically accessible since the homogeneous background 
value for $h$ is negligible.
Even if a background value is generated, 
the masses of SM particles are lighter than the 
Higgs during preheating as long as \il{h \ll \hcrit}, in which
\beq
    \hcrit ~=~ \sqrt{\frac{g^2\phi^2 + \xih R}{\abs{\Lh}}}
    \label{eq:hcrit}
\eeq
is the Higgs value at the barrier separating the 
false and true vacua, \ie, the local maximum of the potential.

Therefore, the dominant decay channel of the Higgs is into top quarks\footnote{
The branching ratios are different
from the well-known case in the electroweak vacuum since the Higgs
field value and its mass (given by the inflaton/curvature coupling) in
our setup are effectively independent.
} at a rate 
\beq
    \Gamma_h ~\simeq~ \frac{3y_t^2 m_h}{16\pi} 
    \label{eq:Gammah}
\eeq
in the rest frame, where $y_t$ is the top Yukawa coupling,
evaluated at the Higgs mass scale, and we have denoted 
\beq
\label{eq:mh}
    m_h ~=~ \sqrt{g^2\phi^2 + \xih R}
\eeq
as the effective Higgs mass.  In general, we should
include a Lorentz factor \il{\gamma = \omega_{h_k}/m_h} 
which suppresses the decay rate for relativistic particles.
The tachyonic instability produces particles with physical 
momenta \il{k/a\lesssim \phiamp{}^2\sqrt{\xieff\Lphi/2}},
which translates to \il{\gamma\simeq \mathcal{O}(1)} and does not
appreciably suppress the rate.  Likewise, the parametric instability 
with \il{\ggOL \gg 1} produces particles with
\il{k/a\lesssim [\Lphi\phiamp{}^2\sqrt{g^2/\Lphi}]^{1\hspace{-0.1mm}/\hspace{-0.1mm}2}},
which is non-relativistic.  The exception is small couplings \il{\ggOL \ll 1},
for which Higgs production is dominated by particles with 
a Lorentz factor \il{\gamma \simeq k/(am_h) \simeq \sqrt{\Lphi/g^2}}.
While we include the full effect of this time-dilation on the decay 
rate in our numerical computations,  our analytical calculations are 
performed assuming that \il{\gamma\simeq\max(1,\sqrt{\Lphi/g^2})}.

Naturally, as perturbative decays exponentially suppress the 
number density, they work against the non-perturbative production processes
above.  The field dependence of the effective
Higgs mass $m_h$ implies an interesting chronology of events.
If the parametric resonance is the dominant production mechanism, then 
a burst of particles is produced as $\phi$ passes through the 
origin.  Meanwhile, decays are strongest when $\phi$ is maximally
displaced.  The result is that rather than $n_{h_k}$ maintaining 
a constant value during its adiabatic evolution, it is exponentially
suppressed at the rate $\Gamma_h$.  
However, due to its dependence on evolving background quantities,
the decay rate is a non-trivial function of time.  
The effect on the number density is to dissipate it as 
\beq
\log n_{h_k} ~\propto~ -\int dt\, \frac{\Gamma_h}{\gamma} 
~=~ -\int dx \,\frac{a\Gamma_h}{\gamma\sqrt{\Lphi}\varphiamp}
\label{eq:decayexponent}
\eeq
where the integration is performed over times when decays are kinematically possible.  

An important limit is that in which 
the non-minimal couplings \il{\xih,\xiphi} are negligible, 
as this also represents the system for times \il{x\gtrsim \xxi}.
The pivotal observation is that the effective decay rate scales
as \il{a\Gamma_h\propto |\varphi|} and thus has 
\emph{no overall scale-factor dependence,  allowing a direct competition
between the production and decay rates} to determine the fate of the 
electroweak vacuum;  this contrasts with massive 
preheating, in which the decay exponent carries a logarithmic
time dependence.  (We refer to Appendix~\ref{sec:quadratic}
for a more thorough comparison.)
Explicitly, the time-averaged conformal decay 
rate $\langle a\Gamma_h\rangle$ is given by
\beq
    \frac{\expt{a\Gamma_h}}{\sqrt{\Lphi}\varphiamp} ~=~ \frac{3y_t^2}{16\pi}\sqrt{\frac{g^2}{\Lphi}} \expt{\frac{\abs{\varphi}}{\varphiamp}}
    ~\approx~ 0.036 y_t^2 \sqrt{\frac{g^2}{\Lphi}} \ ,
    \label{eq:Gammahavg}
\eeq
where we have assumed the regime \il{\ggOL \gtrsim 1}.
Indeed, we find that the number density of Higgs particles does not 
grow for couplings exceeding
\beq
    \frac{g^2}{\Lphi} ~\gtrsim~ 
    2.8 \times 10^3 \left(\frac{\mumax}{0.24}\right)^{\!2}\!\left(\frac{y_t}{0.5}\right)^{\!-4} \ .
    \label{eq:nogrowth}
\eeq
Remarkably, this result suggests that electroweak vacuum metastability 
can be achieved in massless preheating.  Moreover, metastability is given
not by an upper bound but by a \emph{lower bound} on the quartic 
coupling (see also Refs.~\cite{Choi:2019osi,Lebedev:2021zdh}).

In the opposite limit, where $g$ is negligible and the curvature
term dominates, neither the production or perturbative decays 
evolve exponentially.  In particular, we have 
\il{\expt{a\Gamma_h} \propto \log a} and 
a growth rate that is dominant at early times.  The result is 
that perturbative decays do not quell the early phase of 
tachyonic production, but if the vacuum survives,
decays increasingly dissipate the fluctuations as the system evolves.

Looking back to Fig.~\ref{fig:ngrowth}, we have demonstrated
the effect of the perturbative decays on particle production 
by plotting $n_{h_k}$ both with (blue curves) and 
without (grayscale curves) the decays incorporated.  
Clearly, in the absence of decays, the phase-space density remains 
constant when the inflaton is away from \il{\phi\approx 0}, 
as the $n_{h_k}$ correspond to an adiabatic invariant.
However, when the Higgs decays are turned on, the curves show a 
dissipation during the adiabatic evolution.  The dissipation
does not always appear to be at a constant rate, reflecting 
that in our numerical computations, the decay rates are 
not averaged as in Eq.~\eqref{eq:Gammahavg} but are given 
instantaneously as a function of the effective Higgs mass $m_h$.

Before concluding this section, we put aside the Higgs for a moment
and make some remarks regarding the quantum fluctuations 
of the inflaton field.
An extensive part of the discussion in this section 
can be applied by analogy to the production of inflaton particles.
These particles are generated through the inflaton self-coupling, so that
the effective masses [analogous to Eq.~\eqref{eq:omegahk}] are 
\beq
\omega_{\phi_k}^2 ~=~ \frac{k^2}{a^2} + 3\Lphi\phi^2,
\label{eq:omegaphik}
\eeq
in the Jordan frame, or also in the Einstein frame after neglecting 
higher-order terms in $\mathcal{O}(a^{-2})$. 
In fact, production via this inflaton ``self-resonance'' is understood
by the simple replacement \il{\ggOL = 3} in the 
analysis of Sec.~\ref{subsec:Parametric}~\cite{Greene:1997fu}.
For example, from our result for the Higgs number
density $n_{h}$ in Eq.~\eqref{eq:nhbroad},
we can estimate the inflaton number density $n_{\phi}$
with appropriate replacements.
Interestingly, the characteristic growth exponent \il{\mumax \approx 0.036} for 
inflaton production corresponds to the weakest of the \il{\ggOL \geq 1} regime.
In Sec.~\ref{sec:Backreaction} below, we shall find
that this self-resonance and its properties are extremely material to the 
discussion of vacuum destabilization and the end of preheating.


\section{Backreaction and Vacuum Destabilization\label{sec:Backreaction}}


While Sec.~\ref{sec:NoBackreaction} provides the groundwork for 
our study of electroweak vacuum metastability, an essential 
ingredient is still absent from our analysis.  
In particular, recalling our assumption that
the Higgs field has a negligible background value \il{h\approx 0},
the term in the potential which threatens to destabilize the vacuum
(the self-interaction \il{\lambda_h h^4/4}) has not yet entered
our analysis.  In order to incorporate the unstable vacuum,
we must consider not only the growth of Higgs fluctuations but 
also the \emph{backreaction} that these fluctuations have on the system.
As discussed in Sec.~\ref{sec:Intro}, although the fluctuations 
in massless preheating appear to grow unimpeded, they will
inevitably grow large enough to either trigger vacuum decay or 
disrupt the background inflaton evolution through backreaction.

A complete treatment of the backreaction and non-linearities that
appear, especially in the later stages of preheating, requires 
numerical lattice simulations.  However, we find that, in 
order to probe the Higgs stability, or
estimate the onset of the non-linear stage, 
the one-loop Hartree approximation~\cite{Boyanovsky:1994me}
is sufficient.  That is, we assume the factorization 
\il{h^4 \rightarrow 6\langle h^2\rangle h^2 - 3\hvar^2} for the quartic term
in the Lagrangian (and an analogous expression for the inflaton), 
where the expectation values are given by
\begin{align}
    \hvar ~&=~ \frac{1}{(2\pi)^3}\int d^3k\abs{h_k}^2  \\
    \phivar ~&=~ \frac{1}{(2\pi)^3}\int d^3k\abs{\phi_k}^2 \ .
\end{align}
In turn, the effective masses for the Higgs and inflaton modes---from 
Eq.~\eqref{eq:omegahk} and Eq.~\eqref{eq:omegaphik}, respectively---are modified
as the variance of the fluctuations grows:
\begin{align}
    \omega_{h_k}^2 ~&=~ \frac{k^2}{a^2} + g^2\phi^2 + \xih R + g^2\phivar + 3\lambda_h\hvar \\
    \omega_{\phi_k}^2 ~&=~ \frac{k^2}{a^2} + 3\Lphi\phi^2 + g^2\hvar + 3\Lphi\phivar  \ .
    \label{eq:omegabackreaction}
\end{align}
The fluctuations also couple to the inflaton background, 
modifying the equation of motion as
\beq
    \ddot{\phi} + 3H\dot{\phi} + \Lphi\phi^3 + \left(3\Lphi\phivar + g^2\hvar\right)\phi ~=~ 0 \ .
    \label{eq:phibackreaction}
\eeq
Given that the motion of $\phi$ drives the particle-production processes,
once these oscillations are disrupted the dynamics of the
type discussed in Sec.~\ref{sec:NoBackreaction} is shut down.
The termination of these processes generally coincides with the energy density 
of the fluctuations growing comparable to that of the inflaton background and thus
corresponds to the end of the linear stage of preheating.

We shall include all of the Hartree terms in our numerical calculations
(as detailed in Appendix~\ref{sec:numerics}), 
but the corrections from the self-interactions $3\Lh\hvar$
and $3\Lphi\phivar$ will tend to carry the most weight for our 
analytical calculations.  We find that from simple 
analytic estimates we can reproduce the salient lattice simulation 
results in the literature.  
For further validation, our numerical analyses are also compared 
to the literature on models for which the inflaton potential is
quadratic in Appendix~\ref{sec:quadratic}. 
For further discussion on the non-linear dynamics and backreaction 
in preheating, we refer the reader to Refs.~\mbox{\cite{Kofman:1997yn,Greene:1997fu}}. 

The main results of Sec.~\ref{sec:NoBackreaction} consist of 
the produced number density $n_h$, so it is necessary to translate 
between these quantities and the variance $\hvar$.  While one can fully express $\hvar$
in terms of Bogoliubov coefficients, the result is a sum of two terms, 
one of which is rapidly oscillating and not important for 
our purpose of producing order-of-magnitude estimates~\cite{Kofman:1997yn}.
As long as the produced Higgs particles are non-relativistic
one can write \il{\hvar \simeq n_{h}/(a^3\omega_h)},
with a similar expression applying to the inflaton or any other relevant fields.
Note that the number density is well defined only when evolving adiabatically,
so the points at which we evaluate $\omega_{h_k}$ 
must remain consistent with this assumption.

\subsection{Onset of Non-Linear Stage\label{subsec:xend}}

In the context of vacuum metastability, a natural concern 
is that particle-production processes arising from scale-invariant
terms do not terminate in the linear stage of preheating.  
That is, these processes terminate only once field fluctuations 
become so large that they significantly backreact 
on the system at some time $\xNL$.  
The longer the linear stage lasts, the more concern for 
vacuum destabilization since the Higgs fluctuations must remain 
sufficiently controlled over this entire duration.

In our model, either the Higgs or inflaton fluctuations may bring 
about an end to the linear stage, but a spectator field
could potentially play this role as well (see Sec.~\ref{sec:Conclusions}
for further discussion along these lines).  
Let us first examine the necessary conditions for 
the Higgs to play this role.
There are two competing effects that influence the 
background inflaton field~\cite{Greene:1997fu}.   
On the one hand, as the inflaton loses energy to particle production,
$\varphiamp$ falls and thereby decreases the
effective frequency of oscillations \il{\sqrt{\Lphi}\, \phiamp}.
On the other hand, as the Higgs variance grows
the inflaton obtains an effective mass \il{\sqrt{g^2 \hvar}}, 
increasing the oscillation frequency.
The inflaton oscillations are disrupted once these two
quantities are similar in magnitude, \ie, once 
\il{g^2\!\hvar\!/(\Lphi\phiamp{}^2)} approaches unity.
Of course, throughout this process the size of $\hvar$ must remain 
controlled \il{\hvar \lesssim \hcrit^2} so that we do not
destabilize the Higgs.  According to Eq.~\eqref{eq:hcrit}, 
this requirement leads to a rather tight constraint on the couplings:
\beq
    \frac{g^2}{\Lphi} ~\gtrsim~ \sqrt{\frac{\abs{\Lh}}{\Lphi}} \ ,
\eeq
which is approximately \il{\ggOL \gtrsim 10^4} for our benchmark
parameter choice \il{\Lphi = 10^{-10}}.
However, we have already found that in this regime 
the rate of perturbative decays dominates over that of particle production
[see previous discussion regarding Eq.~\eqref{eq:nogrowth}].
We can thus conclude that, in the absence of backreaction
sourced by any other fields, the linear stage of \emph{preheating 
must be terminated by the inflaton fluctuations} and not the Higgs.

Unfortunately, the inflaton fluctuations grow slowly with the
weak characteristic exponent \il{\mumax \approx 0.036} (refer to the
end of Sec.~\ref{subsec:PerturbativeDecays}), so this can take a 
significant amount of time.  The analysis above applies just as well
to the inflaton variance $\phivar$ with the replacements 
\il{g^2 \to 3 \Lphi} and \il{\Gamma_h \rightarrow \Gamma_\phi}, 
where $\Gamma_{\phi}$ is the model-dependent inflaton decay rate.
Then, the onset of the non-linear (NL) stage occurs at
\beq
    \xNL ~\simeq~ -\frac{1}{4\mumax}W_{-1}\hspace{-1mm}\left[-\sqrt{\frac{3}{8}}\frac{9\Lphi^2}{(2\pi)^6}\right] \ ,
\label{eq:xNL}
\eeq
where $W_{-1}$ denotes the negative branch of the 
Lambert $W$-function
and we have neglected $\Gamma_{\phi}$.  Using the parameters above 
we find \il{\xNL \approx 413}.  Remarkably, this figure is consistent to 
less than $2\%$ error with lattice-simulation results~\cite{Khlebnikov:1996mc} 
that give \il{\xNL = 76 - 14.3\log\Lphi\approx 405}.

\subsection{Vacuum Destabilization\label{subsec:DestabTime}}

\begin{figure}[t]
    \centering
    \includegraphics[keepaspectratio,width=0.48\textwidth]{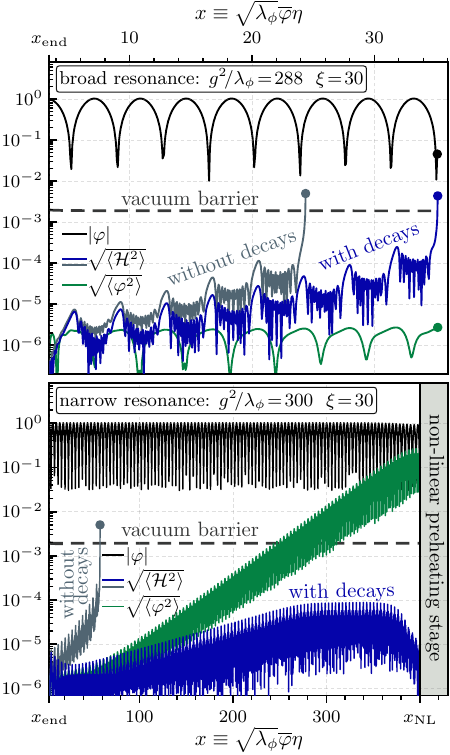}
    \caption{Evolution of the conformal variance for the Higgs
    \il{\chvar} and inflaton \il{\langle \varphi^2\rangle}, 
    with their backreaction on the system included through 
    the one-loop Hartree approximation.  
    If the variance grows to the extent that $\omegaHk^2$ is
    dominated by the (negative) \il{3\Lh\chvar} term,
    the Higgs experiences a runaway tachyonic growth that rapidly 
    destabilizes the electroweak vacuum, causing the sharp divergence in the
    curves.  A point marker at the end of a curve indicates destabilization,
    and the gray curves show $\chvar$
    in the absence of decays.  Identically to Fig.~\ref{fig:ngrowth}, 
    the top (bottom) panels correspond to broad (narrow) parametric 
    resonance, respectively.  In the latter, $\langle\varphi^2\rangle$
    grows sufficiently to end the linear preheating 
    stage at \il{\xNL\approx 400}, consistent with lattice 
    results in the literature~\cite{Khlebnikov:1996mc}.
    }
\label{fig:hk2}
\end{figure}

In the above, we have established a clear criterion for model
viability: for a given choice of couplings,
if the electroweak vacuum survives for a duration longer 
than $\xNL$, then it survives the linear stage of preheating.
The transient non-linear stage that follows tends to asymptote to
thermal equilibrium, during which resonant particle production stops
and the energy density is redistributed among the modes to approach
a thermal distribution.  We expect that if metastability is 
maintained until $\xNL$ then it also survives the non-linear stage;
we briefly discuss these issues and how massless preheating contrasts
with other scenarios in the conclusions in Sec.~\ref{sec:Conclusions}.

Our task then shifts to calculating the vacuum decay times
$\xdec$ as a function of the model parameters.
A robust method to conservatively estimate $\xdec$ 
is as follows~\mbox{\cite{Ema:2016kpf, Enqvist:2016mqj}}.  As the Higgs
fluctuations grow, the effective mass term associated with the Higgs 
self-interaction \il{3\lambda_h\hvar} becomes negative and large enough 
to make some Higgs modes tachyonic over a time interval $\Delta t$.
These modes grow exponentially in this interval at a rate
\il{\sqrt{3|\lambda_h|\hvar}}.  In turn, this amplifies $\hvar$,
which then makes the tachyonic masses even larger---\ie, the system enters a
positive feedback loop which rapidly destabilizes the vacuum.
We can estimate $\xdec$ as the time at which the growth exponent
becomes larger than $\mathcal{O}(1)$:
\beq
    \sqrt{\frac{3\hvar}{\phiamp{}^2}\frac{|\Lh|}{\Lphi}}\,\Delta x ~\gtrsim~ 1 \ ,
    \label{eq:xdeccondition}
\eeq
and we have used that \il{\Delta t = \Delta x/(\sqrt{\Lphi}\,\phiamp)}.

Alternatively, in situations where the Higgs modes oscillate 
slowly relative to the background inflaton field, the simpler condition
\il{3\Lh \hvar \gtrsim m_h^2} is appropriate, in which we evaluate the
effective Higgs mass $m_h$ [of Eq.~\eqref{eq:mh}] at the amplitude 
of the inflaton oscillations \il{\phi = \phiamp}.

The process of vacuum destabilization as it unfolds
is illustrated in Fig.~\ref{fig:hk2} by 
including backreaction in our numerical calculations.
The panels reflect the same parameter choices made for 
Fig.~\ref{fig:ngrowth} in the previous section, where we focus 
on the broad \il{\ggOL = 2n^2} and narrow \il{\ggOL = 2n^2 + n}
parametric resonances within the same coupling band (with \il{n=12}). 
The curvature coupling \il{\xieff = 30} is subdominant but non-negligible.
Both the Higgs and inflaton variance are shown, 
and the former shows the results both with (blue) and without (gray) Higgs 
decays.  In the top panel, the perturbative decays 
delay destabilization but it ultimately occurs
at \il{\xdec\approx 35}.  The runaway growth of the Higgs 
fluctuations is observed toward the endpoint of each curve, as they
finally cross the vacuum barrier \il{\chvar \geq (a\hcrit)^2}.
By contrast, in the lower panel we see that the narrow resonance 
would lead to destabilization in the absence of decays, 
after not much longer \il{\xdec\approx 55}.  However, 
we see that \emph{accounting for perturbative decays prevents vacuum 
destabilization}.  Moreover, we observe the onset of 
the non-linear stage at $\xNL$ induced by the growth of inflaton 
fluctuations, as $\cphivar$ grows sufficiently large.
As expected, this stage occurs at \il{\xNL\approx 400}.

Let us now survey $\xdec$ over the parameter space of couplings
and produce analytical estimates in each region.  We shall work largely 
in parallel to Sec.~\ref{sec:NoBackreaction} and employ the results 
from that section.  That is, we shall first cover the \il{\xieff = 0} 
and \il{g = 0} limits separately and then progress to the mixed 
case, providing a general picture of vacuum metastability 
at the end of the section.

\begin{figure*}[tb]
    \includegraphics[keepaspectratio,width=0.99\textwidth]{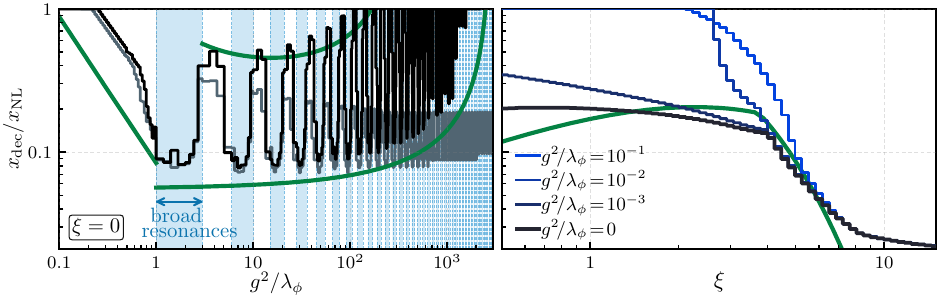}
    \caption{The decay time $\xdec$ of the electroweak vacuum 
    as a function of $\ggOL$ (left panel) and $\xieff$ (right panel), 
    normalized by the onset time $\xNL$ of the non-linear stage.
    The vacuum survives preheating for \il{\xdec \geq \xNL}.
    In the left panel, the results are shown both with 
    (black curve) and without (gray curve) perturbative Higgs decays.
    Our analytical estimates in this panel (green curves)
    correspond to the different regimes of Eq.~\eqref{eq:xdecquartic}.  
    In particular, the two green curves for \il{\ggOL \gtrsim 1} 
    show the broad and narrow regimes, evaluated using \il{\mumax \approx 0.24} 
    and Eq.~\eqref{eq:narrowfit}, respectively.   
    As these $\mumax$ correspond to the minimum and maximum 
    growth exponents, they estimate the envelope of the highly
    non-monotonic $\xdec$. 
    In the right panel, the analytical estimates are also shown
    by green curves and correspond to Eq.~\eqref{eq:xdeccurvature1}
    and Eq.~\eqref{eq:xdeccurvature2}, plotted over their regions of validity.
    }
\label{fig:slices}
\end{figure*}

\subsubsection{From Parametric Instability\label{subsubsec:QuarticDecay}}

Let us first consider destabilization of the vacuum 
from the parametric resonance.  The effective mass of a Higgs mode
\il{\omega^2_{h_k} \approx k^2/a^2 + g^2\phi^2 + 3\Lh\hvar} 
may first become tachyonic once the inflaton field 
passes through \il{\phi=0}.  We can estimate $\xdec$ 
by following the discussion surrounding
Eq.~\eqref{eq:xdeccondition}, but we must first calculate 
the Higgs variance.  Using that \il{\hvar \simeq n_h/(a^3\omega_h)}, 
with Eq.~\eqref{eq:nhbroad} and Eq.~\eqref{eq:nhnarrow} for the different
coupling regimes, we find 
\beq
    \hvar ~\simeq~
    \begin{cases}
        \displaystyle{\frac{\Lphi\phiamp{}^2}{T^{3\hspace{-0.2mm}/\hspace{-0.2mm}2}}\sqrt{\frac{\mumax}{2x}}e^{2\muh x}} & \text{for } \ggOL \lesssim 1 \\[6pt]
        \displaystyle{\frac{\Lphi\phiamp{}^3}{16\pi^3\abs{\phi}} 
        \left(\frac{g^2}{8\Lphi}\right)^{\!\!\frac{1}{4}} \!\frac{e^{2\muh x}}{\sqrt{\mumax x}}}
            & \text{for } \ggOL \gtrsim 1
    \end{cases} \ ,
    \label{eq:hvarquartic}
\eeq
where we have defined the effective rate
\beq
    \muh ~\equiv~ \mumax - \frac{1}{2\gamma}\frac{\expt{a\Gamma_h}}{\sqrt{\Lphi}\varphiamp}
    \label{eq:muh}
\eeq
in terms of the conformal decay rate of Eq.~\eqref{eq:Gammahavg}.
The Lorentz factor 
\il{\gamma\simeq\max(1,\sqrt{\Lphi\hspace{-0.2mm}/\hspace{-0.4mm}g^2})} 
accounts for the dilation of relativistic Higgs decays, which 
is negligible in the \il{\ggOL \gtrsim 1} regime 
but important for \il{\ggOL \lesssim 1}.
Finally, the vacuum decay time is found in both regimes:
\beq
    \xdec \simeq\hspace{0.3mm}
    \begin{cases}
        \displaystyle{\frac{-1}{4\mu_h}W_{-1}\hspace{-1mm}\left[\frac{-9\Lh^2}{32\pi^2}\frac{F_1^2\mu_h}{T\mumax}\right]} & \!\!\!\text{for } \ggOL \lesssim 1 \\[9pt]
        \displaystyle{\frac{-1}{4\mu_h}W_{-1}\hspace{-1mm}\left[\frac{-9\Lh^2}{(2\pi)^6}\frac{\mu_h}{\mumax}\right]} & \!\!\!\text{for } \ggOL \gtrsim 1 
    \end{cases} \, . 
    \label{eq:xdecquartic}
\eeq
The \il{\ggOL \gtrsim 1} result is based on the destabilization condition
in Eq.~\eqref{eq:xdeccondition}.  Meanwhile, for \il{\ggOL \lesssim 1}
the Higgs modes oscillate slowly relative to the inflaton background, 
so that \il{3\Lh\hvar \gtrsim g^2\phiamp{}^2} is an appropriate condition.
Additionally, note that $\mumax$ depends on the regime of
the resonance, as given explicitly in Sec.~\ref{subsec:Parametric}.

Manifestly, the effective rate $\mu_h$ shows the competition 
between particle production and decays.  
For large enough coupling, the decays can eventually dominate,
because $\mumax$ is globally bounded from above; the decay rate
of produced Higgs particles scales with the coupling, but
the rate of particle production does not.

Let us compute the critical value of the coupling for which
\il{\xdec \geq \xNL}, \ie, the smallest coupling for which 
the metastable vacuum survives preheating.
We shall separately consider the couplings 
corresponding to the broad and narrow resonances.  
In the broad regime, we find \il{\ggOL \gtrsim 2.4\times 10^3}.
Likewise, using Eq.~\eqref{eq:narrowfit} for $\mumax$, for
the narrow resonances we find \il{\ggOL \gtrsim 213}.
The dramatic gap between these thresholds reflects the 
difference in the strength of the resonance between these two regimes.
Indeed, confirming our earlier observations in Eq.~\eqref{eq:nogrowth},
we find that perturbative decays of the Higgs stabilize the electroweak vacuum 
for sufficiently large quartic couplings, \emph{and these couplings
significantly differ based on the regime of the parametric resonance.}

To paint a more complete picture and confirm
our analytical results, we solve the mode equations numerically 
within the one-loop Hartree approximation and 
scan continuously over the range of couplings $\ggOL$;
this is shown in the left-hand panel of Fig.~\ref{fig:slices}.  
The black and gray curves show 
the numerical results with and without perturbative decays, 
respectively.  In line with our calculations in Eq.~\eqref{eq:xdecquartic},
we find that $\xdec$ is highly \emph{non-monotonic} 
with respect to the quartic coupling.
Indeed, this behavior originates from the structure of the 
resonance bands (in Fig.~\ref{fig:bands}),
in which the minima (maxima) of the oscillations correspond 
to the broad (narrow) resonance regimes, respectively.  
Therefore, evaluating our analytical result in 
Eq.~\eqref{eq:xdecquartic} at the broad and narrow $\mumax$ values
estimates the envelope of the $\xdec$ oscillations.
This envelope is shown by the thick green curves at \il{\ggOL \geq 1}
in Fig.~\ref{fig:slices}.
We also show our analytical estimate for the \il{\ggOL \leq 1} region.  
Meanwhile, the pivotal role of the perturbative decays is highlighted 
by the gray curve, which shows the numerical $\xdec$ with perturbative
decays turned off.  Indeed, in the absence of decays, 
the vacuum is stabilized only for very small coupling \il{\ggOL \lesssim 0.25}.

\subsubsection{From Tachyonic Instability}

Likewise, let us consider the pure curvature coupling limit,
in which \il{g=0}.  We once again confine our discussion to \il{\xieff > 0}
since otherwise the vacuum decays.
From Sec.~\ref{sec:NoBackreaction}, we know that the growth
of fluctuations from this tachyonic source is very different from
the parametric instability.  Most notably, the tachyonic production 
ceases once \il{x \gtrsim \xxi}, even for the zero mode.
We estimate the variance again using \il{\hvar \simeq n_h/(a^3\omega_h)},
and since the effective mass is curvature dominated we have
\il{\omega_{h_k}^2 \simeq \xieff R \simeq \xieff\Lphi\phiamp{}^4}.
Using Eq.~\eqref{eq:nhtachyonic} then yields
\beq
    \hvar ~\simeq~ \frac{3^{15/8}}{8}\frac{\xieff^{1/4}\Lphi\phiamp{}^4}{(8\pi)^{3/2}}
    \left(\frac{x}{x_0}\right)^{\!\!4\sqrt{\frac{2\xieff}{3\sqrt{3}}}}
\eeq
for \il{x < \xxi}.  Afterward, there is no particle production, so $n_h$
is fixed and \il{\omega_h\propto \phiamp{}^2\propto 1/a^2}.  The variance 
then redshifts as \il{\hvar \approx n_h/(a^3\omega_h) \propto 1/a},
which is slower than the scaling one would find in the quartic-dominated case.

The fact that $\hvar$ redshifts in this way is a crucial observation.
If the ratio \il{\hvar/\hcrit^2} grows at times \il{x\gtrsim \xxi} 
due to its redshifting behavior, the vacuum may decay even after particle 
production shuts down.  For the quartic interaction, this ratio would
be fixed.  However, for the curvature interaction, the variance redshifts 
more slowly \il{\hvar \propto 1/a}, and the barrier 
redshifts more rapidly \il{\hcrit^2 \simeq \xih R/|\Lh| \propto 1/a^4}.  
Consequently, the ratio redshifts as \il{\hvar/\hcrit^2 \propto a^3}
and grows.  Even if the Higgs remains stable throughout the tachyonic phase,
it destabilizes once the fluctuations grow beyond the extent of
the barrier \il{\hvar \gtrsim \hcrit^2};  this leads to a vacuum decay at
\beq
    \xdec ~\simeq~ \frac{8\sqrt{\pi}\xieff^{3/4}}{3^{1/8}|\Lh|^{1/3}}\left(\frac{x_0}{\xxi}\right)^{\!\!\frac{4}{3}\!\sqrt{\!\frac{2\xieff}{3\sqrt{3}}}} \ ,
    \label{eq:xdeccurvature1}
\eeq
which is valid for \il{\xdec > \xxi}, or equivalently \il{\xieff \lesssim 16}.
Note that the presence of even a small quartic coupling may
shield the vacuum from this effect, since the quartic term will 
eventually dominate \il{g^2\phi^2 \gtrsim \xieff R} and 
halt the growth of \il{\hvar/\hcrit^2}, which happens once
\il{x\gtrsim \sqrt{12\xieff\Lphi\hspace{-0.2mm}/\hspace{-0.3mm}g^2}}.

Otherwise, destabilization occurs during the tachyonic phase
and the condition in Eq.~\eqref{eq:xdeccondition} can be used 
again to produce an estimate.  This task is now more complicated
since backreaction is not the only source of tachyonicity.  
The vacuum could decay in one of two ways:
(i) by the $3\Lh\hvar$ term
or (ii) directly by the tachyonic mass of the curvature term.
A coupling of order \il{\xieff\gtrsim\mathcal{O}(10)} is sufficient
to produce the latter, and in this case the vacuum decays rapidly.  
For smaller $\xieff$, however, the false vacuum can survive much longer.
Using Eq.~\eqref{eq:xdeccondition} with the approximation 
that \il{\Delta x \simeq T/2} leads to 
\beq
    \xdec ~\simeq~ \frac{256\sqrt{2}\pi^{3/2}\xieff^{3/4}}{9\cdot 3^{7/8}|\Lh| T^2}\left(\frac{\xxi}{x_0}\right)^{\!\!1-4\sqrt{\!\frac{2\xieff}{3\sqrt{3}}}} \ .
    \label{eq:xdeccurvature2}
\eeq
We consider this result consistent with our assumptions only
if Eq.~\eqref{eq:xdeccurvature2} gives an earlier decay time
than Eq.~\eqref{eq:xdeccurvature1};  in terms of the curvature coupling, 
this implies a region of validity \il{\xieff \gtrsim 4}
for Eq.~\eqref{eq:xdeccurvature2}.

Our numerical computations of the vacuum decay time $\xdec$ 
are shown in the right panel in Fig.~\ref{fig:slices}.  
The pure curvature coupling case corresponds to
the black curve and the analytical estimates constitute the thick
green curve.  We have also included the numerical results for several 
small quartic couplings in the range \il{10^{-3} \lesssim \ggOL \lesssim 10^{-1}}.
These results are in line with our rough estimates and confirm that the 
presence even of a small quartic coupling can prevent decay of the false vacuum.
We have neglected \il{\xieff \gtrsim \mathcal{O}(10)} in our 
analytical estimates since these correspond to rapid destabilization,
for which the precise values of $\xdec$ are not essential.

\subsubsection{The Mixed Case}

Let us finally consider the general mixed case in which both 
couplings $\xieff$ and $\ggOL$ are non-zero.
As we have observed in Sec.~\ref{subsec:MixedCase}, if the strength of
the curvature coupling is at least comparable to $\ggOL$ 
the system undergoes a sequence of distinct phases of particle production.
In the context of vacuum stability, this means that the electroweak
vacuum must initially survive a phase of tachyonic production until 
\il{\xxi\simeq [6\xieff/\sqrt{\ggOL}]^{1\hspace{-0.2mm}/\hspace{-0.2mm}2}}
and then survive the parametric instability until non-linear 
dynamics set in at $\xNL$.  In what follows below, we shall assume 
such a scenario for our analytical calculations.

Rather than concern ourselves directly with $\xdec$, we produce an 
estimate of the constraint for vacuum metastability
by examining where \il{\hvar\gtrsim \hcrit^2}.  In the \il{\xieff > 0} region 
we can write $\hvar$ using Eq.~\eqref{eq:mixedn} together with 
\il{\hvar\simeq n_h/(a^3\omega_h)}.  We arrive at the expression
\beq
    \hvar ~\simeq~ \frac{\Lphi\varphiamp}{8\pi^3\sqrt{\xieff}}\left(\frac{T}{2x}\sqrt{\frac{g^2}{2\Lphi}}\right)^{\!\!\!3/2}
    \!\!e^{\frac{12\sqrt{2}\pi\xieff}{T\sqrt{\!\ggOL}}\left(\!\frac{1}{x_0}-\frac{1}{x}\!\right)} \ .
    \label{eq:mixedhvar}
\eeq
Owing to the exponential growth of $\hvar$, the constraint on 
our parameter space has only a logarithmic sensitivity to 
$\hcrit^2$ and the coefficient in Eq.~\eqref{eq:mixedhvar}.  
That said, the constraint is sensitive to the rate of perturbative
decays, and we should reintroduce this rate as we calculate our result.
Along these lines, we find 
\begin{align}
    \xieff  ~&\lesssim~  \frac{x_0 T}{12\sqrt{2} \pi}  \sqrt{\frac{g^2}{\Lphi}} 
    \left[\log\mathcal{C}_+ + \frac{\expt{a\Gamma_h}}{\sqrt{\Lphi}\varphiamp}\xdec\right]  \ ,
     \label{eq:xiconstraintP}
\end{align}
where all of the logarithmic terms have been 
collected into the quantity $\log \mathcal{C}_+$.  
The dependence on $\xdec$ and $\langle a\Gamma_h\rangle$ are important. 
To manage $\xdec$, a natural assumption is that \il{\xdec \lesssim \xxi} in the region we
are evaluating, since otherwise the tachyonic effect is not 
strong enough to destabilize the vacuum.  We therefore have
at most \il{\xdec \simeq (6\xieff/\sqrt{\ggOL})^{1/2}}.
We can solve the more complicated inequality that results for $\xieff$
(and we shall do this numerically below),
but based on our observations of the tachyonic effect in this section
\il{\xdec=\mathcal{O}(10)} serves as an appropriate simplifying approximation.  
As discussed in Sec.~\ref{subsec:PerturbativeDecays}, the 
perturbative decays arising from the quartic interaction are 
generally more relevant, and these scale parametrically as \il{\Gamma_h\propto \sqrt{\ggOL}}.
These considerations lead to a constraint of roughly 
\il{\xieff \lesssim \mathcal{O}(0.1)\ggOL} to ensure metastability.

\begin{figure}[tb]
    \centering
    \includegraphics[keepaspectratio,width=0.49\textwidth]{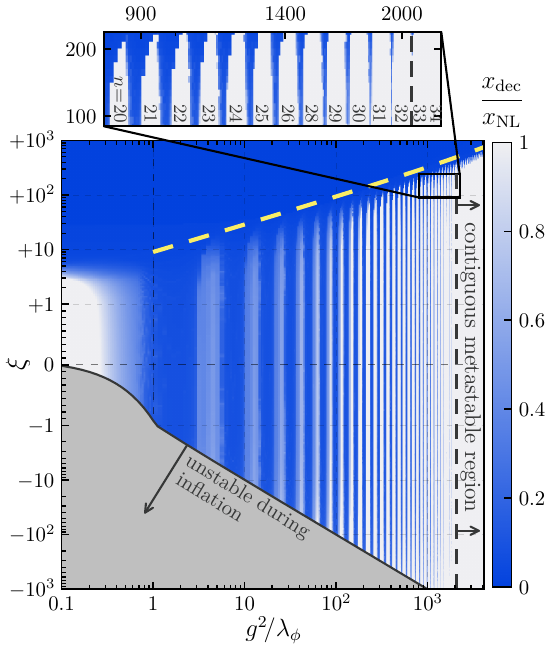}
    \caption{The decay time $\xdec$ of the electroweak vacuum
    computed numerically and normalized by $\xNL$. The 
    gray region is excluded due to violating Eq.~\eqref{eq:stabilityinflation}.
    The white regions show ``islands of (meta)stability'' in which 
    the false vacuum survives preheating.  
    Along the $\ggOL$ direction, the pattern 
    reflects the band structure in Sec.~\ref{subsec:Parametric}, 
    with the least stable regions centered around \il{\ggOL = 2n^2} 
    (for \il{n\in\mathbb{N}}).
    For larger $\ggOL$, perturbative Higgs decays enlarge 
    the metastable regions until they become contiguous at
    \il{\ggOL \gtrsim 2\times 10^3} 
    [in agreement with Eq.~\eqref{eq:xdecquartic}].
    The effect of the curvature interaction is to form an envelope
    over the metastable regions, and the yellow curve 
    shows our estimate [from Eq.~\eqref{eq:xiconstraintP}] 
    for \il{\xieff > 0}.
    }
\label{fig:xdec}
\end{figure}

An interesting feature of the mixed-coupling scenario 
is that it opens the possibility of metastability in the 
\il{\xieff < 0} region, which is ruled out if \il{\ggOL = 0}.
A similar procedure to the above is followed to produce a bound on the 
\il{\xieff < 0} region, the details of which are included in
Appendix~\ref{sec:tachyonic}.  The constraint is expressed as
\begin{eqnarray}
& \hskip -0.32cm -\xieff - \xiphi \frac{g^2}{\Lphi} \lesssim 
    \frac{ \small \log\mathcal{C}_- 
       - \frac{\sqrt{6}
      	H_1(3,0)}{3T\sqrt{-\xi}}\frac{g^2}{\lambda_\phi}(x_{\rm dec}^2- x_0^2)
      + (\xdec-x_0)\frac{\langle a\Gamma_h\rangle   }{\sqrt{\Lphi}\varphiamp}}{\frac{8}{T}\sqrt{\frac{6}{-\xieff}}H_1(3,0)\log\!\big(\frac{\xdec}{x_0}\big)}, \nonumber \\
     \label{eq:xiconstraintN}
\end{eqnarray}
and we have again taken \il{\xdec=\mathcal{O}(10)} and packaged
the logarithmic terms in a quantity $\mathcal{C}_-$.
The function giving \il{H_1(3,0)\approx 0.76} is 
defined in Appendix~\ref{sec:tachyonic}.
Note that for \il{\xieff \gg 1} the term proportional 
to $\ggOL$ on the left-hand side is subdominant and the remaining terms
balance for \il{|\xieff| \sim \ggOL}.  
Therefore, neglecting numerical coefficients, the constraint becomes  
\il{\xieff \lesssim \mathcal{O}(g^2/\Lphi)}.  The negative
curvature coupling thus yields a weaker bound, which is 
expected based on our observations from Sec.~\ref{subsec:MixedCase}.
Note that the inflation constraint \il{\xih - \xiphi\ggOL \gg 1/12} 
in Eq.~\eqref{eq:stabilityinflation} yields a bound that is roughly
coincident.

A more complete picture of the constraints is achieved by numerical 
calculations, and we display these in Fig.~\ref{fig:xdec}.
For this figure, we have calculated the time of vacuum decay $\xdec$ over the 
\il{\{\ggOL,\xieff\}} parameter space, 
making the simplifying assumption that \il{\xiphi = 0}.
Note that the gray region excludes the subset of couplings 
that destabilizes the vacuum during inflation.
Additionally, note that the curvature coupling is shown on a 
log scale over the negative and positive axes, with the exception 
of the range \il{-1 \leq \xieff \leq +1} where the scale is linear.
The numerics run until a time $\xNL$
so that the white regions effectively show where \il{\xdec/\xNL \geq 1} 
and thus where the vacuum remains metastable throughout preheating.

Echoing the behavior observed in Fig.~\ref{fig:slices},
we find that over the vast majority 
of parameter space, the metastable regions
are not connected.   Instead, we find a large number of disjoint
``islands of (meta)stability'' scattered 
throughout.  Indeed, \emph{the constraint for 
metastability cannot be fully expressed as a simple bound on the couplings}---the 
fate of the electroweak vacuum depends on $\ggOL$ in a highly monotonic way.
Naturally, the most unstable regions appear where the characteristic 
exponent $\mumax$ is maximized---\ie, around the broad resonances 
\il{\ggOL = 2n^2} (for \il{n\in\mathbb{N}}).
Conversely, the most stable regions appear around the narrow resonances,
where $\mumax$ is minimized [see Eq.~\eqref{eq:narrowfit}].
As $\ggOL$ is taken to larger values, the metastable regions
grow and start to merge, as shown by the magnified inset panel.  
The integers written over the metastable regions
correspond to the couplings of the narrow resonances written in 
the form \il{\ggOL = 2n^2+n}.
The couplings eventually become large enough for perturbative
decays to stabilize the Higgs, regardless of the resonance regime,
forming a contiguous metastable region for \il{\ggOL \gtrsim 2\times 10^3}.  
This observation agrees with our analytical estimates in 
Eq.~\eqref{eq:nogrowth} and Eq.~\eqref{eq:xdecquartic} for the broad regime.
In principle, the only limitation in taking larger couplings is that we
do not ruin the flatness of the inflaton potential, which (as
briefly discussed in Sec.~\ref{subsec:InflationaryRegime}) roughly
requires \il{\ggOL \ll 10^5}.

Beyond these observations, we notice that the interplay of the 
couplings is also clarified in Fig.~\ref{fig:xdec}.
In particular, the figure shows that the Higgs-curvature interaction 
has the effect of cutting off the metastable regions and 
imposing an envelope that scales with the couplings.
This envelope is what we have analytically estimated in Eq.~\eqref{eq:xiconstraintP}
and Eq.~\eqref{eq:xiconstraintN}.   
We have plotted the numerical solution to the inequality 
in Eq.~\eqref{eq:xiconstraintP} using a yellow-dashed curve.
This curve is indeed consistent with the approximate bound 
that we estimated \il{\xieff \lesssim \mathcal{O}(0.1)\ggOL}.  
We have not included the analogous curve for the \il{\xieff < 0} 
region since this closely coincides with the constraint (the gray region) 
on the stability of the vacuum during inflation.
On the whole, the interplay between the two interactions
effectively extends the range of metastability for the 
Higgs-curvature coupling, shielded by the presence of a 
similarly large Higgs-inflaton coupling.


\section{Conclusions and Discussion\label{sec:Conclusions}}


Some amount of degeneracy often exists in inflationary models in the sense
that observational constraints may be satisfied over a degenerate
subspace of the model parameters.  Undoubtedly, one reason
that studies of post-inflationary dynamics are essential is that
this dynamics may break such degeneracies by leading to qualitatively different
preheating histories, subject to a qualitatively different 
set of possible constraints. 
For example, as we have encountered in this paper, the stability of the Higgs
during inflation can be ensured by either a direct coupling to the 
inflaton or a non-minimal coupling to gravity.  
The dynamical roles played by these interactions are similar 
during inflation but differ dramatically during preheating, in which the 
non-trivial interplay between these interactions implies a 
rich structure of metastable regions and associated constraints on 
the Higgs couplings.  Indeed, the broader motivation for
this work is to explore dynamics that may reveal 
independent probes for models of early-universe cosmology.

In this paper, we have examined the massless preheating 
dynamics that arises in models composed of scale-invariant 
interactions, in which the inflaton potential is effectively 
quartic after inflation.  Most notably, we have focused on 
the implications these models have for electroweak vacuum metastability.  
We have provided constraints on the couplings for 
which the Higgs remains stabilized during and after inflation.
Among these couplings, we have included the possibility
that the Higgs and the inflaton have non-minimal couplings to gravity.  
While our study is motivated by addressing the metastability
of the vacuum,  our results are readily generalized 
to other approximately scale-invariant models. 

Several comments are in order.   First, 
while we have considered non-minimal gravitational
couplings for both the Higgs and inflaton fields, the Higgs coupling
$\xih$ has received the bulk of our attention.  
And while the effects of $\xiphi$ have been included in our analysis
through the effective curvature coupling $\xieff$ [in Eq.~\eqref{eq:xieff}],
our treatment can be extended in several ways.
For example, we have not taken into account the higher-order effects 
on the field fluctuations that appear as a result of \il{\xiphi\neq 0}
corrections to the background-field solutions.   For this reason,
our results are applicable only for \il{|\xiphi| \lesssim 1}, 
which is the scope assumed for this paper.
These higher-order effects are challenging to incorporate analytically
in massless preheating.  Notably, even the zeroth-order background
solution [the elliptic function in Eq.~\eqref{eq:varphi}]
is considerably more complicated than the sinusoidal form
found in massive preheating.  It would be worthwhile to study
the effects of \il{\xiphi \approx - \mathcal{O}(1)} without 
resorting to such approximations for the background equations of motion.

Second, our study is performed under the assumption
that there is no new physics beyond the SM that
significantly alters the renormalization-group evolution 
of the Higgs quartic coupling.  This assumption could be reasonable, 
as there are no hints of new physics in collider 
or weakly interacting massive particle dark-matter searches.  
Either way, the cosmological history 
following the preheating stage---\ie, the reheating epoch and,
more precisely, the inflaton decay channels---should be consistent 
with this assumption.   The reheating epoch could then be realized in 
a number of ways.  For instance, the inflaton could couple 
to the right-handed neutrino, which is responsible for 
neutrino masses and leptogenesis~\cite{Fukugita:1986hr}. 
In this case, reheating could be completed by the decay of inflaton 
to right-handed neutrinos and their subsequent decay to SM particles.  
Alternatively, if the inflaton is stable, it also 
opens the possibility that the inflaton is a candidate for 
dark matter~\cite{Almeida:2018oid,Lebedev:2021zdh}
or dark radiation~\cite{Babichev:2020yeo},
provided that preheating converts most of the inflaton energy 
density into SM degrees of freedom.  Indeed, the details and 
variations on these possibilities ultimately depend on the 
sign and size of the inflaton mass-squared term. 

Along another direction, some comments are in order regarding
the stage of preheating that occurs after $\xNL$ in our study.
After entering the non-linear stage of preheating, 
the energy density held in fluctuations 
is redistributed among the modes by rescattering processes.  
In general, the Higgs can destabilize during this time, and 
this possibility has been addressed for massive preheating 
in Ref.~\cite{Ema:2016kpf}, based on the results of 
Refs.~\mbox{\cite{Harigaya:2013vwa, Mukaida:2015ria}}.
In massive preheating, this destabilization arises because 
the effective Higgs mass induced by the inflaton-Higgs coupling redshifts
as \il{g^2\phi^2 \propto 1/a^3}, while that induced by the Higgs 
self-coupling redshifts as \il{3\Lh\hvar \propto 1/a^2}.
In other words, 
the tachyonic contribution continues to grow relative to 
the stabilizing mass term and can eventually trigger decay
of the vacuum, even though it was stabilized at the end
of the linear preheating stage.
The details of the thermal Higgs mass after preheating and 
the evolution of the background temperature thus become 
important to ensure metastability as thermalization begins.
That said, in massless preheating, these two mass 
contributions redshift at the same rate \il{g^2\phi^2 \sim 3\Lh\hvar \sim 1/a^2},
so these same concerns over destabilization are not as relevant.
Nevertheless, the stages of evolution leading to thermalization
are, of course, interesting in the context of massless preheating for a 
host of other reasons.  Furthermore, other considerations such as
the thermal corrections to the vacuum tunneling probability become important
during the reheating epoch, such as those studied in Ref.~\cite{Rose:2015lna}. 

Thermal effects could be important even before $\xNL$,
through direct decay channels (if any) of the inflaton into SM particles 
or through the Higgs decay products.  Whether the Higgs decay products 
become quickly thermalized is a nontrivial issue, since 
the particle-production rate can greatly exceed the thermalization rate.  
In this case, one may not necessarily assume that the Higgs decay products form 
a thermal plasma, even if the thermal relaxation time is much shorter than the 
age of the Universe at that moment.  Nevertheless, if instant thermalization 
of the decay products is assumed, we can discuss thermal backreaction on the Higgs dynamics. 
If the thermal mass of the Higgs becomes comparable to the zero-temperature effective mass, 
our analysis would be modified, and we would expect broader regions in Fig.~\ref{fig:xdec} 
to become (meta)stable.  Our preliminary study shows that the effect of the Higgs 
thermal mass is insignificant over a wide range of the parameter space. 
However, it is important to correctly account for the full evolution 
of the decay products without the thermalization assumption, which we leave
for future work.

As we recall from Sec.~\ref{subsec:xend}, the time $\xNL$ at which
the non-linear stage of preheating begins is central to our results.
We concluded, in our minimal model realization, that the inflaton
self-resonance is responsible for the onset of the non-linear stage,
as its fluctuations grow from the quartic self-resonance.
However, a natural extension is to consider the presence of 
a spectator field that does not couple to the Higgs.  
If such a spectator field is likewise subject to particle 
production, then $\xNL$ could be triggered by the growth of 
the spectator fluctuations instead; this is not difficult to 
arrange, considering that the characteristic exponent 
for the inflaton self-resonance is rather small at \il{\mumax\approx 0.036}.
A reduction in $\xNL$ would imply that our metastable
regions, such as those that appear in Fig.~\ref{fig:xdec}, grow in extent.
This enhancement could be substantial.  For example,
taking a quartic coupling between the spectator and the inflaton,
with the maximal exponent \il{\mumax \approx 0.24}, we find 
that \il{\xNL\approx 63}, roughly a factor of 6 smaller
than the inflaton self-resonance.  

Moreover, the possibility of spectator fields opens other avenues 
of exploration.  For instance, such fields could generate important
cosmological observables such as non-Gaussianities in the density 
perturbations~\cite{Chambers:2008gu}.  
Additionally, spectator fields could naturally serve 
as dark-matter candidates.  The corresponding relic abundance 
generated from preheating would depend non-trivially on the 
couplings and spins of the spectator fields.  It is worth noting
that such dark-matter production from the thermal bath with a generic 
equation of state---including the radiation-like equation of
state relevant to our study---has been a topic 
of recent interest~\cite{Garcia:2020eof,Garcia:2020wiy}.

There are also a number of possible extensions that would be interesting
to explore in the context of multi-field inflation.  For instance,
in models which allow for a sizeable angular inflaton velocity 
at the end of inflation~\cite{Cline:2019fxx,Kawasaki:2020xyf}, 
the inflaton-dependent modulation of the effective Higgs masses could be suppressed.
As a result, the rate of particle production could be suppressed as well. 
On the other hand, the onset of the non-linear stage of preheating 
would also occur at a later time $\xNL$.  The end result of the competition
between these two effects, and the more complicated preheating dynamics altogether, 
requires a dedicated study~\cite{spiraling}.
An alternative multi-field extension could involve
hybrid inflation~\cite{Linde:1993cn}.  In this case,
vacuum stability during inflation is ensured 
by the inflaton-Higgs coupling as usual, but this 
coupling will not oscillate much during the preheating 
phase since most of the energy is transferred to the 
so-called waterfall fields~\cite{GarciaBellido:1997wm}.
These features imply that the electroweak vacuum would be
relatively stable during both inflation and preheating.

It would also be interesting to consider the presence 
of additional terms that break the scale invariance of the 
theory, as these can have rich dynamical implications.  
While the breaking of the scale invariance due to the non-minimal 
coupling terms are relevant only around the end of inflation, 
effects of lower-dimensional terms become more and more important 
at a later time.  For instance, a non-vanishing inflaton mass 
$\mphi$ might alter the parametric resonance once the inflaton 
oscillations reach a sufficiently small amplitude~\cite{Greene:1997fu}. 
In particular, for couplings \il{g^2/\lambda_\phi \gtrsim \lambda_\phi/\mphi^2} 
the resonance becomes stochastic once \il{\phiamp \lesssim \mphi/\!\sqrt{\Lphi}}. 
The parametric resonance then increasingly behaves as in massive 
preheating (see Appendix~\ref{sec:quadratic}), with the stochastic resonance 
giving way to the narrow resonance once $\bar\phi\lesssim \mphi/g$. 
Not only does this evolution change the spectrum of produced 
particles and the production rates, the introduction of \il{\mphi\neq 0} 
also allows the resonance to terminate at some time before $\xend$.
A small inflaton mass can thus modify our constraints 
for electroweak metastability in a non-trivial way.

Yet another possible extension of our work concerns 
the gravity formulation.  In this paper, given 
the non-minimal couplings of our scalar fields to gravity, 
we have observed some of the physical distinctions that may 
appear between the metric and Palatini formulations.
Still, a more general extension could be explored along the lines of the
Einstein-Cartan formulation in the presence of Holst and Nieh-Yan 
terms~\mbox{\cite{Shaposhnikov:2020gts,Shaposhnikov:2020frq,Shaposhnikov:2020aen}}. 
The Einstein-Cartan theory generalizes our treatment, reproducing 
the metric and Palatini formulations in certain limits and
allowing for a continuous interpolation between them.

All in all, if the metastability of the electroweak
vacuum may be ensured in models from which massless preheating
emerges, we may adopt a renewed interest in observables that 
are sensitive to the details of preheating.
A well-suited example consists of gravitational waves (GWs)
produced during this epoch~\cite{Figueroa:2017vfa}.
Massless preheating is distinctive in this sense
because the amplitude of the produced gravitational radiation
does not dissipate and the frequency of the waves does not redshift.
While these properties may present challenges for the 
observational prospects of GWs from such scenarios, several
methods show promise~\cite{Aggarwal:2020olq}.  For example, 
radio telescopes may be used to examine the distortion in 
the CMB from by-products of high-frequency GWs interacting 
with background magnetic fields~\cite{Domcke:2020yzq}. 
Indeed, such observations could even be used to probe broader
early-universe properties, such as the 
energy scale for inflation~\cite{Cai:2021gju}.


\begin{acknowledgments}
TT thanks Kazunori Nakayama for useful discussions regarding Ref.~\cite{Ema:2016kpf}.
The research activities of CSS, JK, and TT were supported in part by the IBS 
under project code IBS-R018-D1 and those of JK were supported 
in part by the Science and Technology Research Council (STFC) 
under the Consolidated Grant ST/T00102X/1.
This work was completed in part at the Aspen Center for Physics,
which is supported by the National Science Foundation under Grant PHY-1607611. 
The opinions and conclusions expressed herein are those of the authors, 
and do not represent any funding agencies.
\end{acknowledgments}


\appendix


\section{Comparison to Massive Preheating\label{sec:quadratic}}


In this appendix, we provide a brief overview of the ``massive 
preheating'' dynamics, in which our Lagrangian takes the 
same form as Eq.~\eqref{eq:model}, but the potential is 
instead given in the Jordan frame by
\beq\label{eq:VJquad}
\VJ(\phi,h) ~=~ \frac{1}{2}\mphi^2\phi^2 + \frac{1}{2}g^2\phi^2h^2 + \frac{1}{4}\Lh h^4 \ ,
\eeq
where $\mphi$ is the inflaton mass.
Note that previous studies in the literature have examined 
electroweak vacuum metastability for this model in 
the \il{\xiphi = 0} limit~\mbox{\cite{Ema:2016kpf,Ema:2017loe}},
and we refer the reader to those papers for details beyond the scope
of our summary.

All other features of the model beyond the mass term 
are assumed to remain the same.  
After the end of inflation, when \il{|\xiphi|\phi^2 \ll 1} 
the inflaton potential is approximately quadratic and the field evolves
according to
\beq
    \ddot{\phi} + 3H\dot{\phi} + \mphi^2\phi ~=~ 0 \ .
\eeq
The inflaton amplitude then redshifts as \il{\phiamp\propto 1/a^{3/2}}
and [in line with Eq.~\eqref{eq:eqnofstate}] the cosmological
equation of state is matter-like.  The inflaton field evolution 
is sinusoidal:
\beq\label{eq:phiquad}
    \varphi(x) ~=~ \varphiamp\cos(x - x_0) \ ,
\eeq
where in analogy to Sec.~\ref{subsec:InflatonEvolution}
we have defined \il{\varphi = a^{3/2}\phi} and the dimensionless
time \il{x\equiv mt}.  The constant $x_0$ is fixed by the
conditions at the end of inflation, and we shall assume 
\il{\phiend = \mathcal{O}(1)}.  The scale factor evolves as
\beq
    a(x) ~=~ \frac{1}{2}\big(\sqrt{3}\varphiamp x\big)^{\!2/3}
\eeq
and thus for consistency \il{\xend = 2\sqrt{2/3}\phiend^{-3/2}}.

To compute the growth of fluctuations we write
the equations of motion for the Higgs modes in the Einstein frame.
By defining \il{\mathcal{H}_k \equiv a^{3/2}\Omega^{-1/2}|_{h = 0}h_k}, 
we can remove the damping terms, and up to $\mathcal{O}(a^{-3})$ we have
\beq
    \mathcal{H}_k'' + \omega_{\mathcal{H}_k}^2\mathcal{H}_k ~=~ 0 \ ,
    \label{eq:quadmodeeqns}
\eeq
in which the effective masses are 
\begin{align}
    \omegaHk^2 ~&=~ \frac{\kappa^2}{a^2} + \Big(\frac{g^2}{\mphi^2} + \xih\Big)\frac{\varphi^2}{a^3}
    + \xieff\Big(\frac{\varphi^2}{a^3} - \frac{\varphi'^2}{a^3}\Big) \nn \\
    ~&=~ \frac{\kappa^2}{a^2} + \frac{\varphiamp{}^2}{a^3}\Big[\xieff\sin^2(x-x_0) \\
    &\hspace*{22mm} + \Big(\frac{g^2}{\mphi^2} + \xieff + \xih\Big)\cos^2(x-x_0)\Big] \nn \ .
\end{align}
We have defined the rescaled momenta \il{\kappa \equiv k/m}, and in the 
second line we have employed Eq.~\eqref{eq:phiquad}.
Additionally, in analogy to Eq.~\eqref{eq:xieff} we have defined the
effective curvature coupling 
\beq
    \xieff \equiv  \xih + \xiphi - 6\theta\xih\xiphi - \frac{3}{8} \ .
\eeq

The rate of cosmological expansion is much smaller than that of the
inflaton oscillations, so that the mode equations in Eq.~\eqref{eq:quadmodeeqns} 
approximately take the Mathieu form \il{\mathcal{H}_k'' + \{A_k - 2q\cos[2(x-x_0)]\}\mathcal{H}_k = 0},
with the slowly varying parameters given by
\begin{align}
    A_k ~&\equiv~ \frac{\kappa^2}{a^2} + \frac{\phiamp{}^2}{2}\!\left(\frac{g^2}{\mphi^2} + \xih\right) \nn \\
    q ~&\equiv -\frac{\phiamp{}^2}{4}\!\left(\frac{g^2}{\mphi^2} + 2\xieff + \xih\right) \ .
\end{align}

An important observation is that the contributions 
to the effective Higgs mass from the quartic interaction 
\il{g^2\phi^2 \propto 1/a^3} and non-minimal coupling \il{\xih R\propto 1/a^3}
\emph{redshift identically} in the massive preheating scenario.
Because much of the complexity we found in our study of massless
preheating arose from the mismatch between these two terms,
the massive preheating scenario is simpler in this respect.
For example, whether particle production for a given mode 
occurs through the tachyonic or parametric instability is 
determined by the sign of \il{A_k - 2|q|}.  Namely, 
\il{A_k - 2|q| > 0} yields the parametric instability, and
\il{A_k - 2|q| < 0} yields the tachyonic instability.
We can write these explicitly as
\beq
    A_k - 2|q| ~=~ \frac{\kappa^2}{2} + \phiamp{}^2
    \begin{cases}
        \left(-\xieff \right) & \text{for } q < 0 \\
        \left( \frac{g^2}{\mphi^2} + \xieff + \xih \right) & \text{for } q > 0 
    \end{cases}
    \ .
\eeq
Therefore, unlike in the massless preheating scenario, the type 
of particle production that drives the zero mode is a fixed time-independent 
property based on the couplings.  Notably, in the \il{g,\xiphi \rightarrow 0}
limit we recover the well-known result that \il{3/16 < \xih < 3/8} 
is necessary for the complete absence of tachyonic 
production~\cite{Ema:2016kpf}.

Another crucial distinction between the massless and massive
preheating dynamics is found by reexamining the perturbative Higgs decays.
In particular, using the rest-frame decay rate $\Gamma_h$ in 
Eq.~\eqref{eq:Gammah}, the decays introduce a dissipative exponent
\il{-\int dx\,\Gamma_h/\gamma} that suppresses the growth of the Higgs number density.
In terms of the Mathieu parameters the effective Higgs mass
is given by \il{m_h^2 = A_0 - 2q\cos[2(x-x_0)]} so that
\beq
    -\int dx\,\Gamma_h ~\propto~ \int dx \sqrt{A_0 - 2q\cos{[2(x-x_0)]}} \ ,
\eeq
where the integration region is over the preheating times \il{x \geq \xend}
for which \il{m_h^2 \geq 0} and we have taken \il{\gamma = 1}.
Because this integral has an identical form to the 
accumulated phase of the zero mode $\Theta_0$
[refer to the general quantity defined in Eq.~\eqref{eq:tachyonicformula}], 
we can employ previous calculations for $\Theta_k$ 
available in the literature~\cite{Dufaux:2006ee}.
The decay exponent amounts to a sum over each non-tachyonic 
oscillation period \il{-\int dx\Gamma_h \propto \int dx\,\Theta_0},
and we can extract the time dependence by
noting that the integrand scales approximately as 
\il{\Theta_0\propto \smash{\sqrt{|q|}} \propto 1/x} so that 
\beq
    -\int dx\,\Gamma_h ~\propto~ \sqrt{2|q|_{\rm end}}\log\left(\frac{x}{\xend}\right) \ .
    \label{eq:decayexponentquad}
\eeq

\begin{figure}[tb]
    \centering
    \includegraphics[keepaspectratio,width=0.49\textwidth]{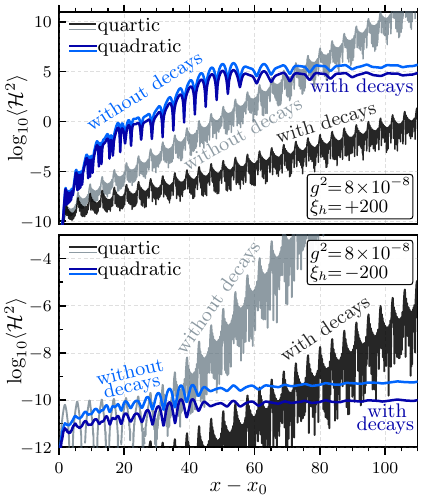}
    \caption{A comparison of the Higgs variance $\chvar$
    between massless and massive preheating, \ie, between the
    quartic and quadratic inflaton potentials, respectively, taking
    \il{\mphi = 10^{-5}} and \il{\Lphi = 10^{-10}}.
    The backreaction effect is not included in this figure.
    The top and bottom panels show a difference in sign for the 
    curvature coupling $\xih$.   We have also included curves that
    show the result if perturbative decays are turned off.  Notably,
    this effect is minimal in massive preheating but significant
    in the massless case.  The contrast between the preheating 
    scenarios is also evident in the short-lived growth of 
    fluctuations for the massive case which ceases for \il{|q| \ll \frac{1}{4}}.
    Note that although $x$ and $\mathcal{H}$ have the same qualitative 
    meaning in both preheating scenarios their explicit
    definitions differ due to the difference in background cosmology.
    }
\label{fig:ngrowthquad}
\end{figure}

The logarithmic time-dependence of the decay exponent in 
Eq.~\eqref{eq:decayexponentquad} demonstrates a critical distinction 
between the massive and massless preheating scenarios.
For the latter, the time-dependence of the decay exponent
was found [in Eq.~\eqref{eq:decayexponent}
and Eq.~\eqref{eq:Gammahavg}] to be linear, making it possible for the perturbative decays
to efficiently counter the growth of Higgs fluctuations, even shutting them
down for sufficiently large quartic coupling.
By contrast, the logarithmic time-dependence in massive preheating
shows that \emph{the capacity for perturbative decays to 
stabilize the electroweak vacuum is relatively negligible}.
This distinction makes the features of the massless preheating dynamics somewhat unique.

In order to further illustrate this comparison, we have plotted 
the evolution of the Higgs variance \il{\chvar} in Fig.~\ref{fig:ngrowthquad}
for both the quadratic and quartic scenarios,
using the quartic coupling \il{\ggOmm = \ggOL = 800} and curvature couplings
\il{\xih = \pm 200}.  The inflaton mass \il{\mphi = 10^{-5}} and
coupling \il{\Lphi = 10^{-10}} are taken in the respective cases.
In contrast to the unimpeded growth that arises from the quartic potential, 
we observe the end of particle production that occurs in the quadratic case 
once \il{|q| \lesssim 1/4}.  And comparing the curves, which show the 
results with perturbative decays both included and not included,
we see manifestly the relative importance of perturbative decays in these scenarios.
Indeed, although the couplings are the same for each of the comparisons
in Fig.~\ref{fig:ngrowthquad}, the results for vacuum stability can 
greatly differ.  For instance, upon including backreaction one finds 
that, for the couplings in the bottom panel, the false vacuum survives in
the quadratic case, but it decays in the quartic case.  Notwithstanding,
owing to the complex pattern of metastability regions for the quartic case,
we can change this outcome with only a small change in the coupling.

\begin{figure}[tb]
    \centering
    \includegraphics[keepaspectratio,width=0.49\textwidth]{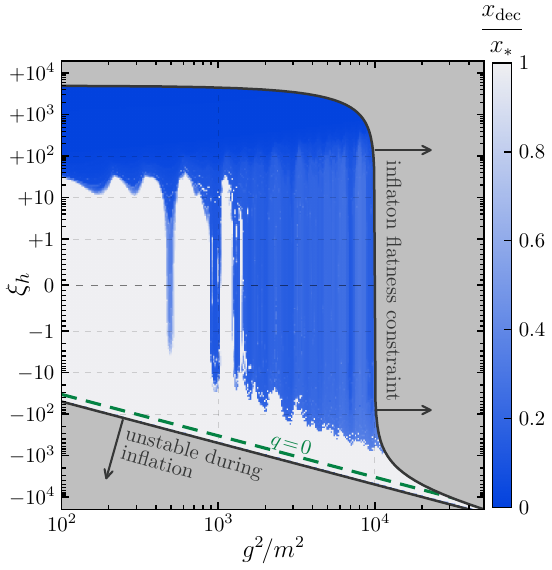}
    \caption{The decay time $\xdec$ of the electroweak vacuum in the 
    massive preheating scenario (for comparison with Fig.~\ref{fig:xdec}).
    The results are given relative to a fiducial
    time \il{x_* \approx 2\times 10^3} for which the resonance
    is deep in the narrow regime over the parameter space shown.
    Our most immediate observation is that the metastable 
    region is connected, which contrasts with the 
    more complex disjoint regions found in the massless 
    preheating scenario.  The features appear consistent with the 
    bounds found in previous studies in the literature~\cite{Ema:2016kpf,Ema:2017loe}.
    }
\label{fig:xdecquad}
\end{figure}

An analytical study of the preheating dynamics and vacuum decay times 
is found in the literature for the unmixed case~\cite{Ema:2016kpf}, 
while the mixed case is studied mostly numerically in Ref.~\cite{Ema:2017loe}.
For the purpose of our comparison we shall focus on numerical results for $\xdec$.
Along these lines, we provide an analog to Fig.~\ref{fig:xdec} 
for the massive preheating scenario in Fig.~\ref{fig:xdecquad}.
Note that in order to make a clear comparison between this 
result and the massless preheating result in Fig.~\ref{fig:xdec},
we have constructed Fig.~\ref{fig:xdecquad} using the same
axes and axis scales.   
The gray exclusion regions correspond either to the Higgs
destabilizing during inflation \il{\xih \lesssim -g^2\!/(2\mphi^2)} or the flatness of the inflaton potential
being ruined by quantum corrections \il{\xih \gtrsim \frac{1}{2}(10^{-6} - g^2)/\mphi^2}.
The line showing \il{q = 0} is significant since along it the effective
masses of the Higgs modes do not oscillate in time and 
the resonant particle production vanishes.

The most immediate observation we make in Fig.~\ref{fig:xdecquad}
is that the structure of the metastable region is relatively 
simple in comparison to Fig.~\ref{fig:xdec}.  For \il{\ggOmm \gtrsim 10^3},
the upper bound of the metastable region is roughly linear but 
flattens for smaller values of $\ggOmm$.  These results appear consistent 
with both Ref.~\cite{Ema:2016kpf} and Ref.~\cite{Ema:2017loe}.
In particular, in Ref.~\cite{Ema:2016kpf}, bounds on the unmixed couplings
were given approximately as \il{g^2/m^2\lesssim 10^{3}} and \il{\xieff \lesssim 10},
as reflected in our figure.


\section{Details for Tachyonic Production\label{sec:tachyonic}} 


In this appendix, we provide details of our calculations 
for which the non-minimal curvature couplings 
are dominant.   Let us restate for convenience the general form given in 
Eq.~\eqref{eq:tachyonicformula} for the tachyonic growth 
in phase-space density $n_{h_k}$ for a given mode after
passing through a tachyonic phase $j$ times~\cite{Dufaux:2006ee}:
\beq
    n_{h_k} ~=~ e^{2 j X_k} \left(2\cos\Theta_k\right)^{2(j-1)} \ .
    \label{eq:tachyonicformulaappendix}
\eeq
We recall that \il{X_k \equiv \int dt\,\Omega_{h_k}} 
denotes the accumulated exponent when the effective mass of 
the mode is tachyonic and \il{\Theta_k\equiv \int dt\,\omega_{h_k}} 
denotes the accumulated phase during the non-tachyonic intervals.  
Our primary objective is to evaluate these factors so that we may
compute the variance of fluctuations $\hvar$ and, hence, the vacuum
destabilization time for a given parameter set.
In what follows, although the curvature terms have an overall time dependence, 
we neglect this within a single tachyonic interval.  

Let us define functions $H_1(\alpha,\beta)$ and $H_2(\alpha,\beta)$ that
will hold the oscillatory part of $X_k$ and $\Theta_k$.
The way these functions are paired depends on the sign of the coupling.
Namely, for the accumulated phase we have
\beq
    \Theta_k = 
    \begin{cases}
        2\sqrt{+\rh - \kappa^2} H_1(\alpha, \beta) & \text{for } \xieff > 0 \\
        2\sqrt{-\rh + \kappa^2} H_2(\alpha, \beta) & \text{for } \xieff <  0
     \end{cases} \ ,
\eeq
and for the growth exponent we have
\beq
    X_k = 
    \begin{cases}
        2\sqrt{+\rh - \kappa^2} H_2(\alpha,\beta) & \text{for } \xieff > 0  \\
        2\sqrt{-\rh + \kappa^2} H_1(\alpha,\beta) & \text{for } \xieff < 0 
    \end{cases}
\eeq
and we have introduced the parameter combinations
\beq
    \alpha ~\equiv~ \frac{3\rh}{\rh - \kappa^2} - 2\rphi\beta \ , 
    \qquad  \ 
    \beta ~\equiv~ \frac{\ggOL}{\rh - \kappa^2}  \ .
    \label{eq:params}
\eeq
Meanwhile, the oscillatory functions are defined as
\begin{align}
    H_1 (\alpha, \beta) &\equiv  
    \int_{0}^{\xcross} \hspace{-0.0mm}  dx \sqrt{\beta\left(\frac{\varphi}{\varphiamp}\right)^{\!\!2} + \alpha\left(\frac{\varphi}{\varphiamp}\right)^{\!\!4} - 1} \nn \\
    H_2 (\alpha, \beta) &\equiv  
    \int_{\xcross}^{T\!/4} \hspace{-1.3mm}dx \sqrt{1 - \beta\left(\frac{\varphi}{\varphiamp}\right)^{\!\!2} - \alpha\left(\frac{\varphi}{\varphiamp}\right)^{\!\!4}} 
    \label{eq:Hfuncs}
\end{align}
where $\varphi$ refers to the analytical solution for the
background inflaton field in Eq.~\eqref{eq:varphi}.
Note that for our purposes here we take \il{x_0 = 0}.
The quantity $\xcross$ is the crossing time that separates tachyonic 
and non-tachyonic phases of evolution for
the mode, which can be expressed in terms of the inverse elliptic function as
\beq
    \xcross(\alpha ,\beta) ~\equiv~   
    \text{arccn}\!\left(\!\sqrt{\frac{\!-\beta + \sqrt{4 \alpha + \beta^2}}{2\alpha}}, 
    \!\frac{1}{\sqrt{2}}\!\right) \ .
\eeq

For the remainder of this appendix we shall focus solely
on the accumulated exponent $X_k$ rather than the phase $\Theta_k$.
While the latter is relevant for the detailed distribution of $n_{h_k}$,
it is irrelevant to the overall growth.
For clarity, we divide the following analysis in a manner parallel to 
Sec.~\ref{sec:NoBackreaction} and Sec.~\ref{sec:Backreaction} in the paper,
giving additional support to our derivation of results in the main text.

\subsection{Pure Curvature Coupling (\il{g = 0})}

In the limit of pure curvature coupling, our parameters in
Eq.~\eqref{eq:params} reduce to \il{\alpha = 3\rh/(\rh-\kappa^2)} and \il{\beta = 0},
and the tachyonic phase appears only for \il{\alpha > 1}.  
As in the main body of the paper, we ignore the \il{\xieff < 0} 
region, since the vacuum is not stable.
Identifying the limits \il{H_1 (1, 0) = 0} 
and \il{H_2(1,0) = \sqrt{2}} along with
\begin{align}
    H_1 (\alpha \rightarrow \infty,0) ~&\simeq~  \frac{2\pi^{3/2}\sqrt{\alpha}}{\Gamma(\frac{1}{4})^2} \nn \\
    H_2 (\alpha \rightarrow \infty,0) ~&\simeq~ \sqrt{\frac{\pi}{2\sqrt{\alpha}}}\frac{\Gamma(\frac{5}{4})}{\Gamma(\frac{7}{4})} \ ,
\end{align}
we can construct the approximate fits
\begin{align}
    H_1 (\alpha, 0) ~&\simeq~ \frac{2 \pi^{3/2}\sqrt{\alpha -1}}{\Gamma(\frac{1}{4})^2} - \frac{0.59 (\alpha -1 )^{1/3}}{1 + 0.23 \alpha} \nn \\
    H_2(\alpha, 0) ~&\simeq~ 
    \sqrt{\frac{\pi}{2\sqrt{\alpha}}}\frac{\Gamma(\frac{5}{4})}{\Gamma(\frac{7}{4})}
    \left(1 + \frac{\frac{2}{\sqrt{\pi}} \frac{\Gamma(7/4)}{\Gamma(5/4)} - 1}{\alpha^2}\right) \ .
    \label{eq:appendixfit1}
\end{align}
In the left panel in Fig.~\ref{fig:appendixfit} we have plotted these
fits (shown by solid curves) and the numerically integrated functions 
(shown by point markers) for comparison.  
By inspection of Eq.~\eqref{eq:params}, we see that for \il{\xieff > 0} 
we are confined to the range \il{\alpha \in [3,\infty)}.
Likewise, for the \il{\xieff < 0} region \il{\alpha \in (0,3]} is required.  
The result in Eq.~\eqref{eq:Xk} for the tachyonic Higgs spectrum 
is then found directly using the leading terms in Eq.~\eqref{eq:appendixfit1}.

\subsection{Mixed Case\label{subsec:appendixmixed}}

\begin{figure}[tb]
    \centering
    \includegraphics[keepaspectratio,width=0.49\textwidth]{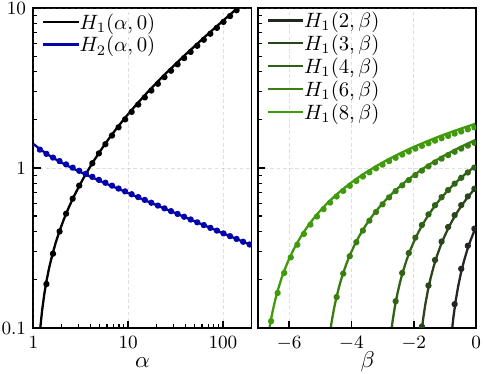}
    \caption{The functions $H_1(\alpha,\beta)$ and $H_2(\alpha,\beta)$ 
    in Eq.~\eqref{eq:Hfuncs} computed numerically (point markers) for 
    comparison to our analytical approximations (shown by solid curves).
    The left panel shows the \il{\beta\rightarrow 0} approximation for
    both functions from Eq.~\eqref{eq:appendixfit1}, and the right panel shows the approximation
    for $H_1(\alpha,\beta)$ derived in Eq.~\eqref{eq:appendixfit2} for several
    values of $\alpha$.  
    }
\label{fig:appendixfit}
\end{figure}

As the mixed case with \il{\xieff > 0} is fully addressed in 
Sec.~\ref{subsec:MixedCase}, let us address the \il{\xieff < 0} region 
in this appendix, \ie, the derivation of the comoving number density Eq.~\eqref{eq:nh0mixedN}.
Note that given Eq.~\eqref{eq:params}, this region corresponds to \il{\beta < 0}.
Additionally, the range of $\alpha$ is extended in the mixed-coupling case 
[according to Eq.~\eqref{eq:params}].
The dependence on both $\alpha$ and $\beta$ is important in what follows,
so the fits above are not sufficient.  But as we are interested in only the
growth exponent $X_k$, we require only the $H_1(\alpha,\beta)$ function
in this regime.  This function has the approximate forms
\begin{align}
    H_1(\alpha,\beta) ~&\simeq~ \frac{H_1 (3, 0)}{\sqrt{3}-1}\frac{\beta + \alpha-1}{\sqrt{\alpha}+1} \nn \\
    ~&\simeq~ \frac{H_1 (3, 0)}{2} \left( \beta + \alpha -1 \right)  \ .
    \label{eq:appendixfit2}
\end{align}
While the first approximation gives a more accurate fit, the 
second is linear in both $\alpha$ and $\beta$ and 
useful for our estimates.
We have plotted the latter approximation 
using solid curves in the right panel of Fig.~\ref{fig:appendixfit}, 
with point markers indicating the numerically integrated result.

Using this approximation, we find that the growth exponent is given
up to \il{\mathcal{O}(\kappa^2)} by 
\begin{align}
    2jX_k ~\simeq~ \frac{4H_1(3,0)}{T}\!\!\int \!dx\, &\frac{\kappa^2 - 2\rh + \frac{g^2}{\Lphi}\left(2\rphi - 1\right)}{\sqrt{\kappa^2 - \rh}}\!  \\
    \simeq~ \frac{-4H_1(3,0)}{T\sqrt{-6\xieff}}\Bigg\{\!\Big(\xiphi\frac{g^2}{\Lphi} + &\xieff\Big)\Bigg[12\log\!\left(\frac{x}{x_0}\right) \!+\! \frac{\kappa^2(x^2 \!-\! x_0^2)}{2\xieff}\Bigg] \nn \\
    \quad \  -\frac{1}{2}\!\Big(\frac{g^2}{\Lphi} + &\kappa^2\Big)\!\Bigg[(x^2\! - x_0^2) \!+\! \frac{\kappa^2(x^4 \!-\! x_0^4)}{24\xieff}\Bigg]\Bigg\} \nn 
\end{align}
where $j$ is the number of times the effective Higgs mass has 
passed through the tachyonic region.  Note that 
Eq.~\eqref{eq:nh0mixedN} is recovered in the \il{\kappa \rightarrow 0} limit.
The growth exponent readily gives the phase-space density
$n_{h_k}$ via Eq.~\eqref{eq:tachyonicformulaappendix} if we
neglect the oscillatory component $\Theta_k$.

Using the saddlepoint method to integrate over the momenta, we 
obtain the Higgs comoving number density
\begin{align}
    n_h \hspace{1.5mm}\simeq\hspace{1.5mm} &\left(\!\!\frac{\sqrt{\Lphi}\varphiamp}{x}\!\right)^{\!\!\!3}\! 
    \left[\!\frac{\frac{T}{4\pi H_1(3,0)}(-6\xieff)^{3/2}}{-24\left(\xiphi\frac{g^2}{2\Lphi} + \xieff\right) - \frac{g^2}{2\Lphi}x^2}\!\right]^{\!3/2} \\
    &\times \left(\frac{x}{x_0}\right)^{\!\!-\frac{8\sqrt{6}H_1(3,0)}{T\sqrt{-\xieff}}\left(\xiphi\frac{g^2}{\Lphi} + \xieff\right)}
  e^{\frac{\sqrt{6}
  		H_1(3,0)}{3T\sqrt{-\xi}}\frac{g^2}{\lambda_\phi}(x^2- x_0^2)} \nn \ ,
\end{align}
which is valid for times \il{x \lesssim \xxi}.  
Using that the Higgs variance is \il{\hvar\simeq n_h/(a^3\omega_h)},
we are finally in a position where we can produce 
the constraint in Eq.~\eqref{eq:xiconstraintN}.


\section{Numerical Methods\label{sec:numerics}}


In this appendix, we summarize the numerical methods used to
compute the evolution of dynamical quantities in this paper.
Note that analogous methods are applied for the comparison to
massive preheating in Appendix~\ref{sec:quadratic}.

We perform our computations in the cosmological time $t$ and 
establish our initial conditions at the end of inflation.
According to the analysis in Sec.~\ref{subsec:InflationaryRegime},
the initial conditions at $t_{\rm end}$ are given by
\begin{align}
    \phi(t_{\rm end}) ~&=~ +\phiend \nn \\
    \dot{\phi}(t_{\rm end}) ~&=~ -\sqrt{V(\phiend)} \ ,
\end{align}
where \il{\phiend = 1} is assumed.
Furthermore, we assume for our numerics that the 
inflaton potential is given by a purely quartic 
potential \il{V(\phi) = \Lphi\phi^4/4}, and we 
have neglected the difference between the canonical and 
non-canonical inflaton field.  While this affects
the initial velocity by an \il{\mathcal{O}(1)} factor, 
we have checked that such changes
have a negligible impact on our results.

The inflationary constraint in Eq.~\eqref{eq:stabilityinflation}
is assumed so that the homogeneous value $h$ for the Higgs 
field is stabilized strongly at the origin.  Even after the 
end of inflation, accounting for $h$ has a negligible influence 
on our results, so we ignore its equation of motion overall.
We solve the equations of motion for the background inflaton field
$\phi$, inflaton fluctuations $\phi_k$, and Higgs fluctuations $h_k$.
As discussed in Sec.~\ref{sec:Backreaction}, the backreaction 
is accounted for using the Hartree approximation 
\il{h^4 \rightarrow 6\langle h^2\rangle h^2 - 3\langle h^2\rangle^2}
for the Higgs self-interaction and using an analogous substitution
for the quartic inflaton potential 
\il{\phi^4 \rightarrow 6\langle\phi^2\rangle\phi^2 - 3\langle\phi^2\rangle^2}~\cite{Boyanovsky:1994me}.

We discretize the momenta linearly over the range \il{k\in [0,\Lambda]}
with \il{N = 300} lattice points.  The momentum cutoff $\Lambda$ is chosen
based on the structure of the parametric and tachyonic instabilities; \ie,
we use the maximal momentum of each unstable band.  For example, with \il{\xih = 0}
and \il{\ggOL \gg 1} our numerics use a cutoff near the maximum of the 
resonance band \il{\Lambda = \sqrt{\Lphi}\varphiamp (\ggOL)^{1/4}}.

The equations of motion are then given by
\begin{align}
    \ddot{\phi} + 3H\dot{\phi} + \Lphi \phi^3 + 3\Lphi\phivar\phi + g^2\hvar\phi ~&=~ 0 \\
    \label{eq:phimodes}
    \ddot{\phi}_k + 3H\dot{\phi}_k + \omega_{\phi_k}^2 \phi_k ~&=~ 0  \\
    \label{eq:hmodes}
    \ddot{h}_k + (3H + \Gamma_{h_k})\dot{h}_k + \omega_{h_k}^2 h_k ~&=~ 0  \ ,
\end{align}
where the mode energies are
\begin{align}
    \omega_{\phi_k}^2 ~&=~ \frac{k^2}{a^2} + 3\Lphi\phi^2 + 3\Lphi\phivar + g^2\hvar \nn \\
    \omega_{h_k}^2 ~&=~ \frac{k^2}{a^2} + g^2\phi^2 + \xih R + g^2\phivar + 3\Lh\hvar \ .
\end{align}

The Hubble parameter is computed using the first Friedmann equation 
\il{H^2 = \rhotot/3} for the total energy density $\rhotot$.  Note that $\rhotot$ includes 
both the energy density of the inflaton background \il{\rho_{\phi} = \dot{\phi}^2/2 + V(\phi)}
and that of the field fluctuations, although the latter 
do not provide a significant contribution until the system
approaches the non-linear stage of preheating.
We evolve these equations of motion until the Higgs destabilizes or
the onset of non-linear dynamics is reached at \il{\xNL \approx 400}.

The decay rates $\Gamma_{h_k}$ for the Higgs modes are calculated 
in the same way as Eq.~\eqref{eq:Gammah}, except in our numerical
calculations we generally write \il{\Gamma_{h_k} = \Gamma_h/\gamma},
where the Lorentz factor $\gamma$ carries some momentum dependence.

The initial conditions for the field fluctuations in 
Eq.~\eqref{eq:phimodes} and Eq.~\eqref{eq:hmodes} are taken 
as Gaussian such that \il{\phi_k = 1/\sqrt{2\omega_{\phi_k}}} and 
\il{h_k = 1/\sqrt{2\omega_{h_k}}} at the initial time.
However, this initial condition is UV divergent and sensitive to the 
cutoff scale $\Lambda$ introduced in our discretization.
Employing adiabatic renormalization~\cite{Postma:2017hbk}, we 
calculate the variance so that it is initially vanishing:
\beq
    \langle h^2\rangle ~\equiv~ \frac{1}{2\pi^2}\int_0^{\Lambda}\! dk\, k^2 \Big[\abs{h_k(t)}^2 - \frac{1}{a^2}\abs{h_k(t_{\rm end})}^2\Big] \ .
\eeq
The scaling of the counterterm with $a$ is chosen so that
it scales with the physical momentum \il{k/a}.

Finally, we incorporate the running of the Higgs four-point coupling $\Lh(\mu)$
by evaluating the renormalization scale at \il{\smash{\mu = \sqrt{\hvar}}}.
The effect of the backreaction is most relevant at energies near
\il{h_{\rm SM} \approx 10^{10}\,\GeV} where, based on SM measurements, 
the coupling changes sign.  Therefore, in our numerics we 
take the approximate
\beq
    \Lh ~=~ 0.01\,\text{sgn}\left(h_{\rm SM} - \sqrt{\hvar}\right) \ ,
\eeq
which is similar to that used in other numerical 
studies~\cite{Ema:2016kpf,Enqvist:2016mqj} of electroweak vacuum metastability.

\bibliography{references}

\begin{thebibliography}{85}%
\makeatletter
\providecommand \@ifxundefined [1]{%
 \@ifx{#1\undefined}
}%
\providecommand \@ifnum [1]{%
 \ifnum #1\expandafter \@firstoftwo
 \else \expandafter \@secondoftwo
 \fi
}%
\providecommand \@ifx [1]{%
 \ifx #1\expandafter \@firstoftwo
 \else \expandafter \@secondoftwo
 \fi
}%
\providecommand \natexlab [1]{#1}%
\providecommand \enquote  [1]{``#1''}%
\providecommand \bibnamefont  [1]{#1}%
\providecommand \bibfnamefont [1]{#1}%
\providecommand \citenamefont [1]{#1}%
\providecommand \href@noop [0]{\@secondoftwo}%
\providecommand \href [0]{\begingroup \@sanitize@url \@href}%
\providecommand \@href[1]{\@@startlink{#1}\@@href}%
\providecommand \@@href[1]{\endgroup#1\@@endlink}%
\providecommand \@sanitize@url [0]{\catcode `\\12\catcode `\$12\catcode
  `\&12\catcode `\#12\catcode `\^12\catcode `\_12\catcode `\%12\relax}%
\providecommand \@@startlink[1]{}%
\providecommand \@@endlink[0]{}%
\providecommand \url  [0]{\begingroup\@sanitize@url \@url }%
\providecommand \@url [1]{\endgroup\@href {#1}{\urlprefix }}%
\providecommand \urlprefix  [0]{URL }%
\providecommand \Eprint [0]{\href }%
\providecommand \doibase [0]{http://dx.doi.org/}%
\providecommand \selectlanguage [0]{\@gobble}%
\providecommand \bibinfo  [0]{\@secondoftwo}%
\providecommand \bibfield  [0]{\@secondoftwo}%
\providecommand \translation [1]{[#1]}%
\providecommand \BibitemOpen [0]{}%
\providecommand \bibitemStop [0]{}%
\providecommand \bibitemNoStop [0]{.\EOS\space}%
\providecommand \EOS [0]{\spacefactor3000\relax}%
\providecommand \BibitemShut  [1]{\csname bibitem#1\endcsname}%
\let\auto@bib@innerbib\@empty
\bibitem [{\citenamefont {Zyla}\ \emph {et~al.}(2020)\citenamefont {Zyla} \emph
  {et~al.}}]{Zyla:2020zbs}%
  \BibitemOpen
  \bibfield  {author} {\bibinfo {author} {\bibfnamefont {P.~A.}\ \bibnamefont
  {Zyla}} \emph {et~al.} (\bibinfo {collaboration} {Particle Data Group}),\
  }\href {\doibase 10.1093/ptep/ptaa104} {\bibfield  {journal} {\bibinfo
  {journal} {PTEP}\ }\textbf {\bibinfo {volume} {2020}},\ \bibinfo {pages}
  {083C01} (\bibinfo {year} {2020})}\BibitemShut {NoStop}%
\bibitem [{\citenamefont {Degrassi}\ \emph {et~al.}(2012)\citenamefont
  {Degrassi}, \citenamefont {Di~Vita}, \citenamefont {Elias-Miro},
  \citenamefont {Espinosa}, \citenamefont {Giudice}, \citenamefont {Isidori},\
  and\ \citenamefont {Strumia}}]{Degrassi:2012ry}%
  \BibitemOpen
  \bibfield  {author} {\bibinfo {author} {\bibfnamefont {G.}~\bibnamefont
  {Degrassi}}, \bibinfo {author} {\bibfnamefont {S.}~\bibnamefont {Di~Vita}},
  \bibinfo {author} {\bibfnamefont {J.}~\bibnamefont {Elias-Miro}}, \bibinfo
  {author} {\bibfnamefont {J.~R.}\ \bibnamefont {Espinosa}}, \bibinfo {author}
  {\bibfnamefont {G.~F.}\ \bibnamefont {Giudice}}, \bibinfo {author}
  {\bibfnamefont {G.}~\bibnamefont {Isidori}}, \ and\ \bibinfo {author}
  {\bibfnamefont {A.}~\bibnamefont {Strumia}},\ }\href {\doibase
  10.1007/JHEP08(2012)098} {\bibfield  {journal} {\bibinfo  {journal} {JHEP}\
  }\textbf {\bibinfo {volume} {08}},\ \bibinfo {pages} {098} (\bibinfo {year}
  {2012})},\ \Eprint {http://arxiv.org/abs/1205.6497} {arXiv:1205.6497
  [hep-ph]} \BibitemShut {NoStop}%
\bibitem [{\citenamefont {Bezrukov}\ \emph {et~al.}(2012)\citenamefont
  {Bezrukov}, \citenamefont {Kalmykov}, \citenamefont {Kniehl},\ and\
  \citenamefont {Shaposhnikov}}]{Bezrukov:2012sa}%
  \BibitemOpen
  \bibfield  {author} {\bibinfo {author} {\bibfnamefont {F.}~\bibnamefont
  {Bezrukov}}, \bibinfo {author} {\bibfnamefont {M.~Y.}\ \bibnamefont
  {Kalmykov}}, \bibinfo {author} {\bibfnamefont {B.~A.}\ \bibnamefont
  {Kniehl}}, \ and\ \bibinfo {author} {\bibfnamefont {M.}~\bibnamefont
  {Shaposhnikov}},\ }\href {\doibase 10.1007/JHEP10(2012)140} {\bibfield
  {journal} {\bibinfo  {journal} {JHEP}\ }\textbf {\bibinfo {volume} {10}},\
  \bibinfo {pages} {140} (\bibinfo {year} {2012})},\ \Eprint
  {http://arxiv.org/abs/1205.2893} {arXiv:1205.2893 [hep-ph]} \BibitemShut
  {NoStop}%
\bibitem [{\citenamefont {Alekhin}\ \emph {et~al.}(2012)\citenamefont
  {Alekhin}, \citenamefont {Djouadi},\ and\ \citenamefont
  {Moch}}]{Alekhin:2012py}%
  \BibitemOpen
  \bibfield  {author} {\bibinfo {author} {\bibfnamefont {S.}~\bibnamefont
  {Alekhin}}, \bibinfo {author} {\bibfnamefont {A.}~\bibnamefont {Djouadi}}, \
  and\ \bibinfo {author} {\bibfnamefont {S.}~\bibnamefont {Moch}},\ }\href
  {\doibase 10.1016/j.physletb.2012.08.024} {\bibfield  {journal} {\bibinfo
  {journal} {Phys. Lett. B}\ }\textbf {\bibinfo {volume} {716}},\ \bibinfo
  {pages} {214} (\bibinfo {year} {2012})},\ \Eprint
  {http://arxiv.org/abs/1207.0980} {arXiv:1207.0980 [hep-ph]} \BibitemShut
  {NoStop}%
\bibitem [{\citenamefont {Buttazzo}\ \emph {et~al.}(2013)\citenamefont
  {Buttazzo}, \citenamefont {Degrassi}, \citenamefont {Giardino}, \citenamefont
  {Giudice}, \citenamefont {Sala}, \citenamefont {Salvio},\ and\ \citenamefont
  {Strumia}}]{Buttazzo:2013uya}%
  \BibitemOpen
  \bibfield  {author} {\bibinfo {author} {\bibfnamefont {D.}~\bibnamefont
  {Buttazzo}}, \bibinfo {author} {\bibfnamefont {G.}~\bibnamefont {Degrassi}},
  \bibinfo {author} {\bibfnamefont {P.~P.}\ \bibnamefont {Giardino}}, \bibinfo
  {author} {\bibfnamefont {G.~F.}\ \bibnamefont {Giudice}}, \bibinfo {author}
  {\bibfnamefont {F.}~\bibnamefont {Sala}}, \bibinfo {author} {\bibfnamefont
  {A.}~\bibnamefont {Salvio}}, \ and\ \bibinfo {author} {\bibfnamefont
  {A.}~\bibnamefont {Strumia}},\ }\href {\doibase 10.1007/JHEP12(2013)089}
  {\bibfield  {journal} {\bibinfo  {journal} {JHEP}\ }\textbf {\bibinfo
  {volume} {12}},\ \bibinfo {pages} {089} (\bibinfo {year} {2013})},\ \Eprint
  {http://arxiv.org/abs/1307.3536} {arXiv:1307.3536 [hep-ph]} \BibitemShut
  {NoStop}%
\bibitem [{\citenamefont {Bednyakov}\ \emph {et~al.}(2015)\citenamefont
  {Bednyakov}, \citenamefont {Kniehl}, \citenamefont {Pikelner},\ and\
  \citenamefont {Veretin}}]{Bednyakov:2015sca}%
  \BibitemOpen
  \bibfield  {author} {\bibinfo {author} {\bibfnamefont {A.~V.}\ \bibnamefont
  {Bednyakov}}, \bibinfo {author} {\bibfnamefont {B.~A.}\ \bibnamefont
  {Kniehl}}, \bibinfo {author} {\bibfnamefont {A.~F.}\ \bibnamefont
  {Pikelner}}, \ and\ \bibinfo {author} {\bibfnamefont {O.~L.}\ \bibnamefont
  {Veretin}},\ }\href {\doibase 10.1103/PhysRevLett.115.201802} {\bibfield
  {journal} {\bibinfo  {journal} {Phys. Rev. Lett.}\ }\textbf {\bibinfo
  {volume} {115}},\ \bibinfo {pages} {201802} (\bibinfo {year} {2015})},\
  \Eprint {http://arxiv.org/abs/1507.08833} {arXiv:1507.08833 [hep-ph]}
  \BibitemShut {NoStop}%
\bibitem [{\citenamefont {Andreassen}\ \emph {et~al.}(2018)\citenamefont
  {Andreassen}, \citenamefont {Frost},\ and\ \citenamefont
  {Schwartz}}]{Andreassen:2017rzq}%
  \BibitemOpen
  \bibfield  {author} {\bibinfo {author} {\bibfnamefont {A.}~\bibnamefont
  {Andreassen}}, \bibinfo {author} {\bibfnamefont {W.}~\bibnamefont {Frost}}, \
  and\ \bibinfo {author} {\bibfnamefont {M.~D.}\ \bibnamefont {Schwartz}},\
  }\href {\doibase 10.1103/PhysRevD.97.056006} {\bibfield  {journal} {\bibinfo
  {journal} {Phys. Rev. D}\ }\textbf {\bibinfo {volume} {97}},\ \bibinfo
  {pages} {056006} (\bibinfo {year} {2018})},\ \Eprint
  {http://arxiv.org/abs/1707.08124} {arXiv:1707.08124 [hep-ph]} \BibitemShut
  {NoStop}%
\bibitem [{\citenamefont {Chigusa}\ \emph {et~al.}(2018)\citenamefont
  {Chigusa}, \citenamefont {Moroi},\ and\ \citenamefont
  {Shoji}}]{Chigusa:2018uuj}%
  \BibitemOpen
  \bibfield  {author} {\bibinfo {author} {\bibfnamefont {S.}~\bibnamefont
  {Chigusa}}, \bibinfo {author} {\bibfnamefont {T.}~\bibnamefont {Moroi}}, \
  and\ \bibinfo {author} {\bibfnamefont {Y.}~\bibnamefont {Shoji}},\ }\href
  {\doibase 10.1103/PhysRevD.97.116012} {\bibfield  {journal} {\bibinfo
  {journal} {Phys. Rev. D}\ }\textbf {\bibinfo {volume} {97}},\ \bibinfo
  {pages} {116012} (\bibinfo {year} {2018})},\ \Eprint
  {http://arxiv.org/abs/1803.03902} {arXiv:1803.03902 [hep-ph]} \BibitemShut
  {NoStop}%
\bibitem [{\citenamefont {Markkanen}\ \emph {et~al.}(2018)\citenamefont
  {Markkanen}, \citenamefont {Rajantie},\ and\ \citenamefont
  {Stopyra}}]{Markkanen:2018pdo}%
  \BibitemOpen
  \bibfield  {author} {\bibinfo {author} {\bibfnamefont {T.}~\bibnamefont
  {Markkanen}}, \bibinfo {author} {\bibfnamefont {A.}~\bibnamefont {Rajantie}},
  \ and\ \bibinfo {author} {\bibfnamefont {S.}~\bibnamefont {Stopyra}},\ }\href
  {\doibase 10.3389/fspas.2018.00040} {\bibfield  {journal} {\bibinfo
  {journal} {Front. Astron. Space Sci.}\ }\textbf {\bibinfo {volume} {5}},\
  \bibinfo {pages} {40} (\bibinfo {year} {2018})},\ \Eprint
  {http://arxiv.org/abs/1809.06923} {arXiv:1809.06923 [astro-ph.CO]}
  \BibitemShut {NoStop}%
\bibitem [{\citenamefont {East}\ \emph {et~al.}(2017)\citenamefont {East},
  \citenamefont {Kearney}, \citenamefont {Shakya}, \citenamefont {Yoo},\ and\
  \citenamefont {Zurek}}]{East:2016anr}%
  \BibitemOpen
  \bibfield  {author} {\bibinfo {author} {\bibfnamefont {W.~E.}\ \bibnamefont
  {East}}, \bibinfo {author} {\bibfnamefont {J.}~\bibnamefont {Kearney}},
  \bibinfo {author} {\bibfnamefont {B.}~\bibnamefont {Shakya}}, \bibinfo
  {author} {\bibfnamefont {H.}~\bibnamefont {Yoo}}, \ and\ \bibinfo {author}
  {\bibfnamefont {K.~M.}\ \bibnamefont {Zurek}},\ }\href {\doibase
  10.1103/PhysRevD.95.023526} {\bibfield  {journal} {\bibinfo  {journal} {Phys.
  Rev. D}\ }\textbf {\bibinfo {volume} {95}},\ \bibinfo {pages} {023526}
  (\bibinfo {year} {2017})},\ \Eprint {http://arxiv.org/abs/1607.00381}
  {arXiv:1607.00381 [hep-ph]} \BibitemShut {NoStop}%
\bibitem [{\citenamefont {Fumagalli}\ \emph {et~al.}(2020)\citenamefont
  {Fumagalli}, \citenamefont {Renaux-Petel},\ and\ \citenamefont
  {Ronayne}}]{Fumagalli:2019ohr}%
  \BibitemOpen
  \bibfield  {author} {\bibinfo {author} {\bibfnamefont {J.}~\bibnamefont
  {Fumagalli}}, \bibinfo {author} {\bibfnamefont {S.}~\bibnamefont
  {Renaux-Petel}}, \ and\ \bibinfo {author} {\bibfnamefont {J.~W.}\
  \bibnamefont {Ronayne}},\ }\href {\doibase 10.1007/JHEP02(2020)142}
  {\bibfield  {journal} {\bibinfo  {journal} {JHEP}\ }\textbf {\bibinfo
  {volume} {02}},\ \bibinfo {pages} {142} (\bibinfo {year} {2020})},\ \Eprint
  {http://arxiv.org/abs/1910.13430} {arXiv:1910.13430 [hep-ph]} \BibitemShut
  {NoStop}%
\bibitem [{\citenamefont {Espinosa}\ \emph {et~al.}(2008)\citenamefont
  {Espinosa}, \citenamefont {Giudice},\ and\ \citenamefont
  {Riotto}}]{Espinosa:2007qp}%
  \BibitemOpen
  \bibfield  {author} {\bibinfo {author} {\bibfnamefont {J.~R.}\ \bibnamefont
  {Espinosa}}, \bibinfo {author} {\bibfnamefont {G.~F.}\ \bibnamefont
  {Giudice}}, \ and\ \bibinfo {author} {\bibfnamefont {A.}~\bibnamefont
  {Riotto}},\ }\href {\doibase 10.1088/1475-7516/2008/05/002} {\bibfield
  {journal} {\bibinfo  {journal} {JCAP}\ }\textbf {\bibinfo {volume} {05}},\
  \bibinfo {pages} {002} (\bibinfo {year} {2008})},\ \Eprint
  {http://arxiv.org/abs/0710.2484} {arXiv:0710.2484 [hep-ph]} \BibitemShut
  {NoStop}%
\bibitem [{\citenamefont {Lebedev}\ and\ \citenamefont
  {Westphal}(2013)}]{Lebedev:2012sy}%
  \BibitemOpen
  \bibfield  {author} {\bibinfo {author} {\bibfnamefont {O.}~\bibnamefont
  {Lebedev}}\ and\ \bibinfo {author} {\bibfnamefont {A.}~\bibnamefont
  {Westphal}},\ }\href {\doibase 10.1016/j.physletb.2012.12.069} {\bibfield
  {journal} {\bibinfo  {journal} {Phys. Lett. B}\ }\textbf {\bibinfo {volume}
  {719}},\ \bibinfo {pages} {415} (\bibinfo {year} {2013})},\ \Eprint
  {http://arxiv.org/abs/1210.6987} {arXiv:1210.6987 [hep-ph]} \BibitemShut
  {NoStop}%
\bibitem [{\citenamefont {Felder}\ \emph
  {et~al.}(2001{\natexlab{a}})\citenamefont {Felder}, \citenamefont
  {Garcia-Bellido}, \citenamefont {Greene}, \citenamefont {Kofman},
  \citenamefont {Linde},\ and\ \citenamefont {Tkachev}}]{Felder:2000hj}%
  \BibitemOpen
  \bibfield  {author} {\bibinfo {author} {\bibfnamefont {G.~N.}\ \bibnamefont
  {Felder}}, \bibinfo {author} {\bibfnamefont {J.}~\bibnamefont
  {Garcia-Bellido}}, \bibinfo {author} {\bibfnamefont {P.~B.}\ \bibnamefont
  {Greene}}, \bibinfo {author} {\bibfnamefont {L.}~\bibnamefont {Kofman}},
  \bibinfo {author} {\bibfnamefont {A.~D.}\ \bibnamefont {Linde}}, \ and\
  \bibinfo {author} {\bibfnamefont {I.}~\bibnamefont {Tkachev}},\ }\href
  {\doibase 10.1103/PhysRevLett.87.011601} {\bibfield  {journal} {\bibinfo
  {journal} {Phys. Rev. Lett.}\ }\textbf {\bibinfo {volume} {87}},\ \bibinfo
  {pages} {011601} (\bibinfo {year} {2001}{\natexlab{a}})},\ \Eprint
  {http://arxiv.org/abs/hep-ph/0012142} {arXiv:hep-ph/0012142} \BibitemShut
  {NoStop}%
\bibitem [{\citenamefont {Felder}\ \emph
  {et~al.}(2001{\natexlab{b}})\citenamefont {Felder}, \citenamefont {Kofman},\
  and\ \citenamefont {Linde}}]{Felder:2001kt}%
  \BibitemOpen
  \bibfield  {author} {\bibinfo {author} {\bibfnamefont {G.~N.}\ \bibnamefont
  {Felder}}, \bibinfo {author} {\bibfnamefont {L.}~\bibnamefont {Kofman}}, \
  and\ \bibinfo {author} {\bibfnamefont {A.~D.}\ \bibnamefont {Linde}},\ }\href
  {\doibase 10.1103/PhysRevD.64.123517} {\bibfield  {journal} {\bibinfo
  {journal} {Phys. Rev. D}\ }\textbf {\bibinfo {volume} {64}},\ \bibinfo
  {pages} {123517} (\bibinfo {year} {2001}{\natexlab{b}})},\ \Eprint
  {http://arxiv.org/abs/hep-th/0106179} {arXiv:hep-th/0106179} \BibitemShut
  {NoStop}%
\bibitem [{\citenamefont {Dufaux}\ \emph {et~al.}(2006)\citenamefont {Dufaux},
  \citenamefont {Felder}, \citenamefont {Kofman}, \citenamefont {Peloso},\ and\
  \citenamefont {Podolsky}}]{Dufaux:2006ee}%
  \BibitemOpen
  \bibfield  {author} {\bibinfo {author} {\bibfnamefont {J.~F.}\ \bibnamefont
  {Dufaux}}, \bibinfo {author} {\bibfnamefont {G.~N.}\ \bibnamefont {Felder}},
  \bibinfo {author} {\bibfnamefont {L.}~\bibnamefont {Kofman}}, \bibinfo
  {author} {\bibfnamefont {M.}~\bibnamefont {Peloso}}, \ and\ \bibinfo {author}
  {\bibfnamefont {D.}~\bibnamefont {Podolsky}},\ }\href {\doibase
  10.1088/1475-7516/2006/07/006} {\bibfield  {journal} {\bibinfo  {journal}
  {JCAP}\ }\textbf {\bibinfo {volume} {07}},\ \bibinfo {pages} {006} (\bibinfo
  {year} {2006})},\ \Eprint {http://arxiv.org/abs/hep-ph/0602144}
  {arXiv:hep-ph/0602144} \BibitemShut {NoStop}%
\bibitem [{\citenamefont {Kofman}\ \emph {et~al.}(1994)\citenamefont {Kofman},
  \citenamefont {Linde},\ and\ \citenamefont {Starobinsky}}]{Kofman:1994rk}%
  \BibitemOpen
  \bibfield  {author} {\bibinfo {author} {\bibfnamefont {L.}~\bibnamefont
  {Kofman}}, \bibinfo {author} {\bibfnamefont {A.~D.}\ \bibnamefont {Linde}}, \
  and\ \bibinfo {author} {\bibfnamefont {A.~A.}\ \bibnamefont {Starobinsky}},\
  }\href {\doibase 10.1103/PhysRevLett.73.3195} {\bibfield  {journal} {\bibinfo
   {journal} {Phys. Rev. Lett.}\ }\textbf {\bibinfo {volume} {73}},\ \bibinfo
  {pages} {3195} (\bibinfo {year} {1994})},\ \Eprint
  {http://arxiv.org/abs/hep-th/9405187} {arXiv:hep-th/9405187} \BibitemShut
  {NoStop}%
\bibitem [{\citenamefont {Kofman}\ \emph {et~al.}(1997)\citenamefont {Kofman},
  \citenamefont {Linde},\ and\ \citenamefont {Starobinsky}}]{Kofman:1997yn}%
  \BibitemOpen
  \bibfield  {author} {\bibinfo {author} {\bibfnamefont {L.}~\bibnamefont
  {Kofman}}, \bibinfo {author} {\bibfnamefont {A.~D.}\ \bibnamefont {Linde}}, \
  and\ \bibinfo {author} {\bibfnamefont {A.~A.}\ \bibnamefont {Starobinsky}},\
  }\href {\doibase 10.1103/PhysRevD.56.3258} {\bibfield  {journal} {\bibinfo
  {journal} {Phys. Rev. D}\ }\textbf {\bibinfo {volume} {56}},\ \bibinfo
  {pages} {3258} (\bibinfo {year} {1997})},\ \Eprint
  {http://arxiv.org/abs/hep-ph/9704452} {arXiv:hep-ph/9704452} \BibitemShut
  {NoStop}%
\bibitem [{\citenamefont {Greene}\ \emph {et~al.}(1997)\citenamefont {Greene},
  \citenamefont {Kofman}, \citenamefont {Linde},\ and\ \citenamefont
  {Starobinsky}}]{Greene:1997fu}%
  \BibitemOpen
  \bibfield  {author} {\bibinfo {author} {\bibfnamefont {P.~B.}\ \bibnamefont
  {Greene}}, \bibinfo {author} {\bibfnamefont {L.}~\bibnamefont {Kofman}},
  \bibinfo {author} {\bibfnamefont {A.~D.}\ \bibnamefont {Linde}}, \ and\
  \bibinfo {author} {\bibfnamefont {A.~A.}\ \bibnamefont {Starobinsky}},\
  }\href {\doibase 10.1103/PhysRevD.56.6175} {\bibfield  {journal} {\bibinfo
  {journal} {Phys. Rev. D}\ }\textbf {\bibinfo {volume} {56}},\ \bibinfo
  {pages} {6175} (\bibinfo {year} {1997})},\ \Eprint
  {http://arxiv.org/abs/hep-ph/9705347} {arXiv:hep-ph/9705347} \BibitemShut
  {NoStop}%
\bibitem [{\citenamefont {Herranen}\ \emph {et~al.}(2015)\citenamefont
  {Herranen}, \citenamefont {Markkanen}, \citenamefont {Nurmi},\ and\
  \citenamefont {Rajantie}}]{Herranen:2015ima}%
  \BibitemOpen
  \bibfield  {author} {\bibinfo {author} {\bibfnamefont {M.}~\bibnamefont
  {Herranen}}, \bibinfo {author} {\bibfnamefont {T.}~\bibnamefont {Markkanen}},
  \bibinfo {author} {\bibfnamefont {S.}~\bibnamefont {Nurmi}}, \ and\ \bibinfo
  {author} {\bibfnamefont {A.}~\bibnamefont {Rajantie}},\ }\href {\doibase
  10.1103/PhysRevLett.115.241301} {\bibfield  {journal} {\bibinfo  {journal}
  {Phys. Rev. Lett.}\ }\textbf {\bibinfo {volume} {115}},\ \bibinfo {pages}
  {241301} (\bibinfo {year} {2015})},\ \Eprint
  {http://arxiv.org/abs/1506.04065} {arXiv:1506.04065 [hep-ph]} \BibitemShut
  {NoStop}%
\bibitem [{\citenamefont {Ema}\ \emph {et~al.}(2016)\citenamefont {Ema},
  \citenamefont {Mukaida},\ and\ \citenamefont {Nakayama}}]{Ema:2016kpf}%
  \BibitemOpen
  \bibfield  {author} {\bibinfo {author} {\bibfnamefont {Y.}~\bibnamefont
  {Ema}}, \bibinfo {author} {\bibfnamefont {K.}~\bibnamefont {Mukaida}}, \ and\
  \bibinfo {author} {\bibfnamefont {K.}~\bibnamefont {Nakayama}},\ }\href
  {\doibase 10.1088/1475-7516/2016/10/043} {\bibfield  {journal} {\bibinfo
  {journal} {JCAP}\ }\textbf {\bibinfo {volume} {10}},\ \bibinfo {pages} {043}
  (\bibinfo {year} {2016})},\ \Eprint {http://arxiv.org/abs/1602.00483}
  {arXiv:1602.00483 [hep-ph]} \BibitemShut {NoStop}%
\bibitem [{\citenamefont {Kohri}\ and\ \citenamefont
  {Matsui}(2016)}]{Kohri:2016wof}%
  \BibitemOpen
  \bibfield  {author} {\bibinfo {author} {\bibfnamefont {K.}~\bibnamefont
  {Kohri}}\ and\ \bibinfo {author} {\bibfnamefont {H.}~\bibnamefont {Matsui}},\
  }\href {\doibase 10.1103/PhysRevD.94.103509} {\bibfield  {journal} {\bibinfo
  {journal} {Phys. Rev. D}\ }\textbf {\bibinfo {volume} {94}},\ \bibinfo
  {pages} {103509} (\bibinfo {year} {2016})},\ \Eprint
  {http://arxiv.org/abs/1602.02100} {arXiv:1602.02100 [hep-ph]} \BibitemShut
  {NoStop}%
\bibitem [{\citenamefont {Enqvist}\ \emph {et~al.}(2016)\citenamefont
  {Enqvist}, \citenamefont {Karciauskas}, \citenamefont {Lebedev},
  \citenamefont {Rusak},\ and\ \citenamefont {Zatta}}]{Enqvist:2016mqj}%
  \BibitemOpen
  \bibfield  {author} {\bibinfo {author} {\bibfnamefont {K.}~\bibnamefont
  {Enqvist}}, \bibinfo {author} {\bibfnamefont {M.}~\bibnamefont
  {Karciauskas}}, \bibinfo {author} {\bibfnamefont {O.}~\bibnamefont
  {Lebedev}}, \bibinfo {author} {\bibfnamefont {S.}~\bibnamefont {Rusak}}, \
  and\ \bibinfo {author} {\bibfnamefont {M.}~\bibnamefont {Zatta}},\ }\href
  {\doibase 10.1088/1475-7516/2016/11/025} {\bibfield  {journal} {\bibinfo
  {journal} {JCAP}\ }\textbf {\bibinfo {volume} {11}},\ \bibinfo {pages} {025}
  (\bibinfo {year} {2016})},\ \Eprint {http://arxiv.org/abs/1608.08848}
  {arXiv:1608.08848 [hep-ph]} \BibitemShut {NoStop}%
\bibitem [{\citenamefont {Ema}\ \emph {et~al.}(2017)\citenamefont {Ema},
  \citenamefont {Karciauskas}, \citenamefont {Lebedev},\ and\ \citenamefont
  {Zatta}}]{Ema:2017loe}%
  \BibitemOpen
  \bibfield  {author} {\bibinfo {author} {\bibfnamefont {Y.}~\bibnamefont
  {Ema}}, \bibinfo {author} {\bibfnamefont {M.}~\bibnamefont {Karciauskas}},
  \bibinfo {author} {\bibfnamefont {O.}~\bibnamefont {Lebedev}}, \ and\
  \bibinfo {author} {\bibfnamefont {M.}~\bibnamefont {Zatta}},\ }\href
  {\doibase 10.1088/1475-7516/2017/06/054} {\bibfield  {journal} {\bibinfo
  {journal} {JCAP}\ }\textbf {\bibinfo {volume} {06}},\ \bibinfo {pages} {054}
  (\bibinfo {year} {2017})},\ \Eprint {http://arxiv.org/abs/1703.04681}
  {arXiv:1703.04681 [hep-ph]} \BibitemShut {NoStop}%
\bibitem [{\citenamefont {Gross}\ \emph {et~al.}(2016)\citenamefont {Gross},
  \citenamefont {Lebedev},\ and\ \citenamefont {Zatta}}]{Gross:2015bea}%
  \BibitemOpen
  \bibfield  {author} {\bibinfo {author} {\bibfnamefont {C.}~\bibnamefont
  {Gross}}, \bibinfo {author} {\bibfnamefont {O.}~\bibnamefont {Lebedev}}, \
  and\ \bibinfo {author} {\bibfnamefont {M.}~\bibnamefont {Zatta}},\ }\href
  {\doibase 10.1016/j.physletb.2015.12.014} {\bibfield  {journal} {\bibinfo
  {journal} {Phys. Lett. B}\ }\textbf {\bibinfo {volume} {753}},\ \bibinfo
  {pages} {178} (\bibinfo {year} {2016})},\ \Eprint
  {http://arxiv.org/abs/1506.05106} {arXiv:1506.05106 [hep-ph]} \BibitemShut
  {NoStop}%
\bibitem [{\citenamefont {Kasuya}\ and\ \citenamefont
  {Kawasaki}(1996)}]{Kasuya:1996np}%
  \BibitemOpen
  \bibfield  {author} {\bibinfo {author} {\bibfnamefont {S.}~\bibnamefont
  {Kasuya}}\ and\ \bibinfo {author} {\bibfnamefont {M.}~\bibnamefont
  {Kawasaki}},\ }\href {\doibase 10.1016/S0370-2693(96)01216-6} {\bibfield
  {journal} {\bibinfo  {journal} {Phys. Lett. B}\ }\textbf {\bibinfo {volume}
  {388}},\ \bibinfo {pages} {686} (\bibinfo {year} {1996})},\ \Eprint
  {http://arxiv.org/abs/hep-ph/9603317} {arXiv:hep-ph/9603317} \BibitemShut
  {NoStop}%
\bibitem [{\citenamefont {Felder}\ \emph {et~al.}(1999)\citenamefont {Felder},
  \citenamefont {Kofman},\ and\ \citenamefont {Linde}}]{Felder:1998vq}%
  \BibitemOpen
  \bibfield  {author} {\bibinfo {author} {\bibfnamefont {G.~N.}\ \bibnamefont
  {Felder}}, \bibinfo {author} {\bibfnamefont {L.}~\bibnamefont {Kofman}}, \
  and\ \bibinfo {author} {\bibfnamefont {A.~D.}\ \bibnamefont {Linde}},\ }\href
  {\doibase 10.1103/PhysRevD.59.123523} {\bibfield  {journal} {\bibinfo
  {journal} {Phys. Rev. D}\ }\textbf {\bibinfo {volume} {59}},\ \bibinfo
  {pages} {123523} (\bibinfo {year} {1999})},\ \Eprint
  {http://arxiv.org/abs/hep-ph/9812289} {arXiv:hep-ph/9812289} \BibitemShut
  {NoStop}%
\bibitem [{\citenamefont {Bezrukov}\ \emph {et~al.}(2009)\citenamefont
  {Bezrukov}, \citenamefont {Gorbunov},\ and\ \citenamefont
  {Shaposhnikov}}]{Bezrukov:2008ut}%
  \BibitemOpen
  \bibfield  {author} {\bibinfo {author} {\bibfnamefont {F.}~\bibnamefont
  {Bezrukov}}, \bibinfo {author} {\bibfnamefont {D.}~\bibnamefont {Gorbunov}},
  \ and\ \bibinfo {author} {\bibfnamefont {M.}~\bibnamefont {Shaposhnikov}},\
  }\href {\doibase 10.1088/1475-7516/2009/06/029} {\bibfield  {journal}
  {\bibinfo  {journal} {JCAP}\ }\textbf {\bibinfo {volume} {06}},\ \bibinfo
  {pages} {029} (\bibinfo {year} {2009})},\ \Eprint
  {http://arxiv.org/abs/0812.3622} {arXiv:0812.3622 [hep-ph]} \BibitemShut
  {NoStop}%
\bibitem [{\citenamefont {Garcia-Bellido}\ \emph {et~al.}(2009)\citenamefont
  {Garcia-Bellido}, \citenamefont {Figueroa},\ and\ \citenamefont
  {Rubio}}]{GarciaBellido:2008ab}%
  \BibitemOpen
  \bibfield  {author} {\bibinfo {author} {\bibfnamefont {J.}~\bibnamefont
  {Garcia-Bellido}}, \bibinfo {author} {\bibfnamefont {D.~G.}\ \bibnamefont
  {Figueroa}}, \ and\ \bibinfo {author} {\bibfnamefont {J.}~\bibnamefont
  {Rubio}},\ }\href {\doibase 10.1103/PhysRevD.79.063531} {\bibfield  {journal}
  {\bibinfo  {journal} {Phys. Rev. D}\ }\textbf {\bibinfo {volume} {79}},\
  \bibinfo {pages} {063531} (\bibinfo {year} {2009})},\ \Eprint
  {http://arxiv.org/abs/0812.4624} {arXiv:0812.4624 [hep-ph]} \BibitemShut
  {NoStop}%
\bibitem [{\citenamefont {Mukaida}\ and\ \citenamefont
  {Nakayama}(2013)}]{Mukaida:2012bz}%
  \BibitemOpen
  \bibfield  {author} {\bibinfo {author} {\bibfnamefont {K.}~\bibnamefont
  {Mukaida}}\ and\ \bibinfo {author} {\bibfnamefont {K.}~\bibnamefont
  {Nakayama}},\ }\href {\doibase 10.1088/1475-7516/2013/03/002} {\bibfield
  {journal} {\bibinfo  {journal} {JCAP}\ }\textbf {\bibinfo {volume} {03}},\
  \bibinfo {pages} {002} (\bibinfo {year} {2013})},\ \Eprint
  {http://arxiv.org/abs/1212.4985} {arXiv:1212.4985 [hep-ph]} \BibitemShut
  {NoStop}%
\bibitem [{\citenamefont {Repond}\ and\ \citenamefont
  {Rubio}(2016)}]{Repond:2016sol}%
  \BibitemOpen
  \bibfield  {author} {\bibinfo {author} {\bibfnamefont {J.}~\bibnamefont
  {Repond}}\ and\ \bibinfo {author} {\bibfnamefont {J.}~\bibnamefont {Rubio}},\
  }\href {\doibase 10.1088/1475-7516/2016/07/043} {\bibfield  {journal}
  {\bibinfo  {journal} {JCAP}\ }\textbf {\bibinfo {volume} {07}},\ \bibinfo
  {pages} {043} (\bibinfo {year} {2016})},\ \Eprint
  {http://arxiv.org/abs/1604.08238} {arXiv:1604.08238 [astro-ph.CO]}
  \BibitemShut {NoStop}%
\bibitem [{\citenamefont {Fan}\ \emph {et~al.}(2021)\citenamefont {Fan},
  \citenamefont {Lozanov},\ and\ \citenamefont {Lu}}]{Fan:2021otj}%
  \BibitemOpen
  \bibfield  {author} {\bibinfo {author} {\bibfnamefont {J.}~\bibnamefont
  {Fan}}, \bibinfo {author} {\bibfnamefont {K.~D.}\ \bibnamefont {Lozanov}}, \
  and\ \bibinfo {author} {\bibfnamefont {Q.}~\bibnamefont {Lu}},\ }\href
  {\doibase 10.1007/JHEP05(2021)069} {\bibfield  {journal} {\bibinfo  {journal}
  {JHEP}\ }\textbf {\bibinfo {volume} {05}},\ \bibinfo {pages} {069} (\bibinfo
  {year} {2021})},\ \Eprint {http://arxiv.org/abs/2101.11008} {arXiv:2101.11008
  [hep-ph]} \BibitemShut {NoStop}%
\bibitem [{\citenamefont {Buchbinder}\ \emph {et~al.}(1992)\citenamefont
  {Buchbinder}, \citenamefont {Odintsov},\ and\ \citenamefont
  {Shapiro}}]{Buchbinder:1992rb}%
  \BibitemOpen
  \bibfield  {author} {\bibinfo {author} {\bibfnamefont {I.~L.}\ \bibnamefont
  {Buchbinder}}, \bibinfo {author} {\bibfnamefont {S.~D.}\ \bibnamefont
  {Odintsov}}, \ and\ \bibinfo {author} {\bibfnamefont {I.~L.}\ \bibnamefont
  {Shapiro}},\ }\href@noop {} {\emph {\bibinfo {title} {{Effective action in
  quantum gravity}}}}\ (\bibinfo {year} {1992})\BibitemShut {NoStop}%
\bibitem [{\citenamefont {Akrami}\ \emph {et~al.}(2020)\citenamefont {Akrami}
  \emph {et~al.}}]{Planck:2018jri}%
  \BibitemOpen
  \bibfield  {author} {\bibinfo {author} {\bibfnamefont {Y.}~\bibnamefont
  {Akrami}} \emph {et~al.} (\bibinfo {collaboration} {Planck}),\ }\href
  {\doibase 10.1051/0004-6361/201833887} {\bibfield  {journal} {\bibinfo
  {journal} {Astron. Astrophys.}\ }\textbf {\bibinfo {volume} {641}},\ \bibinfo
  {pages} {A10} (\bibinfo {year} {2020})},\ \Eprint
  {http://arxiv.org/abs/1807.06211} {arXiv:1807.06211 [astro-ph.CO]}
  \BibitemShut {NoStop}%
\bibitem [{\citenamefont {Spokoiny}(1984)}]{Spokoiny:1984bd}%
  \BibitemOpen
  \bibfield  {author} {\bibinfo {author} {\bibfnamefont {B.~L.}\ \bibnamefont
  {Spokoiny}},\ }\href {\doibase 10.1016/0370-2693(84)90587-2} {\bibfield
  {journal} {\bibinfo  {journal} {Phys. Lett. B}\ }\textbf {\bibinfo {volume}
  {147}},\ \bibinfo {pages} {39} (\bibinfo {year} {1984})}\BibitemShut
  {NoStop}%
\bibitem [{\citenamefont {Futamase}\ and\ \citenamefont
  {Maeda}(1989)}]{Futamase:1987ua}%
  \BibitemOpen
  \bibfield  {author} {\bibinfo {author} {\bibfnamefont {T.}~\bibnamefont
  {Futamase}}\ and\ \bibinfo {author} {\bibfnamefont {K.-i.}\ \bibnamefont
  {Maeda}},\ }\href {\doibase 10.1103/PhysRevD.39.399} {\bibfield  {journal}
  {\bibinfo  {journal} {Phys. Rev. D}\ }\textbf {\bibinfo {volume} {39}},\
  \bibinfo {pages} {399} (\bibinfo {year} {1989})}\BibitemShut {NoStop}%
\bibitem [{\citenamefont {Salopek}\ \emph {et~al.}(1989)\citenamefont
  {Salopek}, \citenamefont {Bond},\ and\ \citenamefont
  {Bardeen}}]{Salopek:1988qh}%
  \BibitemOpen
  \bibfield  {author} {\bibinfo {author} {\bibfnamefont {D.~S.}\ \bibnamefont
  {Salopek}}, \bibinfo {author} {\bibfnamefont {J.~R.}\ \bibnamefont {Bond}}, \
  and\ \bibinfo {author} {\bibfnamefont {J.~M.}\ \bibnamefont {Bardeen}},\
  }\href {\doibase 10.1103/PhysRevD.40.1753} {\bibfield  {journal} {\bibinfo
  {journal} {Phys. Rev. D}\ }\textbf {\bibinfo {volume} {40}},\ \bibinfo
  {pages} {1753} (\bibinfo {year} {1989})}\BibitemShut {NoStop}%
\bibitem [{\citenamefont {Fakir}\ and\ \citenamefont
  {Unruh}(1990)}]{Fakir:1990eg}%
  \BibitemOpen
  \bibfield  {author} {\bibinfo {author} {\bibfnamefont {R.}~\bibnamefont
  {Fakir}}\ and\ \bibinfo {author} {\bibfnamefont {W.~G.}\ \bibnamefont
  {Unruh}},\ }\href {\doibase 10.1103/PhysRevD.41.1783} {\bibfield  {journal}
  {\bibinfo  {journal} {Phys. Rev. D}\ }\textbf {\bibinfo {volume} {41}},\
  \bibinfo {pages} {1783} (\bibinfo {year} {1990})}\BibitemShut {NoStop}%
\bibitem [{\citenamefont {Bezrukov}\ and\ \citenamefont
  {Shaposhnikov}(2008)}]{Bezrukov:2007ep}%
  \BibitemOpen
  \bibfield  {author} {\bibinfo {author} {\bibfnamefont {F.~L.}\ \bibnamefont
  {Bezrukov}}\ and\ \bibinfo {author} {\bibfnamefont {M.}~\bibnamefont
  {Shaposhnikov}},\ }\href {\doibase 10.1016/j.physletb.2007.11.072} {\bibfield
   {journal} {\bibinfo  {journal} {Phys. Lett. B}\ }\textbf {\bibinfo {volume}
  {659}},\ \bibinfo {pages} {703} (\bibinfo {year} {2008})},\ \Eprint
  {http://arxiv.org/abs/0710.3755} {arXiv:0710.3755 [hep-th]} \BibitemShut
  {NoStop}%
\bibitem [{\citenamefont {Salvio}\ and\ \citenamefont
  {Strumia}(2014)}]{Salvio:2014soa}%
  \BibitemOpen
  \bibfield  {author} {\bibinfo {author} {\bibfnamefont {A.}~\bibnamefont
  {Salvio}}\ and\ \bibinfo {author} {\bibfnamefont {A.}~\bibnamefont
  {Strumia}},\ }\href {\doibase 10.1007/JHEP06(2014)080} {\bibfield  {journal}
  {\bibinfo  {journal} {JHEP}\ }\textbf {\bibinfo {volume} {06}},\ \bibinfo
  {pages} {080} (\bibinfo {year} {2014})},\ \Eprint
  {http://arxiv.org/abs/1403.4226} {arXiv:1403.4226 [hep-ph]} \BibitemShut
  {NoStop}%
\bibitem [{\citenamefont {Csaki}\ \emph {et~al.}(2014)\citenamefont {Csaki},
  \citenamefont {Kaloper}, \citenamefont {Serra},\ and\ \citenamefont
  {Terning}}]{Csaki:2014bua}%
  \BibitemOpen
  \bibfield  {author} {\bibinfo {author} {\bibfnamefont {C.}~\bibnamefont
  {Csaki}}, \bibinfo {author} {\bibfnamefont {N.}~\bibnamefont {Kaloper}},
  \bibinfo {author} {\bibfnamefont {J.}~\bibnamefont {Serra}}, \ and\ \bibinfo
  {author} {\bibfnamefont {J.}~\bibnamefont {Terning}},\ }\href {\doibase
  10.1103/PhysRevLett.113.161302} {\bibfield  {journal} {\bibinfo  {journal}
  {Phys. Rev. Lett.}\ }\textbf {\bibinfo {volume} {113}},\ \bibinfo {pages}
  {161302} (\bibinfo {year} {2014})},\ \Eprint {http://arxiv.org/abs/1406.5192}
  {arXiv:1406.5192 [hep-th]} \BibitemShut {NoStop}%
\bibitem [{\citenamefont {Kannike}\ \emph {et~al.}(2015)\citenamefont
  {Kannike}, \citenamefont {H\"utsi}, \citenamefont {Pizza}, \citenamefont
  {Racioppi}, \citenamefont {Raidal}, \citenamefont {Salvio},\ and\
  \citenamefont {Strumia}}]{Kannike:2015apa}%
  \BibitemOpen
  \bibfield  {author} {\bibinfo {author} {\bibfnamefont {K.}~\bibnamefont
  {Kannike}}, \bibinfo {author} {\bibfnamefont {G.}~\bibnamefont {H\"utsi}},
  \bibinfo {author} {\bibfnamefont {L.}~\bibnamefont {Pizza}}, \bibinfo
  {author} {\bibfnamefont {A.}~\bibnamefont {Racioppi}}, \bibinfo {author}
  {\bibfnamefont {M.}~\bibnamefont {Raidal}}, \bibinfo {author} {\bibfnamefont
  {A.}~\bibnamefont {Salvio}}, \ and\ \bibinfo {author} {\bibfnamefont
  {A.}~\bibnamefont {Strumia}},\ }\href {\doibase 10.1007/JHEP05(2015)065}
  {\bibfield  {journal} {\bibinfo  {journal} {JHEP}\ }\textbf {\bibinfo
  {volume} {05}},\ \bibinfo {pages} {065} (\bibinfo {year} {2015})},\ \Eprint
  {http://arxiv.org/abs/1502.01334} {arXiv:1502.01334 [astro-ph.CO]}
  \BibitemShut {NoStop}%
\bibitem [{\citenamefont {Palatini}(1919)}]{Palatini:1919}%
  \BibitemOpen
  \bibfield  {author} {\bibinfo {author} {\bibfnamefont {A.}~\bibnamefont
  {Palatini}},\ }\href {\doibase 10.1007/BF03014670} {\bibfield  {journal}
  {\bibinfo  {journal} {Rendiconti del Circolo Matematico di Palermo
  (1884-1940)}\ }\textbf {\bibinfo {volume} {43}},\ \bibinfo {pages} {203}
  (\bibinfo {year} {1919})}\BibitemShut {NoStop}%
\bibitem [{\citenamefont {Einstein}(1925)}]{Einstein:1925}%
  \BibitemOpen
  \bibfield  {author} {\bibinfo {author} {\bibfnamefont {A.}~\bibnamefont
  {Einstein}},\ }\href@noop {} {\bibfield  {journal} {\bibinfo  {journal}
  {Sitzungsber. K. Preuss. Akad. Wiss.}\ }\textbf {\bibinfo {volume} {43}},\
  \bibinfo {pages} {414} (\bibinfo {year} {1925})}\BibitemShut {NoStop}%
\bibitem [{\citenamefont {Bauer}\ and\ \citenamefont
  {Demir}(2008)}]{Bauer:2008zj}%
  \BibitemOpen
  \bibfield  {author} {\bibinfo {author} {\bibfnamefont {F.}~\bibnamefont
  {Bauer}}\ and\ \bibinfo {author} {\bibfnamefont {D.~A.}\ \bibnamefont
  {Demir}},\ }\href {\doibase 10.1016/j.physletb.2008.06.014} {\bibfield
  {journal} {\bibinfo  {journal} {Phys. Lett. B}\ }\textbf {\bibinfo {volume}
  {665}},\ \bibinfo {pages} {222} (\bibinfo {year} {2008})},\ \Eprint
  {http://arxiv.org/abs/0803.2664} {arXiv:0803.2664 [hep-ph]} \BibitemShut
  {NoStop}%
\bibitem [{\citenamefont {Misner}\ \emph {et~al.}(1973)\citenamefont {Misner},
  \citenamefont {Thorne},\ and\ \citenamefont {Wheeler}}]{Misner:1974qy}%
  \BibitemOpen
  \bibfield  {author} {\bibinfo {author} {\bibfnamefont {C.~W.}\ \bibnamefont
  {Misner}}, \bibinfo {author} {\bibfnamefont {K.~S.}\ \bibnamefont {Thorne}},
  \ and\ \bibinfo {author} {\bibfnamefont {J.~A.}\ \bibnamefont {Wheeler}},\
  }\href@noop {} {\emph {\bibinfo {title} {{Gravitation}}}}\ (\bibinfo
  {publisher} {W. H. Freeman},\ \bibinfo {address} {San Francisco},\ \bibinfo
  {year} {1973})\BibitemShut {NoStop}%
\bibitem [{\citenamefont {Dufaux}\ \emph {et~al.}(2010)\citenamefont {Dufaux},
  \citenamefont {Figueroa},\ and\ \citenamefont
  {Garcia-Bellido}}]{Dufaux:2010cf}%
  \BibitemOpen
  \bibfield  {author} {\bibinfo {author} {\bibfnamefont {J.-F.}\ \bibnamefont
  {Dufaux}}, \bibinfo {author} {\bibfnamefont {D.~G.}\ \bibnamefont
  {Figueroa}}, \ and\ \bibinfo {author} {\bibfnamefont {J.}~\bibnamefont
  {Garcia-Bellido}},\ }\href {\doibase 10.1103/PhysRevD.82.083518} {\bibfield
  {journal} {\bibinfo  {journal} {Phys. Rev. D}\ }\textbf {\bibinfo {volume}
  {82}},\ \bibinfo {pages} {083518} (\bibinfo {year} {2010})},\ \Eprint
  {http://arxiv.org/abs/1006.0217} {arXiv:1006.0217 [astro-ph.CO]} \BibitemShut
  {NoStop}%
\bibitem [{\citenamefont {Karamitsos}\ and\ \citenamefont
  {Pilaftsis}(2018)}]{Karamitsos:2017elm}%
  \BibitemOpen
  \bibfield  {author} {\bibinfo {author} {\bibfnamefont {S.}~\bibnamefont
  {Karamitsos}}\ and\ \bibinfo {author} {\bibfnamefont {A.}~\bibnamefont
  {Pilaftsis}},\ }\href {\doibase 10.1016/j.nuclphysb.2017.12.015} {\bibfield
  {journal} {\bibinfo  {journal} {Nucl. Phys. B}\ }\textbf {\bibinfo {volume}
  {927}},\ \bibinfo {pages} {219} (\bibinfo {year} {2018})},\ \Eprint
  {http://arxiv.org/abs/1706.07011} {arXiv:1706.07011 [hep-ph]} \BibitemShut
  {NoStop}%
\bibitem [{\citenamefont {Kallosh}\ \emph {et~al.}(2013)\citenamefont
  {Kallosh}, \citenamefont {Linde},\ and\ \citenamefont
  {Roest}}]{Kallosh:2013yoa}%
  \BibitemOpen
  \bibfield  {author} {\bibinfo {author} {\bibfnamefont {R.}~\bibnamefont
  {Kallosh}}, \bibinfo {author} {\bibfnamefont {A.}~\bibnamefont {Linde}}, \
  and\ \bibinfo {author} {\bibfnamefont {D.}~\bibnamefont {Roest}},\ }\href
  {\doibase 10.1007/JHEP11(2013)198} {\bibfield  {journal} {\bibinfo  {journal}
  {JHEP}\ }\textbf {\bibinfo {volume} {11}},\ \bibinfo {pages} {198} (\bibinfo
  {year} {2013})},\ \Eprint {http://arxiv.org/abs/1311.0472} {arXiv:1311.0472
  [hep-th]} \BibitemShut {NoStop}%
\bibitem [{\citenamefont {Galante}\ \emph {et~al.}(2015)\citenamefont
  {Galante}, \citenamefont {Kallosh}, \citenamefont {Linde},\ and\
  \citenamefont {Roest}}]{Galante:2014ifa}%
  \BibitemOpen
  \bibfield  {author} {\bibinfo {author} {\bibfnamefont {M.}~\bibnamefont
  {Galante}}, \bibinfo {author} {\bibfnamefont {R.}~\bibnamefont {Kallosh}},
  \bibinfo {author} {\bibfnamefont {A.}~\bibnamefont {Linde}}, \ and\ \bibinfo
  {author} {\bibfnamefont {D.}~\bibnamefont {Roest}},\ }\href {\doibase
  10.1103/PhysRevLett.114.141302} {\bibfield  {journal} {\bibinfo  {journal}
  {Phys. Rev. Lett.}\ }\textbf {\bibinfo {volume} {114}},\ \bibinfo {pages}
  {141302} (\bibinfo {year} {2015})},\ \Eprint {http://arxiv.org/abs/1412.3797}
  {arXiv:1412.3797 [hep-th]} \BibitemShut {NoStop}%
\bibitem [{\citenamefont {Carrasco}\ \emph {et~al.}(2015)\citenamefont
  {Carrasco}, \citenamefont {Kallosh},\ and\ \citenamefont
  {Linde}}]{Carrasco:2015pla}%
  \BibitemOpen
  \bibfield  {author} {\bibinfo {author} {\bibfnamefont {J.~J.~M.}\
  \bibnamefont {Carrasco}}, \bibinfo {author} {\bibfnamefont {R.}~\bibnamefont
  {Kallosh}}, \ and\ \bibinfo {author} {\bibfnamefont {A.}~\bibnamefont
  {Linde}},\ }\href {\doibase 10.1007/JHEP10(2015)147} {\bibfield  {journal}
  {\bibinfo  {journal} {JHEP}\ }\textbf {\bibinfo {volume} {10}},\ \bibinfo
  {pages} {147} (\bibinfo {year} {2015})},\ \Eprint
  {http://arxiv.org/abs/1506.01708} {arXiv:1506.01708 [hep-th]} \BibitemShut
  {NoStop}%
\bibitem [{\citenamefont {Renaux-Petel}\ and\ \citenamefont
  {Turzy\'nski}(2016)}]{Renaux-Petel:2015mga}%
  \BibitemOpen
  \bibfield  {author} {\bibinfo {author} {\bibfnamefont {S.}~\bibnamefont
  {Renaux-Petel}}\ and\ \bibinfo {author} {\bibfnamefont {K.}~\bibnamefont
  {Turzy\'nski}},\ }\href {\doibase 10.1103/PhysRevLett.117.141301} {\bibfield
  {journal} {\bibinfo  {journal} {Phys. Rev. Lett.}\ }\textbf {\bibinfo
  {volume} {117}},\ \bibinfo {pages} {141301} (\bibinfo {year} {2016})},\
  \Eprint {http://arxiv.org/abs/1510.01281} {arXiv:1510.01281 [astro-ph.CO]}
  \BibitemShut {NoStop}%
\bibitem [{\citenamefont {Bond}\ \emph {et~al.}(2009)\citenamefont {Bond},
  \citenamefont {Frolov}, \citenamefont {Huang},\ and\ \citenamefont
  {Kofman}}]{Bond:2009xx}%
  \BibitemOpen
  \bibfield  {author} {\bibinfo {author} {\bibfnamefont {J.~R.}\ \bibnamefont
  {Bond}}, \bibinfo {author} {\bibfnamefont {A.~V.}\ \bibnamefont {Frolov}},
  \bibinfo {author} {\bibfnamefont {Z.}~\bibnamefont {Huang}}, \ and\ \bibinfo
  {author} {\bibfnamefont {L.}~\bibnamefont {Kofman}},\ }\href {\doibase
  10.1103/PhysRevLett.103.071301} {\bibfield  {journal} {\bibinfo  {journal}
  {Phys. Rev. Lett.}\ }\textbf {\bibinfo {volume} {103}},\ \bibinfo {pages}
  {071301} (\bibinfo {year} {2009})},\ \Eprint {http://arxiv.org/abs/0903.3407}
  {arXiv:0903.3407 [astro-ph.CO]} \BibitemShut {NoStop}%
\bibitem [{\citenamefont {Rusak}(2020)}]{Rusak:2018kel}%
  \BibitemOpen
  \bibfield  {author} {\bibinfo {author} {\bibfnamefont {S.}~\bibnamefont
  {Rusak}},\ }\href {\doibase 10.1088/1475-7516/2020/05/020} {\bibfield
  {journal} {\bibinfo  {journal} {JCAP}\ }\textbf {\bibinfo {volume} {05}},\
  \bibinfo {pages} {020} (\bibinfo {year} {2020})},\ \Eprint
  {http://arxiv.org/abs/1811.10569} {arXiv:1811.10569 [hep-ph]} \BibitemShut
  {NoStop}%
\bibitem [{\citenamefont {Fu}\ \emph {et~al.}(2017)\citenamefont {Fu},
  \citenamefont {Wu},\ and\ \citenamefont {Yu}}]{Fu:2017iqg}%
  \BibitemOpen
  \bibfield  {author} {\bibinfo {author} {\bibfnamefont {C.}~\bibnamefont
  {Fu}}, \bibinfo {author} {\bibfnamefont {P.}~\bibnamefont {Wu}}, \ and\
  \bibinfo {author} {\bibfnamefont {H.}~\bibnamefont {Yu}},\ }\href {\doibase
  10.1103/PhysRevD.96.103542} {\bibfield  {journal} {\bibinfo  {journal} {Phys.
  Rev. D}\ }\textbf {\bibinfo {volume} {96}},\ \bibinfo {pages} {103542}
  (\bibinfo {year} {2017})},\ \Eprint {http://arxiv.org/abs/1801.04089}
  {arXiv:1801.04089 [gr-qc]} \BibitemShut {NoStop}%
\bibitem [{\citenamefont {Karam}\ \emph {et~al.}(2021)\citenamefont {Karam},
  \citenamefont {Raidal},\ and\ \citenamefont {Tomberg}}]{Karam:2020rpa}%
  \BibitemOpen
  \bibfield  {author} {\bibinfo {author} {\bibfnamefont {A.}~\bibnamefont
  {Karam}}, \bibinfo {author} {\bibfnamefont {M.}~\bibnamefont {Raidal}}, \
  and\ \bibinfo {author} {\bibfnamefont {E.}~\bibnamefont {Tomberg}},\ }\href
  {\doibase 10.1088/1475-7516/2021/03/064} {\bibfield  {journal} {\bibinfo
  {journal} {JCAP}\ }\textbf {\bibinfo {volume} {03}},\ \bibinfo {pages} {064}
  (\bibinfo {year} {2021})},\ \Eprint {http://arxiv.org/abs/2007.03484}
  {arXiv:2007.03484 [astro-ph.CO]} \BibitemShut {NoStop}%
\bibitem [{\citenamefont {Turner}(1983)}]{Turner:1983he}%
  \BibitemOpen
  \bibfield  {author} {\bibinfo {author} {\bibfnamefont {M.~S.}\ \bibnamefont
  {Turner}},\ }\href {\doibase 10.1103/PhysRevD.28.1243} {\bibfield  {journal}
  {\bibinfo  {journal} {Phys. Rev. D}\ }\textbf {\bibinfo {volume} {28}},\
  \bibinfo {pages} {1243} (\bibinfo {year} {1983})}\BibitemShut {NoStop}%
\bibitem [{\citenamefont {Whittaker}\ and\ \citenamefont
  {Watson}(1996)}]{whittaker_watson_1996}%
  \BibitemOpen
  \bibfield  {author} {\bibinfo {author} {\bibfnamefont {E.~T.}\ \bibnamefont
  {Whittaker}}\ and\ \bibinfo {author} {\bibfnamefont {G.~N.}\ \bibnamefont
  {Watson}},\ }\href {\doibase 10.1017/CBO9780511608759} {\emph {\bibinfo
  {title} {A Course of Modern Analysis}}},\ \bibinfo {edition} {4th}\ ed.,\
  Cambridge Mathematical Library\ (\bibinfo  {publisher} {Cambridge University
  Press},\ \bibinfo {year} {1996})\BibitemShut {NoStop}%
\bibitem [{\citenamefont {Lachapelle}\ and\ \citenamefont
  {Brandenberger}(2009)}]{Lachapelle:2008sy}%
  \BibitemOpen
  \bibfield  {author} {\bibinfo {author} {\bibfnamefont {J.}~\bibnamefont
  {Lachapelle}}\ and\ \bibinfo {author} {\bibfnamefont {R.~H.}\ \bibnamefont
  {Brandenberger}},\ }\href {\doibase 10.1088/1475-7516/2009/04/020} {\bibfield
   {journal} {\bibinfo  {journal} {JCAP}\ }\textbf {\bibinfo {volume} {04}},\
  \bibinfo {pages} {020} (\bibinfo {year} {2009})},\ \Eprint
  {http://arxiv.org/abs/0808.0936} {arXiv:0808.0936 [hep-th]} \BibitemShut
  {NoStop}%
\bibitem [{\citenamefont {Choi}\ \emph {et~al.}(2019)\citenamefont {Choi},
  \citenamefont {Kang}, \citenamefont {Lee},\ and\ \citenamefont
  {Yamashita}}]{Choi:2019osi}%
  \BibitemOpen
  \bibfield  {author} {\bibinfo {author} {\bibfnamefont {S.-M.}\ \bibnamefont
  {Choi}}, \bibinfo {author} {\bibfnamefont {Y.-J.}\ \bibnamefont {Kang}},
  \bibinfo {author} {\bibfnamefont {H.~M.}\ \bibnamefont {Lee}}, \ and\
  \bibinfo {author} {\bibfnamefont {K.}~\bibnamefont {Yamashita}},\ }\href
  {\doibase 10.1007/JHEP05(2019)060} {\bibfield  {journal} {\bibinfo  {journal}
  {JHEP}\ }\textbf {\bibinfo {volume} {05}},\ \bibinfo {pages} {060} (\bibinfo
  {year} {2019})},\ \Eprint {http://arxiv.org/abs/1902.03781} {arXiv:1902.03781
  [hep-ph]} \BibitemShut {NoStop}%
\bibitem [{\citenamefont {Lebedev}\ and\ \citenamefont
  {Yoon}(2021)}]{Lebedev:2021zdh}%
  \BibitemOpen
  \bibfield  {author} {\bibinfo {author} {\bibfnamefont {O.}~\bibnamefont
  {Lebedev}}\ and\ \bibinfo {author} {\bibfnamefont {J.-H.}\ \bibnamefont
  {Yoon}},\ }\href@noop {} {\  (\bibinfo {year} {2021})},\ \Eprint
  {http://arxiv.org/abs/2105.05860} {arXiv:2105.05860 [hep-ph]} \BibitemShut
  {NoStop}%
\bibitem [{\citenamefont {Boyanovsky}\ \emph {et~al.}(1995)\citenamefont
  {Boyanovsky}, \citenamefont {de~Vega}, \citenamefont {Holman}, \citenamefont
  {Lee},\ and\ \citenamefont {Singh}}]{Boyanovsky:1994me}%
  \BibitemOpen
  \bibfield  {author} {\bibinfo {author} {\bibfnamefont {D.}~\bibnamefont
  {Boyanovsky}}, \bibinfo {author} {\bibfnamefont {H.~J.}\ \bibnamefont
  {de~Vega}}, \bibinfo {author} {\bibfnamefont {R.}~\bibnamefont {Holman}},
  \bibinfo {author} {\bibfnamefont {D.~S.}\ \bibnamefont {Lee}}, \ and\
  \bibinfo {author} {\bibfnamefont {A.}~\bibnamefont {Singh}},\ }\href
  {\doibase 10.1103/PhysRevD.51.4419} {\bibfield  {journal} {\bibinfo
  {journal} {Phys. Rev. D}\ }\textbf {\bibinfo {volume} {51}},\ \bibinfo
  {pages} {4419} (\bibinfo {year} {1995})},\ \Eprint
  {http://arxiv.org/abs/hep-ph/9408214} {arXiv:hep-ph/9408214} \BibitemShut
  {NoStop}%
\bibitem [{\citenamefont {Khlebnikov}\ and\ \citenamefont
  {Tkachev}(1996)}]{Khlebnikov:1996mc}%
  \BibitemOpen
  \bibfield  {author} {\bibinfo {author} {\bibfnamefont {S.~Y.}\ \bibnamefont
  {Khlebnikov}}\ and\ \bibinfo {author} {\bibfnamefont {I.~I.}\ \bibnamefont
  {Tkachev}},\ }\href {\doibase 10.1103/PhysRevLett.77.219} {\bibfield
  {journal} {\bibinfo  {journal} {Phys. Rev. Lett.}\ }\textbf {\bibinfo
  {volume} {77}},\ \bibinfo {pages} {219} (\bibinfo {year} {1996})},\ \Eprint
  {http://arxiv.org/abs/hep-ph/9603378} {arXiv:hep-ph/9603378} \BibitemShut
  {NoStop}%
\bibitem [{\citenamefont {Fukugita}\ and\ \citenamefont
  {Yanagida}(1986)}]{Fukugita:1986hr}%
  \BibitemOpen
  \bibfield  {author} {\bibinfo {author} {\bibfnamefont {M.}~\bibnamefont
  {Fukugita}}\ and\ \bibinfo {author} {\bibfnamefont {T.}~\bibnamefont
  {Yanagida}},\ }\href {\doibase 10.1016/0370-2693(86)91126-3} {\bibfield
  {journal} {\bibinfo  {journal} {Phys. Lett. B}\ }\textbf {\bibinfo {volume}
  {174}},\ \bibinfo {pages} {45} (\bibinfo {year} {1986})}\BibitemShut
  {NoStop}%
\bibitem [{\citenamefont {Almeida}\ \emph {et~al.}(2019)\citenamefont
  {Almeida}, \citenamefont {Bernal}, \citenamefont {Rubio},\ and\ \citenamefont
  {Tenkanen}}]{Almeida:2018oid}%
  \BibitemOpen
  \bibfield  {author} {\bibinfo {author} {\bibfnamefont {J.~P.~B.}\
  \bibnamefont {Almeida}}, \bibinfo {author} {\bibfnamefont {N.}~\bibnamefont
  {Bernal}}, \bibinfo {author} {\bibfnamefont {J.}~\bibnamefont {Rubio}}, \
  and\ \bibinfo {author} {\bibfnamefont {T.}~\bibnamefont {Tenkanen}},\ }\href
  {\doibase 10.1088/1475-7516/2019/03/012} {\bibfield  {journal} {\bibinfo
  {journal} {JCAP}\ }\textbf {\bibinfo {volume} {03}},\ \bibinfo {pages} {012}
  (\bibinfo {year} {2019})},\ \Eprint {http://arxiv.org/abs/1811.09640}
  {arXiv:1811.09640 [hep-ph]} \BibitemShut {NoStop}%
\bibitem [{\citenamefont {Babichev}\ \emph {et~al.}(2020)\citenamefont
  {Babichev}, \citenamefont {Gorbunov}, \citenamefont {Ramazanov},\ and\
  \citenamefont {Reverberi}}]{Babichev:2020yeo}%
  \BibitemOpen
  \bibfield  {author} {\bibinfo {author} {\bibfnamefont {E.}~\bibnamefont
  {Babichev}}, \bibinfo {author} {\bibfnamefont {D.}~\bibnamefont {Gorbunov}},
  \bibinfo {author} {\bibfnamefont {S.}~\bibnamefont {Ramazanov}}, \ and\
  \bibinfo {author} {\bibfnamefont {L.}~\bibnamefont {Reverberi}},\ }\href
  {\doibase 10.1088/1475-7516/2020/09/059} {\bibfield  {journal} {\bibinfo
  {journal} {JCAP}\ }\textbf {\bibinfo {volume} {09}},\ \bibinfo {pages} {059}
  (\bibinfo {year} {2020})},\ \Eprint {http://arxiv.org/abs/2006.02225}
  {arXiv:2006.02225 [hep-ph]} \BibitemShut {NoStop}%
\bibitem [{\citenamefont {Harigaya}\ and\ \citenamefont
  {Mukaida}(2014)}]{Harigaya:2013vwa}%
  \BibitemOpen
  \bibfield  {author} {\bibinfo {author} {\bibfnamefont {K.}~\bibnamefont
  {Harigaya}}\ and\ \bibinfo {author} {\bibfnamefont {K.}~\bibnamefont
  {Mukaida}},\ }\href {\doibase 10.1007/JHEP05(2014)006} {\bibfield  {journal}
  {\bibinfo  {journal} {JHEP}\ }\textbf {\bibinfo {volume} {05}},\ \bibinfo
  {pages} {006} (\bibinfo {year} {2014})},\ \Eprint
  {http://arxiv.org/abs/1312.3097} {arXiv:1312.3097 [hep-ph]} \BibitemShut
  {NoStop}%
\bibitem [{\citenamefont {Mukaida}\ and\ \citenamefont
  {Yamada}(2016)}]{Mukaida:2015ria}%
  \BibitemOpen
  \bibfield  {author} {\bibinfo {author} {\bibfnamefont {K.}~\bibnamefont
  {Mukaida}}\ and\ \bibinfo {author} {\bibfnamefont {M.}~\bibnamefont
  {Yamada}},\ }\href {\doibase 10.1088/1475-7516/2016/02/003} {\bibfield
  {journal} {\bibinfo  {journal} {JCAP}\ }\textbf {\bibinfo {volume} {02}},\
  \bibinfo {pages} {003} (\bibinfo {year} {2016})},\ \Eprint
  {http://arxiv.org/abs/1506.07661} {arXiv:1506.07661 [hep-ph]} \BibitemShut
  {NoStop}%
\bibitem [{\citenamefont {Delle~Rose}\ \emph {et~al.}(2016)\citenamefont
  {Delle~Rose}, \citenamefont {Marzo},\ and\ \citenamefont
  {Urbano}}]{Rose:2015lna}%
  \BibitemOpen
  \bibfield  {author} {\bibinfo {author} {\bibfnamefont {L.}~\bibnamefont
  {Delle~Rose}}, \bibinfo {author} {\bibfnamefont {C.}~\bibnamefont {Marzo}}, \
  and\ \bibinfo {author} {\bibfnamefont {A.}~\bibnamefont {Urbano}},\ }\href
  {\doibase 10.1007/JHEP05(2016)050} {\bibfield  {journal} {\bibinfo  {journal}
  {JHEP}\ }\textbf {\bibinfo {volume} {05}},\ \bibinfo {pages} {050} (\bibinfo
  {year} {2016})},\ \Eprint {http://arxiv.org/abs/1507.06912} {arXiv:1507.06912
  [hep-ph]} \BibitemShut {NoStop}%
\bibitem [{\citenamefont {Chambers}\ and\ \citenamefont
  {Rajantie}(2008)}]{Chambers:2008gu}%
  \BibitemOpen
  \bibfield  {author} {\bibinfo {author} {\bibfnamefont {A.}~\bibnamefont
  {Chambers}}\ and\ \bibinfo {author} {\bibfnamefont {A.}~\bibnamefont
  {Rajantie}},\ }\href {\doibase 10.1088/1475-7516/2008/08/002} {\bibfield
  {journal} {\bibinfo  {journal} {JCAP}\ }\textbf {\bibinfo {volume} {08}},\
  \bibinfo {pages} {002} (\bibinfo {year} {2008})},\ \Eprint
  {http://arxiv.org/abs/0805.4795} {arXiv:0805.4795 [astro-ph]} \BibitemShut
  {NoStop}%
\bibitem [{\citenamefont {Garcia}\ \emph {et~al.}(2020)\citenamefont {Garcia},
  \citenamefont {Kaneta}, \citenamefont {Mambrini},\ and\ \citenamefont
  {Olive}}]{Garcia:2020eof}%
  \BibitemOpen
  \bibfield  {author} {\bibinfo {author} {\bibfnamefont {M.~A.~G.}\
  \bibnamefont {Garcia}}, \bibinfo {author} {\bibfnamefont {K.}~\bibnamefont
  {Kaneta}}, \bibinfo {author} {\bibfnamefont {Y.}~\bibnamefont {Mambrini}}, \
  and\ \bibinfo {author} {\bibfnamefont {K.~A.}\ \bibnamefont {Olive}},\ }\href
  {\doibase 10.1103/PhysRevD.101.123507} {\bibfield  {journal} {\bibinfo
  {journal} {Phys. Rev. D}\ }\textbf {\bibinfo {volume} {101}},\ \bibinfo
  {pages} {123507} (\bibinfo {year} {2020})},\ \Eprint
  {http://arxiv.org/abs/2004.08404} {arXiv:2004.08404 [hep-ph]} \BibitemShut
  {NoStop}%
\bibitem [{\citenamefont {Garcia}\ \emph {et~al.}(2021)\citenamefont {Garcia},
  \citenamefont {Kaneta}, \citenamefont {Mambrini},\ and\ \citenamefont
  {Olive}}]{Garcia:2020wiy}%
  \BibitemOpen
  \bibfield  {author} {\bibinfo {author} {\bibfnamefont {M.~A.~G.}\
  \bibnamefont {Garcia}}, \bibinfo {author} {\bibfnamefont {K.}~\bibnamefont
  {Kaneta}}, \bibinfo {author} {\bibfnamefont {Y.}~\bibnamefont {Mambrini}}, \
  and\ \bibinfo {author} {\bibfnamefont {K.~A.}\ \bibnamefont {Olive}},\ }\href
  {\doibase 10.1088/1475-7516/2021/04/012} {\bibfield  {journal} {\bibinfo
  {journal} {JCAP}\ }\textbf {\bibinfo {volume} {04}},\ \bibinfo {pages} {012}
  (\bibinfo {year} {2021})},\ \Eprint {http://arxiv.org/abs/2012.10756}
  {arXiv:2012.10756 [hep-ph]} \BibitemShut {NoStop}%
\bibitem [{\citenamefont {Cline}\ \emph {et~al.}(2020)\citenamefont {Cline},
  \citenamefont {Puel},\ and\ \citenamefont {Toma}}]{Cline:2019fxx}%
  \BibitemOpen
  \bibfield  {author} {\bibinfo {author} {\bibfnamefont {J.~M.}\ \bibnamefont
  {Cline}}, \bibinfo {author} {\bibfnamefont {M.}~\bibnamefont {Puel}}, \ and\
  \bibinfo {author} {\bibfnamefont {T.}~\bibnamefont {Toma}},\ }\href {\doibase
  10.1103/PhysRevD.101.043014} {\bibfield  {journal} {\bibinfo  {journal}
  {Phys. Rev. D}\ }\textbf {\bibinfo {volume} {101}},\ \bibinfo {pages}
  {043014} (\bibinfo {year} {2020})},\ \Eprint
  {http://arxiv.org/abs/1909.12300} {arXiv:1909.12300 [hep-ph]} \BibitemShut
  {NoStop}%
\bibitem [{\citenamefont {Kawasaki}\ and\ \citenamefont
  {Ueda}(2021)}]{Kawasaki:2020xyf}%
  \BibitemOpen
  \bibfield  {author} {\bibinfo {author} {\bibfnamefont {M.}~\bibnamefont
  {Kawasaki}}\ and\ \bibinfo {author} {\bibfnamefont {S.}~\bibnamefont
  {Ueda}},\ }\href {\doibase 10.1088/1475-7516/2021/04/049} {\bibfield
  {journal} {\bibinfo  {journal} {JCAP}\ }\textbf {\bibinfo {volume} {04}},\
  \bibinfo {pages} {049} (\bibinfo {year} {2021})},\ \Eprint
  {http://arxiv.org/abs/2011.10397} {arXiv:2011.10397 [hep-ph]} \BibitemShut
  {NoStop}%
\bibitem [{\citenamefont {Kost}\ \emph {et~al.}()\citenamefont {Kost},
  \citenamefont {Shin},\ and\ \citenamefont {Terada}}]{spiraling}%
  \BibitemOpen
  \bibfield  {author} {\bibinfo {author} {\bibfnamefont {J.}~\bibnamefont
  {Kost}}, \bibinfo {author} {\bibfnamefont {C.~S.}\ \bibnamefont {Shin}}, \
  and\ \bibinfo {author} {\bibfnamefont {T.}~\bibnamefont {Terada}},\
  }\href@noop {} {}\bibinfo {note} {$\text{in preparation}$}\BibitemShut
  {NoStop}%
\bibitem [{\citenamefont {Linde}(1994)}]{Linde:1993cn}%
  \BibitemOpen
  \bibfield  {author} {\bibinfo {author} {\bibfnamefont {A.~D.}\ \bibnamefont
  {Linde}},\ }\href {\doibase 10.1103/PhysRevD.49.748} {\bibfield  {journal}
  {\bibinfo  {journal} {Phys. Rev. D}\ }\textbf {\bibinfo {volume} {49}},\
  \bibinfo {pages} {748} (\bibinfo {year} {1994})},\ \Eprint
  {http://arxiv.org/abs/astro-ph/9307002} {arXiv:astro-ph/9307002} \BibitemShut
  {NoStop}%
\bibitem [{\citenamefont {Garcia-Bellido}\ and\ \citenamefont
  {Linde}(1998)}]{GarciaBellido:1997wm}%
  \BibitemOpen
  \bibfield  {author} {\bibinfo {author} {\bibfnamefont {J.}~\bibnamefont
  {Garcia-Bellido}}\ and\ \bibinfo {author} {\bibfnamefont {A.~D.}\
  \bibnamefont {Linde}},\ }\href {\doibase 10.1103/PhysRevD.57.6075} {\bibfield
   {journal} {\bibinfo  {journal} {Phys. Rev. D}\ }\textbf {\bibinfo {volume}
  {57}},\ \bibinfo {pages} {6075} (\bibinfo {year} {1998})},\ \Eprint
  {http://arxiv.org/abs/hep-ph/9711360} {arXiv:hep-ph/9711360} \BibitemShut
  {NoStop}%
\bibitem [{\citenamefont {Shaposhnikov}\ \emph
  {et~al.}(2021{\natexlab{a}})\citenamefont {Shaposhnikov}, \citenamefont
  {Shkerin}, \citenamefont {Timiryasov},\ and\ \citenamefont
  {Zell}}]{Shaposhnikov:2020gts}%
  \BibitemOpen
  \bibfield  {author} {\bibinfo {author} {\bibfnamefont {M.}~\bibnamefont
  {Shaposhnikov}}, \bibinfo {author} {\bibfnamefont {A.}~\bibnamefont
  {Shkerin}}, \bibinfo {author} {\bibfnamefont {I.}~\bibnamefont {Timiryasov}},
  \ and\ \bibinfo {author} {\bibfnamefont {S.}~\bibnamefont {Zell}},\ }\href
  {\doibase 10.1088/1475-7516/2021/02/008} {\bibfield  {journal} {\bibinfo
  {journal} {JCAP}\ }\textbf {\bibinfo {volume} {02}},\ \bibinfo {pages} {008}
  (\bibinfo {year} {2021}{\natexlab{a}})},\ \Eprint
  {http://arxiv.org/abs/2007.14978} {arXiv:2007.14978 [hep-ph]} \BibitemShut
  {NoStop}%
\bibitem [{\citenamefont {Shaposhnikov}\ \emph {et~al.}(2020)\citenamefont
  {Shaposhnikov}, \citenamefont {Shkerin}, \citenamefont {Timiryasov},\ and\
  \citenamefont {Zell}}]{Shaposhnikov:2020frq}%
  \BibitemOpen
  \bibfield  {author} {\bibinfo {author} {\bibfnamefont {M.}~\bibnamefont
  {Shaposhnikov}}, \bibinfo {author} {\bibfnamefont {A.}~\bibnamefont
  {Shkerin}}, \bibinfo {author} {\bibfnamefont {I.}~\bibnamefont {Timiryasov}},
  \ and\ \bibinfo {author} {\bibfnamefont {S.}~\bibnamefont {Zell}},\ }\href
  {\doibase 10.1007/JHEP10(2020)177} {\bibfield  {journal} {\bibinfo  {journal}
  {JHEP}\ }\textbf {\bibinfo {volume} {10}},\ \bibinfo {pages} {177} (\bibinfo
  {year} {2020})},\ \Eprint {http://arxiv.org/abs/2007.16158} {arXiv:2007.16158
  [hep-th]} \BibitemShut {NoStop}%
\bibitem [{\citenamefont {Shaposhnikov}\ \emph
  {et~al.}(2021{\natexlab{b}})\citenamefont {Shaposhnikov}, \citenamefont
  {Shkerin}, \citenamefont {Timiryasov},\ and\ \citenamefont
  {Zell}}]{Shaposhnikov:2020aen}%
  \BibitemOpen
  \bibfield  {author} {\bibinfo {author} {\bibfnamefont {M.}~\bibnamefont
  {Shaposhnikov}}, \bibinfo {author} {\bibfnamefont {A.}~\bibnamefont
  {Shkerin}}, \bibinfo {author} {\bibfnamefont {I.}~\bibnamefont {Timiryasov}},
  \ and\ \bibinfo {author} {\bibfnamefont {S.}~\bibnamefont {Zell}},\ }\href
  {\doibase 10.1103/PhysRevLett.127.169901} {\bibfield  {journal} {\bibinfo
  {journal} {Phys. Rev. Lett.}\ }\textbf {\bibinfo {volume} {126}},\ \bibinfo
  {pages} {161301} (\bibinfo {year} {2021}{\natexlab{b}})},\ \bibinfo {note}
  {[Erratum: Phys.Rev.Lett. 127, 169901 (2021)]},\ \Eprint
  {http://arxiv.org/abs/2008.11686} {arXiv:2008.11686 [hep-ph]} \BibitemShut
  {NoStop}%
\bibitem [{\citenamefont {Figueroa}\ and\ \citenamefont
  {Torrenti}(2017)}]{Figueroa:2017vfa}%
  \BibitemOpen
  \bibfield  {author} {\bibinfo {author} {\bibfnamefont {D.~G.}\ \bibnamefont
  {Figueroa}}\ and\ \bibinfo {author} {\bibfnamefont {F.}~\bibnamefont
  {Torrenti}},\ }\href {\doibase 10.1088/1475-7516/2017/10/057} {\bibfield
  {journal} {\bibinfo  {journal} {JCAP}\ }\textbf {\bibinfo {volume} {10}},\
  \bibinfo {pages} {057} (\bibinfo {year} {2017})},\ \Eprint
  {http://arxiv.org/abs/1707.04533} {arXiv:1707.04533 [astro-ph.CO]}
  \BibitemShut {NoStop}%
\bibitem [{\citenamefont {Aggarwal}\ \emph {et~al.}(2020)\citenamefont
  {Aggarwal} \emph {et~al.}}]{Aggarwal:2020olq}%
  \BibitemOpen
  \bibfield  {author} {\bibinfo {author} {\bibfnamefont {N.}~\bibnamefont
  {Aggarwal}} \emph {et~al.},\ }\href@noop {} {\  (\bibinfo {year} {2020})},\
  \Eprint {http://arxiv.org/abs/2011.12414} {arXiv:2011.12414 [gr-qc]}
  \BibitemShut {NoStop}%
\bibitem [{\citenamefont {Domcke}\ and\ \citenamefont
  {Garcia-Cely}(2021)}]{Domcke:2020yzq}%
  \BibitemOpen
  \bibfield  {author} {\bibinfo {author} {\bibfnamefont {V.}~\bibnamefont
  {Domcke}}\ and\ \bibinfo {author} {\bibfnamefont {C.}~\bibnamefont
  {Garcia-Cely}},\ }\href {\doibase 10.1103/PhysRevLett.126.021104} {\bibfield
  {journal} {\bibinfo  {journal} {Phys. Rev. Lett.}\ }\textbf {\bibinfo
  {volume} {126}},\ \bibinfo {pages} {021104} (\bibinfo {year} {2021})},\
  \Eprint {http://arxiv.org/abs/2006.01161} {arXiv:2006.01161 [astro-ph.CO]}
  \BibitemShut {NoStop}%
\bibitem [{\citenamefont {Cai}\ \emph {et~al.}(2022)\citenamefont {Cai},
  \citenamefont {Guo}, \citenamefont {Ding}, \citenamefont {Fu},\ and\
  \citenamefont {Liu}}]{Cai:2021gju}%
  \BibitemOpen
  \bibfield  {author} {\bibinfo {author} {\bibfnamefont {R.-G.}\ \bibnamefont
  {Cai}}, \bibinfo {author} {\bibfnamefont {Z.-K.}\ \bibnamefont {Guo}},
  \bibinfo {author} {\bibfnamefont {P.-Z.}\ \bibnamefont {Ding}}, \bibinfo
  {author} {\bibfnamefont {C.-J.}\ \bibnamefont {Fu}}, \ and\ \bibinfo {author}
  {\bibfnamefont {J.}~\bibnamefont {Liu}},\ }\href {\doibase
  10.1103/PhysRevD.105.023507} {\bibfield  {journal} {\bibinfo  {journal}
  {Phys. Rev. D}\ }\textbf {\bibinfo {volume} {105}},\ \bibinfo {pages}
  {023507} (\bibinfo {year} {2022})},\ \Eprint
  {http://arxiv.org/abs/2105.00427} {arXiv:2105.00427 [astro-ph.CO]}
  \BibitemShut {NoStop}%
\bibitem [{\citenamefont {Postma}\ and\ \citenamefont {van~de
  Vis}(2017)}]{Postma:2017hbk}%
  \BibitemOpen
  \bibfield  {author} {\bibinfo {author} {\bibfnamefont {M.}~\bibnamefont
  {Postma}}\ and\ \bibinfo {author} {\bibfnamefont {J.}~\bibnamefont {van~de
  Vis}},\ }\href {\doibase 10.1088/1475-7516/2017/05/004} {\bibfield  {journal}
  {\bibinfo  {journal} {JCAP}\ }\textbf {\bibinfo {volume} {05}},\ \bibinfo
  {pages} {004} (\bibinfo {year} {2017})},\ \Eprint
  {http://arxiv.org/abs/1702.07636} {arXiv:1702.07636 [hep-ph]} \BibitemShut
  {NoStop}%
\end{thebibliography}%

\end{document}